%% file: isa_ff.tex
\documentclass[journal=jctcce,manuscript=article]{achemso}
\pdfoutput=1 

\usepackage{caption}
\usepackage{bpchem} 
\usepackage{setspace}
\usepackage{fullpage}
\usepackage{graphicx}
\usepackage{epstopdf}
\usepackage{xspace}
\usepackage{booktabs}
\usepackage{pdflscape}
\usepackage{tablefootnote}
\usepackage[sort&compress,numbers,super]{natbib}
\usepackage{amsmath}
\usepackage{subcaption}
\usepackage{nameref}
\usepackage{xfrac}
\usepackage{multirow}

\newcommand{\super}[1]{\textsuperscript{#1}}

\usepackage[symbol,perpage]{footmisc}

\newcommand{\citeboth}[1]{\citeauthor{#1}\cite{#1}\xspace}
\newcommand*{\citen}{}
\DeclareRobustCommand*{\citen}[1]{%
  \begingroup
    \romannumeral-`\x 
    \setcitestyle{numbers}%
    ref. \cite{#1}%
  \endgroup
}
\newcommand*{\citens}{}
\DeclareRobustCommand*{\citens}[1]{%
  \begingroup
    \romannumeral-`\x 
    \setcitestyle{numbers}%
    refs. \cite{#1}%
  \endgroup
}

\usepackage{letltxmacro}
\LetLtxMacro{\originaleqref}{\eqref}
\renewcommand{\eqref}{eq.~\originaleqref}

\newcommand{\figref}[1]{Figure~\ref{#1}}
\newcommand{\tabref}[1]{Table~\ref{#1}}
\newcommand{\secref}[1]{Section~\ref{#1}}

\newcommand{\isa}{BS-ISA\xspace}
\newcommand{\isaff}{Slater-ISA FF\xspace}
\newcommand{\saptff}{Born-Mayer-IP FF\xspace}
\newcommand{\bmsisaff}{Born-Mayer-sISA FF\xspace}
\newcommand{\ljff}{LJ FF\xspace}
\newcommand{\sapt}{DFT-SAPT (PBE0/AC)\xspace}
\newcommand{\avtz}{aug-cc-pVTZ\xspace}
\newcommand{\A}{\ensuremath{A_{ij}}\xspace}
\newcommand{\B}{\ensuremath{B_{ij}}\xspace}
\newcommand{\C}{\ensuremath{C_{ij,n}}\xspace}
\newcommand{\R}{\ensuremath{r_{ij}}\xspace}

\newcommand{\dhf}{\ensuremath{\delta^{\text{HF}}}\xspace}

\newcommand{\Asr}[1]{\ensuremath{A^{\text{sr}}_{#1}}\xspace}
\newcommand{\Aex}[1]{\ensuremath{A^{\text{exch}}_{#1}}\xspace}
\newcommand{\Ael}[1]{\ensuremath{A^{\text{elst}}_{#1}}\xspace}
\newcommand{\Apen}[1]{\ensuremath{A^{\text{pen}}_{#1}}\xspace}
\newcommand{\Aind}[1]{\ensuremath{A^{\text{ind}}_{#1}}\xspace} 
\newcommand{\Adhf}[1]{\ensuremath{A^{\dhf}_{#1}}\xspace} 

\newcommand{\Bisa}[1]{\ensuremath{B^{\text{ISA}}_{#1}}\xspace}
\newcommand{\Bip}[1]{\ensuremath{B^{\text{IP}}_{#1}}\xspace}

\newcommand{\etot}{\ensuremath{E_{\text{int}}}\xspace}
\newcommand{\erep}{\ensuremath{E^{\text{exch}}}\xspace}
\newcommand{\eelst}{\ensuremath{E^{\text{elst}}}\xspace}
\newcommand{\eind}{\ensuremath{E^{\text{ind}}}\xspace}
\newcommand{\edhf}{\ensuremath{E^{\dhf}}\xspace}
\newcommand{\edisp}{\ensuremath{E^{\text{disp}}}\xspace}

\newcommand{\vtot}{\ensuremath{V_{\text{FF}}}\xspace}
\newcommand{\vrep}{\ensuremath{V^{\text{exch}}}\xspace}
\newcommand{\vcp}{\ensuremath{V^{\text{pen}}}\xspace}
\newcommand{\vsrind}{\ensuremath{V^{\text{ind,sr}}}\xspace}

\newcommand{\velst}{\ensuremath{V^{\text{elst}}}\xspace}
\newcommand{\vind}{\ensuremath{V^{\text{ind}}}\xspace}
\newcommand{\vdhf}{\ensuremath{V^{\dhf}}\xspace}
\newcommand{\vdisp}{\ensuremath{V^{\text{disp}}}\xspace}
\newcommand{\vlr}{\ensuremath{V_{lr}}\xspace}
\newcommand{\vmultipole}{\ensuremath{\sum\limits_{tu}Q_t^iT_{tu}Q_u^j}\xspace}
\newcommand{\vdrude}{\ensuremath{V_{\text{shell}}}\xspace}
\newcommand{\vdrudeind}{\ensuremath{V_{\text{shell}}^{(2)}}\xspace}
\newcommand{\vdrudescf}{\ensuremath{V_{\text{shell}}^{(3-\infty)}}\xspace}

\newcommand{\mse}{\ensuremath{\lVert\text{MSE}\rVert}\xspace}

\author{Mary J. Van Vleet}
\affiliation[UW-Madison]
{Theoretical Chemistry Institute and Department of Chemistry, University of
Wisconsin-Madison, Madison, Wisconsin, 53706, United States}
\author{Alston J. Misquitta}
\affiliation[Queen Mary]
{Department of Physics and Astronomy, Queen Mary University of London, London E1 4NS, United Kingdom}
\author{Anthony J. Stone}
\affiliation[Cambridge]
{Department of Chemistry, University of Cambridge, Cambridge CB2 1EW, United Kingdom}
\author{J.R. Schmidt}
\email{schmidt@chem.wisc.edu}
\affiliation[UW-Madison]
{Theoretical Chemistry Institute and Department of Chemistry, University of
Wisconsin-Madison, Madison, Wisconsin, 53706, United States}

\title{
Beyond Born-Mayer: Improved models for short-range repulsion in ab initio force fields \\
    }

\begin{document}
\maketitle
\onehalfspacing

\begin{abstract}
Short-range repulsion within inter-molecular force fields
is conventionally described by either Lennard-Jones (${A}/{r^{12}}$)
or Born-Mayer ($A\exp(-Br)$) forms. Despite their widespread use,
these simple functional forms are often unable to describe the
interaction energy accurately over a broad range of inter-molecular distances, thus
creating challenges in the development of ab initio force fields and
potentially leading to decreased accuracy and transferability.
Herein, we derive a novel short-range functional form based on a simple 
Slater-like model of overlapping atomic densities and an iterated stockholder atom (ISA)
partitioning of the molecular electron density. 
We demonstrate that this Slater-ISA methodology yields a more accurate, 
transferable, and robust description of the short-range interactions 
at minimal additional computational cost compared to standard
Lennard-Jones or Born-Mayer approaches.  
Finally, we show how this methodology can be adapted to yield the standard Born-Mayer
functional form while still retaining many of the advantages
of the Slater-ISA approach.
\end{abstract}

\begin{section}{Introduction}
\label{sec:intro}

\input{introduction.tex}

\end{section}

\begin{section}{Theory}
\label{sec:theory}

\input{theory.tex}

\end{section}

\begin{section}{Computational Methods}
\label{sec:methods}

\input{methods.tex}

\end{section}

\begin{section}{Results and Discussion}
\label{sec:results}

\input{results.tex}

\end{section}

\begin{section}{Conclusions and Recommendations}
\label{sec:conclusions}

\input{conclusions}

\end{section}

\begin{acknowledgement}

This material is based upon work supported by the National Science
Foundation Graduate Research Fellowship under Grant No. DGE-1256259 and 
by Chemical Sciences, Geosciences and Biosciences
Division, Office of Basic Energy Sciences, Office of Science, U.S. Department
of Energy, under award DE-SC0014059.  
J.R.S is a Camille Dreyfus
Teacher-Scholar. M.V.V. thanks Dr. Jesse McDaniel for helpful discussions.
Computational resources were provided in part by National Science Foundation
Grant CHE-0840494 and using the compute resources and assistance of the
UW-Madison Center for High Throughput Computing (CHTC) in the Department of
Computer Sciences. The CHTC is supported by UW-Madison, the Advanced Computing
Initiative, the Wisconsin Alumni Research Foundation, the Wisconsin Institutes
for Discovery, and the National Science Foundation, and is an active member of
the Open Science Grid, which is supported by the National Science Foundation
and the U.S. Department of Energy's Office of Science.
Compuational resources were also provided in part by the UW Madison Chemistry
Department cluster Phoenix under grant number CHE-0840494.

\end{acknowledgement}

\begin{suppinfo}
Waldman-Hagler analysis of tested \B combination rule(s).
Force field fit quality for the `exact' overlap model. 
Extrapolation algorithm for ISA exponents.
Non-polarizable, point-charge Lennard-Jones Force Fields.
Scale factor tests for Born-Mayer-sISA.
Force field fit qualities and comparisons for Slater-OPT and Born-Mayer-OPT
force fields.
Geometries, ionization potentials, and HOMO values for each monomer species.
Force field parameters for homomonomeric systems for each \isaff, \saptff, and
\bmsisaff. 
Force field fit quality results for homomonomeric systems.
Force field accuracy tests for \ljff.
Weighting function parameter robustness tests for the argon dimer for the
\isaff and the \saptff.
Weighting function parameter robustness tests for the ethane dimer for the \ljff.
\end{suppinfo}

\singlespacing

\renewcommand{\baselinestretch}{1}

\bibliography{library_hardcopy}

\begin{section}{TOC Graphic}
\includegraphics[width=0.9\textwidth]{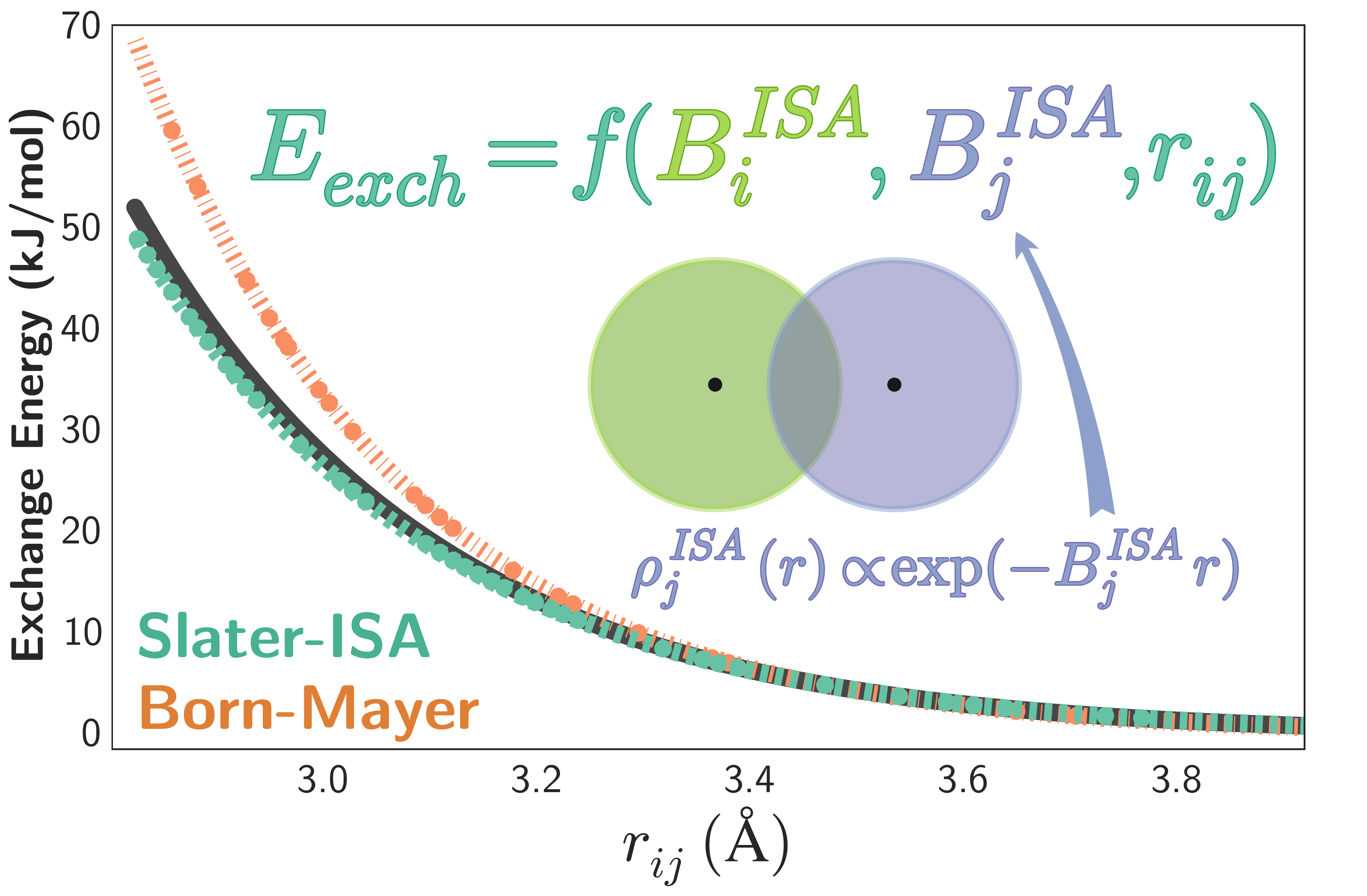}
\end{section}

\end{document}

%% file: introduction.tex
Molecular simulation is an essential tool for interpreting and predicting
the structure, thermodynamics, and dynamics of chemical and
biochemical systems.  The fundamental inputs into these simulations are the
intra- and intermolecular force fields, which provide simple and
computationally efficient descriptions of molecular interactions.
Consequently, the predictive and explanatory power of molecular simulations
depends on the fidelity of the force field to the underlying (exact) potential
energy surface.

In the case of intermolecular interactions, the dominant contributions for
non-reactive systems can be decomposed into the following
physically-meaningful
energy components: electrostatic, exchange-repulsion, induction and dispersion. 
\cite{stone2013theory,margenau1969theory,Riley2010,Stone2007,Dykstra2000}
At large intermolecular distances, where monomer electron overlap can be
neglected, the physics of intermolecular interactions can be described
entirely on the basis of monomer properties (e.g. multipole moments,
polarizabilities), all of which can be calculated with high accuracy from
first principles. 
\cite{Stone1984}
In conjunction with associated distribution schemes that decompose 
molecular monomer properties into atomic contributions,
\cite{stone2013theory,Stone2007,Williams2003,Misquitta2006,Dehez2001,Stone2005,Misquitta2014}
these monomer properties lead to an accurate and computationally efficient
model of `long-range' intermolecular interactions as a sum of atom-atom terms, which
can be straightforwardly included in common molecular simulation packages.

At shorter separations, where the molecular electron density
overlap cannot be neglected, the asymptotic description of
intermolecular interactions breaks down due to the influence of Pauli
repulsion, charge penetration and charge transfer.  These effects
can be quantitatively described using modern electronic
structure methods,\cite{Riley2010,Jeziorski1994,Szalewicz2012,Raghavachari1989,Grimme2011} 
but are far more challenging to model accurately using computationally
inexpensive force fields.
For efficiency and ease of parameterization, most simple force fields use a single
`repulsive' term to model the cumulative influence of (chemically distinct)
short-range interactions. 
These simple models have seen comparatively little progress over the past
eighty years, and the Lennard-Jones\cite{Lennard-Jones1931} (${A}/{r^{12}}$)
and Born-Mayer \cite{Born1932,Buckingham1938}
($A\exp(-Br)$) forms continue as popular descriptions of
short-range effects in standard force fields despite some well-known
limitations (\emph{vide infra}).  

Because the prediction of physical and chemical properties depends on
the choice of short-range interaction model,
\cite{Nezbeda2005,
Galliero2008,Gordon2006,Ruckenstein1997,Galliero2007,Wu2000,Errington1998,McGrath2010,
Parker2015,Sherrill2009,Zgarbova2010,
Bastea2003,Errington1999,Ross1980}
it is essential to develop sufficiently accurate short-range force fields.
This is particularly true in the case of ab initio force field development.
A principle goal
of such a first-principles approach is the reproduction of a calculated
potential energy surface (PES), thus (ideally) yielding accurate predictions
of bulk properties.
\cite{Schmidt2015}
 Substantial deviations between a fitted and calculated PES
lead to non-trivial challenges in the parameterization process, which
in turn can often degrade the quality of property predictions.
The challenge of reproducing an ab initio PES becomes particularly
pronounced at short inter-molecular separations, where many common force field
functional forms are insufficiently accurate.  For example, the popular
Lennard-Jones ($A/{r^{12}}$) functional form is well-known to be substantially
too repulsive at short contacts as compared to the exact
potential.
\cite{Abrahamson1963,Mackerell2004,Parker2015,Sherrill2009,Zgarbova2010}
While the Born-Mayer ($A\exp(-Br)$) functional form is more
physically-justified\cite{Buckingham1938} and fares
better in this regard,\cite{Abrahamson1963} substantial deviations often
persist.\cite{Halgren1992} In addition, parameterization of the Born-Mayer form
is complicated by the strong coupling of the pre-exponential ($A$) and exponent
($B$) parameters, hindering the transferability of the resulting force field.
These considerations, along with the observed sensitivity of
structural and dynamic properties to the treatment of short-range
repulsion,\cite{Nezbeda2005} highlight the need for new approaches
to model short-range repulsive interactions.

Our primary goal in this work is to derive a simple and accurate description
of short-range interactions in molecular systems that improves upon both the
standard Lennard-Jones and Born-Mayer potentials in terms of accuracy,
transferability, and ease of parameterization. Our focus in this work is on ab
initio force field development, and thus we will use the
fidelity of a given force field with respect to an accurate ab initio PES as a
principle metric of force field quality. We note
that other metrics may be more appropriate for the development of empirical
potentials, where Lennard-Jones or Born-Mayer forms may yield highly accurate
`effective' potentials when parameterized against select bulk properties.
Nonetheless, we anticipate that the models proposed in this work may prove
useful for empirical force field development in cases where a more
physically-motivated functional form is necessary.
\cite{Parker2015,Sherrill2009,Zgarbova2010}

The outline of this paper is thus as follows: first, we derive a new functional form
capable of describing short-range repulsion from first principles, and show how the
standard Born-Mayer form follows as an approximation to this more exact model.
Our generalization of the Born-Mayer functional form allows for an
improved description of a variety of short-range effects,
namely electrostatic charge penetration, exchange-repulsion, and density
overlap effects on induction and dispersion. Crucially, we also demonstrate how the associated
atomic exponents can be extracted from
first-principles monomer charge densities via an iterated stockholder atoms
(ISA) density partitioning scheme, thereby reducing the number of required
fitting parameters compared to the Born-Mayer model.
Benchmarking this `Slater-ISA' methodology (functional form and atomic
exponents) against high-level ab initio
calculations and experiment, we find that the approach exhibits increased
accuracy, transferability, and robustness as compared to a typical
Lennard-Jones or Born-Mayer potential.
In addition, we show how the ISA-derived exponents
can be adapted for use within the standard Born-Mayer form (Born-Mayer-sISA),
while still retaining retaining many of the advantages of the Slater-ISA
approach. As such, our methodology also offers an opportunity to dramatically
simplify the development of both empirically-parameterized and ab initio
simulation potentials based upon the standard Born-Mayer form.

%% file: theory.tex




We begin with a formal treatment of the overlap model for the
exchange-repulsion between two isolated atoms, and then extend 
these results to develop a generalized model for the short-range 
interactions in both atomic and molecular systems. 
Finally, we show how the conventional Born-Mayer model can be derived as an
approximation to this more rigorous treatment.

\begin{subsection}{Models for the exchange-repulsion between isolated atoms}
\label{sec:homoatomic_vrep}

It is well known that the exchange-repulsion interaction between two
closed-shell atoms $i$ and $j$ is proportional, or very nearly proportional, 
to the overlap of their respective charge densities:
\cite{Kim1981}
\begin{gather}
\label{eq:atomic_vrep}
\erep_{ij} \approx \vrep_{ij} = K_{ij}(S_{\rho}^{ij})^{\gamma} \\
\label{eq:overlap}
S^{ij}_{\rho} = \int \rho_i(\mathbf{r}) \rho_j(\mathbf{r}) d^3\mathbf{r}.
\end{gather}
Here and throughout, we use $E$ to denote quantum mechanical energies,
and $V$ to denote the corresponding model/force field energies.
Recently two of us have
provided a theoretical justification for this repulsion hypothesis (or overlap
model), and have shown that
$\gamma=1$ provided that asymptotically-correct densities are used to
compute both the atomic densities and $\erep_{ij}$.\cite{Stone2007,Misquitta2015a} 
As this is the case for the calculations in this paper, we assume $\gamma=1$
throughout this work.

The overlap model has frequently been utilized in the literature
and has been found to yield essentially quantitative accuracy for a wide variety of
chemical systems.\cite{Kim1981,Nyeland1986,Ihm1990} 
Prior work exploiting the overlap model has generally followed
one of two strategies. Striving for quantitative accuracy, several
groups have developed approaches to evaluate \eqref{eq:overlap}
via either numerical integration or density fitting of ab-initio molecular
electron densities, $\rho_{i}$ (e.g. SIBFA, GEM, effective fragment potentials).
\cite{Duke2014,Elking2010,Cisneros2006a,Chaudret2014,Chaudret2013,Ohrn2016,
Gresh2007,Gordon2001,Xie2007,Xie2009}
These force fields, while often extremely accurate, lack the simple closed-form analytical expressions that
define standard force fields (such as the Lennard-Jones or Born-Mayer models) and thus are often
much more computationally expensive than conventional models.

In contrast, and similar to our objectives, the overlap model has also been used in the development of standard
force fields. In this case, the molecular electron density as well as the overlap
itself is drastically simplified in order to yield a simple closed-form
expression that can be used within a conventional molecular simulation package.
\cite{Kim1981,Nyeland1986,Ihm1990} 
As we show below, the Born-Mayer model can be `derived' via such an approach.  At the expense of some accuracy, the resulting
overlap-based force fields exhibit high computational efficiency and employ well-known functional forms.

Building on this prior work, our present goal
is to derive rigorous analytical expressions and improved approximations for
both $\rho_{i}$ and \eqref{eq:overlap}, facilitating the construction of ab initio force fields
that exhibit simplicity, high computational efficiency, fidelity to the underlying PES, and (with only trivial modifications) compatibility with standard simulation packages.
We first start with the case of isolated atoms, where it is well-known that the
atomic electron density decays asymptotically as
\begin{align}
\rho_{r \to \infty}(r) \propto r^{2\beta} e^{-2\alpha r}
\end{align}
where the exponent $\alpha = \sqrt{2I}$ is fixed by the vertical ionization potential
$I$, $\beta= -1 + \frac{Q}{\alpha}$, 
and $Q=Z-N+1$
for an
atom with nuclear charge $+Z$ and electronic charge $-N$.
\cite{PatilT-AsymptoticMethods,Hoffmann-Ostenhof1977,Amovilli2006,Misquitta2015a}
The exponential term dominates the asymptotic form of the density,
and the $r$-dependent prefactor may be neglected \cite{Nyeland1986, Ihm1990, Bunge1986, Misquitta2014}.
In this case, the density takes the even simpler form
\begin{align}
\label{eq:rho_tail}
\rho_{r \to \infty}(r) \approx De^{-B r},
\end{align}
where $D$ is a constant that effectively absorbs the 
missing $r$-dependent pre-factor and $B$ is an exponent that is now only approximately equal to 
$2\alpha$. 

In the case of two identical atoms, 
substitution into \eqref{eq:overlap} yields a simple expression for the density overlap, $S_{\rho}$,
\cite{Tai1986,Rosen1931}
\begin{align}
\label{eq:simple_overlap}
\begin{split}
S^{ii}_{\rho} = \frac{\pi D^2}{ B^{3}} P(B r_{ii}) \exp(-B r_{ii}) \\
P(B r_{ii}) = \frac13 (B r_{ii})^2 + B r_{ii} + 1 
\end{split}
\end{align}
as well as (via \eqref{eq:atomic_vrep}) the exchange-repulsion energy\cite{Ihm1990,Rappe1992}:
\begin{align}
\label{eq:homoatomic_vrep}
\vrep_{ii} = \Aex{ii} P(B r_{ii}) \exp(-B r_{ii}).
\end{align}
Here, $r_{ii}$ represents an interatomic distance, and $\Aex{ii}$
indicates a proportionality constant that is typically fit to calculated
values of the exchange-repulsion energy. The only approximations thus far
are the use of the overlap model and the simplified asymptotic form of 
the atomic charge density.
%


For the general case of two hetero-atoms, substitution of
\eqref{eq:rho_tail} into \eqref{eq:overlap} yields the more complicated expression
\cite{Tai1986,Rosen1931}
\begin{align}
\label{eq:complicated_overlap}
\begin{split}
S^{ij}_{\rho} = &
\frac{16\pi D_i D_j \exp(-\{B_i + B_j\}r_{ij}/2)}{(B_i^2-B_j^2)^3r_{ij}}
\times \\
\Bigg [ &
\left(\frac{B_i - B_j}{2}\right)^2 
\bigg(\exp \left(\{B_i-B_j\}\frac{r_{ij}}{2}\right) - \exp \left(-\{B_i-B_j\}\frac{r_{ij}}{2}\right) \bigg) \\
& \qquad \times \left( \left(\frac{B_i + B_j}{2}\right)^2r_{ij}^2 + (B_i + B_j)r_{ij} + 2 \right) \\
& - \left(\frac{B_i + B_j}{2}\right)^2 \exp \left(\{B_i-B_j\}\frac{r_{ij}}{2}\right)
\times \left( \left(\frac{B_i - B_j}{2}\right)^2r_{ij}^2 - (B_i - B_j)r_{ij} + 2 \right) \\
& + \left(\frac{B_i + B_j}{2}\right)^2 \exp \left(-\{B_i-B_j\}\frac{r_{ij}}{2}\right)
\times \left( \left(\frac{B_i - B_j}{2}\right)^2r_{ij}^2 + (B_i - B_j)r_{ij} + 2 \right)
\Bigg ],
\end{split}
\end{align}
which is too
cumbersome to serve as a practical force field functional form.
However, since the above expression reduces to
\eqref{eq:simple_overlap} in the limit $B_i = B_j$, 
and because $|B_i - B_j|$ is small for most atom pairs,
we have found that \eqref{eq:complicated_overlap} may be approximated
using \eqref{eq:simple_overlap} with an \emph{effective} atomic exponent $B$. 
An expansion of \eqref{eq:complicated_overlap} about $B_i = B_j$
suggests that this effective exponent should be given by the arithmetic mean,
$\B = \frac12 (B_i + B_j)$. However, a Waldman-Hagler style analysis \cite{Waldman1993}
(see Supporting Information) suggests instead that a more suitable
exponent is given by the geometric mean combination rule,
\begin{align}
\label{eq:isaff_bij}
B = \B \equiv \sqrt{B_iB_j}.
\end{align}
As shown in the Supporting Information, this approximate overlap model (\eqref{eq:simple_overlap} and
\eqref{eq:isaff_bij}) is of comparable accuracy to the exact overlap from 
\eqref{eq:complicated_overlap}. 
Thus the density overlap and (force field) exchange energies of arbitrary hetero-atoms
take the simple forms
\begin{align}
\label{eq:isaff_overlap}
S^{ij}_{\rho} &= D_{ij} P(B_{ij}, r_{ij}) \exp(-B_{ij}r_{ij}) \\
D_{ij} &= \pi D_i D_j B_{ij}^{-3} \\
P(B_{ij},r_{ij}) &= \frac13 (B_{ij} r_{ij})^2 + B_{ij} r_{ij} + 1
\end{align}
and
\begin{align}
\label{eq:isaff_vrep}
\vrep_{ij} = A^{exch}_{ij} P(B_{ij} r_{ij}) \exp(-B_{ij}r_{ij}).
\end{align}
Due to the connection with the overlap between two s-type Slater orbitals, we refer
to \eqref{eq:isaff_vrep} as the Slater functional form. Note that this expression reduces
to the standard Born-Mayer function by making the further approximation $P(B_{ij}
r_{ij}) = 1$, although it is known\cite{Ihm1990, McDaniel2012} that 
this is a poor approximation with the $B_{ij}$ as defined above.
Instead, as we shall demonstrate in \secref{sec:results}, the
exponents \B need to be scaled for accurate use with a Born-Mayer functional form.

Variants of the polynomial
pre-factor from \eqref{eq:isaff_overlap} have previously been recognized and
used in intermolecular interaction models.
\cite{Buckingham1938,McDaniel2012,Rappe1992}
Early work by \citeboth{Buckingham1938} hypothesized that the functional form
of \eqref{eq:isaff_vrep} would be more accurate than the Born-Mayer
form, though no attempt was made to provide a closed-form expression for
$P$. More recent potentials have incorporated a low-order
polynomial into the exchange repulsion term, either by direct parameterization
\cite{Podeszwa2006, Bukowski2006, Jeziorska2007, Sum2002, Konieczny2015} 
or indirectly by fitting the exchange to ${S_{\rho}}/{r^2}$ rather than
to $S_{\rho}$ itself.
\cite{Kita1976a,Nyeland1986,Ihm1990}
\citeauthor{Kita1976a} have derived (but not tested) \eqref{eq:homoatomic_vrep} for
the homoatomic case.
\cite{Kita1976a}
Recently, and most similar to the spirit of the present work,
York and co-workers have derived a model based upon the overlap of Slater-type orbitals
for use in QM/MM simulations, yielding an expression identical to
\eqref{eq:complicated_overlap}.
\cite{Kuechler2015,Giese2007,Giese2013} 
Those authors treated $D_i$ and $D_j$ as empirical fitting
parameters and estimated atomic exponents ($B_i$ and $B_j$) via atomic-charge dependent functions.
In contrast, we will demonstrate that utilization of the far simpler
functional form from \eqref{eq:isaff_vrep},
in conjunction with exponents calculated from analysis of the first-principles molecular electron density,
yields much higher computational efficiency and simplifies the
parameterization process without significant loss of accuracy.


For an arbitrary pair of interacting atoms, $\Aex{ij}$ can be obtained by
fitting to calculated exchange-repulsion energies. However, assuming that the
overlap proportionality factor $K_{ij}$ is a universal
constant (or, alternatively, separable 
with $K_{ij} = K_iK_j$), then 
\begin{align}
\label{eq:isaff_aij}
\Aex{ij} = \left(K_i\sqrt{\frac{\pi}{B_i^3}}D_i\right)
\left(K_j\sqrt{\frac{\pi} {B_j^3}} D_j\right) \equiv \Aex{i}\Aex{j},
\end{align}
thus providing a combination rule for heteroatomic interaction in
terms of purely atomic quantities. The universality and separability
of $K_{ij}$ are, at present, empirically rather than theoretically justified.
\cite{Day2003,Stone2007,Nobeli1998}
The $\Aex{i}$ can then be obtained, for
example, by a straightforward fitting of calculated ab initio
\emph{homoatomic} exchange-repulsion energies.

\end{subsection}
\begin{subsection}{Models for other short-range interactions between isolated atoms}
\label{sec:heteroatomic_vsr}

Beyond the exchange-repulsion, the density-overlap model may also be used
to model other short-range interaction components, such as the electrostatic
charge penetration energy and the short-range induction and dispersion
energies (that is, the portion modulated by charge overlap).  Indeed, it has 
been demonstrated that the electrostatic charge penetration energy is
approximately proportional to the exchange-repulsion energy, and consequently
to the charge density overlap,
\cite{Stone2007,Misquitta2014}
which has provided a successful basis for modeling the electrostatic charge
penetration energy.
\cite{McDaniel2013,Totton2010}
While the relation between short-range induction and charge overlap
is less clear, 
recent results have demonstrated that the charge-transfer energy, which is the
dominant short-range component of the induction energy,\cite{Misquitta2013} is
approximately proportional to the first-order exchange energy,
\cite{Misquitta2015a,Misquitta2015b}
and prior work has successfully used the overlap hypothesis to
describe the short-range induction.
\cite{Stone2007,Totton2010,McDaniel2013}
We therefore model the electrostatic charge penetration and short-range 
induction interactions as
\begin{align}
\vcp_{ij} &= \Apen{ij} P(B_{ij}, r_{ij}) \exp(-B_{ij}r_{ij}) \\
\intertext{and}
\vsrind_{ij} &= \Aind{ij} P(B_{ij}, r_{ij}) \exp(-B_{ij}r_{ij}).
\end{align}
Aside from the pre-factors \A, these expressions are
identical to that for the exchange-repulsion term.

The behavior of the dispersion interaction at short distances poses a special
challenge. In order to model the short-range dispersion and to resolve the 
unphysical, mathematical divergence of the ${1}/{r^n}$ terms
as $r \rightarrow 0$, \citeauthor{Tang1984} have shown
that the terms in the dispersion expansion should be damped using an 
appropriate incomplete gamma function 
\begin{align}
\label{eq:ttdamp}
f_n(x) &= 1 - e^{-x} \sum \limits_{k=0}^n \frac{(x)^k}{k!} \\
x &= -\frac{d}{dr}\left[\ln \vrep(r)\right] \ r
\end{align}
that accounts for both exchange and charge penetration effects. 
\cite{Tang1984,Tang1992}
Note that the form of this damping factor depends on the model used for
exchange repulsion.
For the Slater functional form (\eqref{eq:isaff_vrep}),
\begin{align}
\label{eq:isaff_ttdamp}
x_{\text{Slater}} &= B_{ij}r_{ij} - \frac{2 B_{ij}^2 r_{ij} + 3 B_{ij} }
{ B_{ij}^2 r_{ij}^2 + 3 B_{ij} r_{ij} + 3} r_{ij}.
\end{align}
Alternatively, if we replace the Slater functional form with the less accurate
Born-Mayer expression, $x$ simplifies to the result originally given by Tang
and Toennies:
\begin{align}
\label{eq:saptff_ttdamp}
x_{\text{Born-Mayer}} &= B_{ij}r_{ij} .
\end{align}

\end{subsection}
\begin{subsection}{Models for short-range interactions between molecules}

The overlap repulsion hypothesis can be extended to molecules
\cite{Stone2007,Wheatley1990,Mitchell2000,Soderhjelm2006,Day2003}
by writing the molecular density $\rho_I$ as a superposition of atomic
densities
\begin{align}
\rho_I(\mathbf{r}) = \sum \limits_{i \in I} \rho_i(\mathbf{r})
\end{align}
where $i$ represents an atom in molecule $I$. In this case, 
\begin{gather}
\vrep_{IJ} = \sum \limits_{i \in I} \sum \limits_{j \in J} \vrep_{ij} \\
\label{eq:molecular_overlap}
\vrep_{ij} = K_{ij}S^{ij}_{\rho} = \int \rho_i(\mathbf{r}) \rho_j(\mathbf{r}) d^3\mathbf{r} .
\end{gather}
Note that the form of \eqref{eq:molecular_overlap} is identical to the
corresponding expression between isolated atoms, but requires partitioning of
the molecular charge density into atom-in-molecule densities, $\rho_i$, each
decaying according to an effective atom-in-molecule density decay exponent,
$B_i$.

In principle, such atom-in-molecule exponents could be estimated from the 
ionization potentials of the corresponding isolated atoms, \cite{McDaniel2013,Rappe1992} 
but this approach neglects the influence of the molecular environment.
A more appealing possibility is to directly evaluate the
atom-in-molecule densities via partitioning of the calculated monomer
densities. Density partitioning has not yet (to our knowledge) been
applied in the context of the overlap model to solve for
\eqref{eq:molecular_overlap}, 
however
several successful efforts in force field development have recently relied on
an atoms-in-molecule approach in order to obtain accurate scaling
relationships for intermolecular force field parameters.\cite{Tkatchenko2012,Tkatchenko2009,Cole2016}
In particular, \citeauthor{Cole2016} utilized a density-derived electrostatic
and chemical (DDEC) partitioning scheme 
\cite{Manz2010,Manz2012}
to generate Lennard-Jones dispersion and short-range repulsion parameters,
though the latter parameters were calculated implicitly by enforcing the
coincidence of the potential minimum and the calculated atomic radius. 

While no unique atom-in-molecule density partitioning scheme exists, an ideal approach
should yield atom-in-molecule densities that strongly resemble those of
isolated atoms, e.g. maximally spherical and asymptotically
exponential.\cite{Yu2011, Levy1984, Misquitta2014, kitaigorodsky2012molecular} 
The recently developed iterated stockholder partitioning of 
\citeauthor{Lillestolen2008} 
obeys this first important constraint of sphericity. 
\cite{Lillestolen2008,Lillestolen2009}
As a non-trivial extension of the original Hirshfeld method,
\cite{Hirshfeld1977}
iterated stockholder atom (ISA) densities are defined as
\begin{align}
\rho_i(\mathbf{r}) = \rho_I(\mathbf{r}) 
\frac{ w_i(\mathbf{r}) }{ \sum \limits_{a \in I} w_a(\mathbf{r}) }
\end{align}
where the converged shape functions $w_i(\mathbf{r})$ are spherical averages of the
atomic densities $\rho_i(\mathbf{r})$:
\begin{align}
w_i(\mathbf{r}) = \langle \rho_i(\mathbf{r}) \rangle_{\text{sph}}.
\end{align}
This formulation ensures, by construction, that the sum of atomic densities
reproduces the overall molecular density. Furthermore, the maximally spherical nature
of the atom-in-molecule densities naturally facilitates a description of short-range
interactions via a simple isotropic site-site model.

\citeauthor{Misquitta2014} have developed a rapidly convergent
implementation of the ISA procedure (\isa\cite{Misquitta2014}) using a basis set expansion
which, in addition to exhibiting good convergence with respect to basis set,
also leads to asymptotically-exponential atomic densities.
Consequently, the \isa method is our preferred density partitioning scheme.
As an example, the spherically-averaged atomic densities for acetone are shown in
\figref{fig:isa}. For simplicity, and because a full treatment
of the anisotropy is beyond the scope of this paper, we subsequently refer to
the spherically-averaged atomic densities (i.e. the shape functions,
$w_i(r)$) as atomic or atom-in-molecule densities.

    \begin{figure}
    \includegraphics[width=0.9\textwidth]{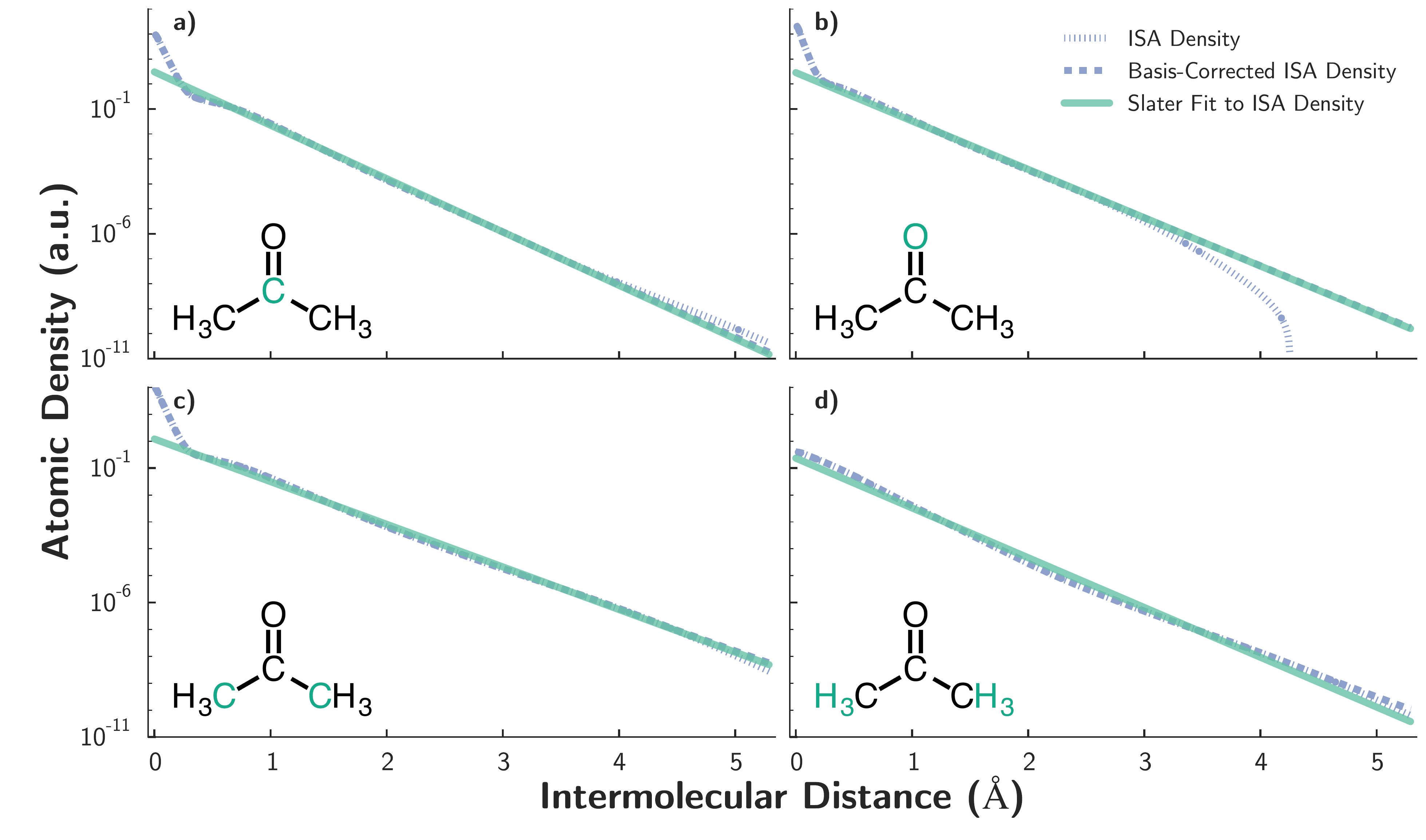}  
    \caption{
        \isa and fitted shape functions for each atom type in acetone: a) carbonyl carbon,
        b) oxygen, c) methyl carbon, d) hydrogen. \isa shape functions (dotted line)
        for each atom type have been obtained at a
        PBE0/aug-cc-pVTZ level of theory. A modified \isa shape
        function (dashed line) corrects the tail-region of the \isa function to
        account for basis set deficiencies in the \isa algorithm. A single Slater
        orbital of the form $D_i^{\text{ISA}}\exp(-\Bisa{i} r)$ (solid line) is fit to the basis-corrected
        \isa shape function, and the obtained $\Bisa{i}$ value is used as an atomic exponent
        in the functional form of \isaff. Results for acetone are typical of
        molecules studied in this work.
    		   }
    \label{fig:isa}
    \end{figure}

From \figref{fig:isa} we see that the ISA atomic shape functions (that is, the
spherically-averaged ISA atoms-in-molecule density) 
decay exponentially outside the core region.
However, note that the exponents governing the spherical density decay, $\Bisa{i}$, 
differ from those of the free atoms. The ISA densities have been observed to account
for electron movement in the molecule, and the consequent density changes
brought about by this movement tend to be manifested in the region of the 
density tails. \cite{Misquitta2014}
The ISA exponents can be obtained by a weighted least-squares fit to the \isa atomic density
(see the \nameref{sec:methods} section for details), with the resulting fitted
atomic densities shown in \figref{fig:isa}. 
Note that even a single exponential is remarkably successful in reproducing the
entirety of the valence atomic density.

Given these fitted ISA exponents, we can now apply
our short-range interaction formalism to polyatomics,
\begin{align}
\label{eq:isaff_sr}
\begin{split}
V^{sr} &= \sum\limits_{ij} \Asr{ij} P(B_{ij}, r_{ij}) \exp(-B_{ij}r_{ij}) \\
P(B_{ij},r_{ij}) &= \frac13 (B_{ij} r_{ij})^2 + B_{ij} r_{ij} + 1 \\
    \Asr{ij} &= \Asr{i}\Asr{j} \\
    \B &= \sqrt{\Bisa{i}\Bisa{j}} \\
\end{split}
\end{align}
where the molecular
short-range energy is now a sum of atom-atom contributions. In conjunction
with appropriately damped atomic dispersion (\eqref{eq:ttdamp} and
\eqref{eq:isaff_ttdamp}), \eqref{eq:isaff_sr}  completely defines our new
short-range force field. We refer to this new functional form and set of atomic
exponents as the \isaff.

\end{subsection}

%% file: methods.tex
To evaluate the \isaff against conventional Born-Mayer and/or Lennard-Jones models, we compare the
ability of each resulting short-range force field to reproduce benchmark ab
initio intermolecular interaction energies for a collection of representative
dimers. Such a metric is directly relevant for ab initio force field
development. Even for an empirically-parameterized force field, however,
fidelity to an accurate ab initio potential should be well correlated with the
highest level of accuracy and transferability achievable with a given
short-range methodology. 

We have developed the \isaff, Born-Mayer, and Lennard-Jones force fields using 
benchmark energies calculated using the symmetry-adapted perturbation theory
based on density-functional theory (DFT-SAPT or SAPT(DFT) 
\cite{Misquitta2002,Misquitta2003,Misquitta2005,Heßelmann2005a,Podeszwa2006a,Heßelmann2002,Heßelmann2003,Heßelmann2002a,Jansen2001}).
DFT-SAPT provides interaction energies that are comparable in accuracy to 
those from CCSD(T) and which are rigorously free from basis set superposition error.
\cite{Podeszwa2005a,Riley2010}
Additionally, at second-order, DFT-SAPT also provides an explicit interaction
energy decomposition into physically-meaningful contributions: 
the electrostatic, exchange-repulsion, induction, and dispersion energies.
This decomposition is vital to the development of models as it allows the
development of separate terms for each type of short-range interaction. 
Terms of third and higher order are estimated using the \dhf correction 
\cite{Jeziorska1987} which contains mainly higher-order induction contributions.
Following prior work,\cite{Yu2011,McDaniel2013} and for the purposes of
fitting to the DFT-SAPT data, we keep the second-order induction term and the
\dhf term separate.

Since the Slater-ISA and Born-Mayer force fields describe only short-range interactions (i.e. those terms which are modulated by
the overlap of the monomer electron densities), they must both be supplemented with additional long-range
terms that describe the electrostatic, polarization, and dispersion
interactions. Here we have chosen a long-range potential of the form 
\begin{align}
\label{eq:elr}
\vlr &= V_{\text{multipole}} + V_{\text{dispersion}} + \vdrude \\
\intertext{where}
V_{\text{multipole}} &= \sum \limits_{ij} \vmultipole
\intertext{includes distributed multipole contributions from each atom up to
quadrupoles,} 
V_{\text{dispersion}} &= - \sum\limits_{ij}\sum\limits_{n=3}^{6} \frac{C_{ij,2n}}{r_{ij}^{2n}} 
\label{eq:dispersion}
\end{align}
describes isotropic dispersion, and 
\vdrude is the polarization energy modeled by Drude oscillators 
\cite{drude1902theory, Lamoureux2003}
as in
\citen{McDaniel2013}.
The accuracy of each of these terms is expected to minimize errors in the
long-range potential, simplifying the comparison between short-range force field functional forms.
Nonetheless, we expect that our results will be qualitatively insensitive to
the particular choice of long-range force field and acknowledge that simpler alternatives may
be preferred for the development of highly efficient simulation potentials.
In the case of the Lennard-Jones force field, we replace \eqref{eq:dispersion} with
the simple $C_{ij,6}/r_{ij}^{6}$ dispersion term that is standard to the
Lennard-Jones model. 

\begin{figure}
      \includegraphics[width=0.6\textwidth]{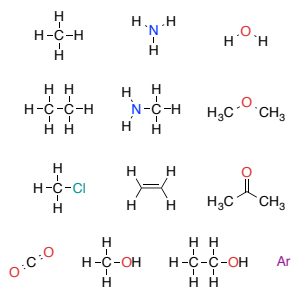}
      \caption{
        The 13 small molecules included in the 91 dimer (13 homomonomeric, 78
        heteromonomeric) test set. Cartesian geometries for all of these
        molecules are given in the Supporting Information.
              }
      \label{fig:molecules}
    \end{figure}

We used a test set consisting of one atom (argon) and 
12 small organic molecules (see \figref{fig:molecules}) from which dimer
potentials could be generated (we will use the term `dimer' to mean two,
potentially dissimilar, interacting molecules or atoms), 
yielding 91 dimer combinations (13 homo-monomeric, 78 hetero-monomeric).
This wide range of systems allowed us to evaluate both the accuracy and
transferability of the Slater-ISA model compared to conventional Born-Mayer
and/or Lennard-Jones models.

A detailed description of this overall methodology is provided below.

\begin{subsection}{Construction of the 91 dimer test set}

Monomer geometries for each of the 13 small molecules were taken from the
experimental NIST
\mbox[CCCBDB] database\cite{Johnson2015NIST} and can be found in the Supporting
Information. For acetone and methyl amine, experimental geometries were
unavailable, and thus
the computational NIST \mbox[CCCBDB] database was used to obtain geometries at
a high level of theory (B3LYP/\avtz for acetone, CCSD(T)/6-311G* for methyl
amine).
For each of the 91 dimers, a training set was constructed
using \sapt interaction energies calculated at 1000 quasi-random dimer
configurations. These configurations were generated using Shoemake's
algorithm,\cite{Shoemake1992} subject to the constraint that the nearest atom pairs
be separated by between 0.75 and 1.3 of the sum of their van der Waals radii. 
This ensured adequate sampling of the potential 
energy surface in the region of the repulsive wall.
The DFT-SAPT interaction energies were evaluated using an asymptotically corrected
PBE0 functional (PBE0/AC) with monomer vertical (first) ionization potentials
computed using the $\Delta$-DFT approach at a PBE0/\avtz level of theory.
Unless otherwise noted, all DFT-SAPT calculations used an \avtz basis set in the 
dimer-centered form with midbond functions (the so-called DC+ form),
and were performed using the MOLPRO2009 software suite.\cite{MOLPRO-WIREs}
The midbond set consisted of a 5s3p1d1f even-tempered basis set with ratios
of 2.5 and centered at $\zeta = 0.5, 0.5, 0.3, 0.3$ for the s,p,d, and f shells,
respectively. This set was placed near the midpoint of the centers of mass of
the two interacting monomers.

A small fraction of DFT-SAPT calculations exhibited unphysical
energies, which were attributed to errors in generating the optimized
effective potential used during the \sapt calculations; these points were
removed from the test set.

\end{subsection}
\begin{subsection}{\isa Calculations}

\isa atomic densities were obtained using CamCASP 5.8
\cite{camcasp5.8, WCMS:WCMS1172, orient4.8}
following the procedure of \citeauthor{Misquitta2014}\cite{Misquitta2014}
For the \isa calculations, an auxiliary basis was constructed from an RI-MP2
\avtz basis set with $s$-functions replaced by the ISA-set2
supplied with the CamCASP program;
CamCASP's ISA-set2 basis was also used for the ISA basis
set.\cite{Misquitta2014} A custom ISA basis set for Ar was used 
(even tempered, $n_{min} = -2, n_{max} = 8$)
\cite{Misquitta2014} 
as no published basis was available.
\isa calculations were performed with the A+DF algorithm, which allows the 
ISA functional to be mixed with some fraction, $\zeta$, of the density-fitting functional.
Following the recommendations of \citeauthor{Misquitta2014}\cite{Misquitta2014},
we have used $\zeta=0.1$ for the multipole moment calculations, 
and $\zeta=0.9$ for the density partitioning used to determine the \B coefficients.

\end{subsection}
\begin{subsection}{Determination of $\Bisa{i}$}

The \isa-derived atomic exponents, $\Bisa{i}$, were obtained from a weighted
least-squares fit to the spherically averaged \isa atomic densities (shape functions),
$w_i(\mathbf{r})$.
In some cases, numerical instabilities and basis-set limitations of the \isa procedure
yielded densities that exhibited non-exponential asymptotic behavior.\cite{Misquitta2014}
To correct for these unphysical densities, we extrapolated the exponential decay
of the valence region to describe the \isa tails also. 
Details of this procedure can be found in the Supporting Information.
The ISA atom-in-molecule exponents were then derived via a log-weighted fit to
the tail-corrected shape-functions $w^a(\mathbf{r})$ for densities within the cutoff
$10^{-2} > w^a > 10^{-20}$ a.u.  This region was chosen to
reproduce the charge density most accurately in the valence regimes most likely to be relevant to
intermolecular interactions.

\end{subsection}
\begin{subsection}{Force Field Functional Forms and Parameterization}
    \label{sec:FF-forms}

The general structure of the force fields \vtot for both the \isaff and the 
Born-Mayer-type models are given by the following equations:

\begin{align}
\begin{split}
\label{eq:bothff}
\vtot &= \sum\limits_{ij} \vrep_{ij} + \velst_{ij} + \vind_{ij} + \vdhf_{ij} +
\vdisp_{ij} \\[10pt]
\vrep_{ij} &= \Aex{ij} P(B_{ij}, r_{ij}) \exp(-B_{ij}r_{ij}) \\
\velst_{ij} &= -\Ael{ij} P(B_{ij}, r_{ij}) \exp(-B_{ij}r_{ij}) + \vmultipole \\
\vind_{ij} &= -\Aind{ij} P(B_{ij}, r_{ij}) \exp(-B_{ij}r_{ij}) + \vdrudeind \\
\vdhf_{ij} &= -\Adhf{ij} P(B_{ij}, r_{ij}) \exp(-B_{ij}r_{ij}) +
\vdrudescf \\
\vdisp_{ij} &= - \sum\limits_{n=3}^{6} f_{2n}(x) \frac{C_{ij,2n}}{r_{ij}^{2n}} \\
\A &= A_iA_j \\
\C &= \sqrt{C_{i,n}C_{j,n}} \\
f_{2n}(x) &= 1 - e^{-x} \sum \limits_{k=0}^{2n} \frac{(x)^k}{k!} \\
\end{split}
\intertext{For the \isaff:}
\begin{split}
\label{eq:isaff}
B_i &= \Bisa{i} \\
\B &= \sqrt{B_iB_j} \\
P(B_{ij},r_{ij}) &= \frac13 (B_{ij} r_{ij})^2 + B_{ij} r_{ij} + 1 \\
x &= B_{ij}r_{ij} - \frac{2 B_{ij}^2 r_{ij} + 3 B_{ij} }
{ B_{ij}^2 r_{ij}^2 + 3 B_{ij} r_{ij} + 3} r_{ij}
\end{split}
\intertext{For all Born-Mayer type models:}
\begin{split}
P(B_{ij},r_{ij}) &= 1 \\
x &= B_{ij}r_{ij}
\end{split}
\intertext{For the \saptff:}
\begin{split}
\label{eq:saptff}
B_i &\equiv \Bip{i} = 2\sqrt{2I_i} \\
\B &= \frac{B_iB_j(B_i + B_j)}{B_i^2 + B_j^2}
\end{split}
\intertext{For the \bmsisaff:}
\begin{split}
    B_i &= 0.84\Bisa{i} \\
    \B &= \sqrt{B_iB_j}
\end{split}
\end{align}

Of the parameters in these force fields, only the coefficients $A_i$ were fit
to reproduce DFT-SAPT dimer energies (details below). All other force field
parameters were derived from first-principles atom or atom-in-molecule properties.
Exponents for the \isaff and the \bmsisaff were derived from \isa calculations,
while exponents for the \saptff were determined from vertical ionization potentials
of the isolated atoms.
Dispersion coefficients (\C) were either used directly from \citen{McDaniel2013} or
were parameterized using analogous methods in the case of argon.
Distributed multipoles $Q_t^i$ for each system were obtained from the \isa-based
distributed multipoles scheme (ISA-DMA) \cite{Misquitta2014}, with the expansion
truncated to rank 2 (quadrupole). 
Note that here, $t=00,10,\dots,22s$ denotes the rank of the multipole in 
the compact notation of \citeboth{stone2013theory}.
(In addition to rank 2 ISA-DMA multipoles, we also tested the use of DMA4
multipoles\cite{Stone2005} 
as well as the use of rank 0 charges obtained from the 
rank truncation or transformation\cite{Ferenczy1997} of either ISA-DMA or DMA4
multipoles; the effect of including a Tang-Toennies damping
factor\cite{McDaniel2013,Tang1984} was studied in all cases.
Each of these alternative long-range electrostatic models proved either
comparably or less accurate for both the \isaff and the \saptff in terms of their
ability to reproduce the DFT-SAPT electrostatic energy, and are not discussed
further.)
Long-range polarization ($V_{shell}$) was modeled using Drude oscillators in a manner
identical to \citen{McDaniel2013}. As in our prior work, during parameterization,
the Drude energy was partitioned into 2\super{nd} (\vdrudeind) and 
higher order (\vdrudescf) contributions, where \vdrudeind is the Drude oscillator
energy due to static charges (excluding intra-molecular contributions), and
\vdrudescf is the difference between the fully converged Drude energy,
$V_{shell}$, and \vdrudeind.  Force field parameters for all homo-monomeric
systems are located in the Supporting Information. 

A weighted least-squares fitting procedure was used to fit $A_i$
parameters to the benchmark \sapt interaction energies on a 
component-by-component basis.
That is, four separate optimizations\cite{McDaniel2013}  were
performed to directly fit \vrep, \velst, \vind, and \vdhf to, respectively, the
following DFT-SAPT quantities (notation as in \citen{Heßelmann2005a}):
\begin{align}
\begin{split}
\erep &\equiv E^{(1)}_{\text{exch}} \\
\eelst &\equiv E^{(1)}_{\text{pol}} \\
\eind &\equiv E^{(2)}_{\text{ind}} + E^{(2)}_{\text{ind-exch}} \\
\edhf &\equiv \delta(\text{HF}).
\end{split}
\intertext{
For \vdisp, no parameters were directly fit to the DFT-SAPT dispersion,
}
\edisp &\equiv E^{(2)}_{\text{disp}} + E^{(2)}_{\text{disp-exch}},
\end{align}
but were instead obtained solely from monomer properties as described above.
Finally, note that no parameters were directly fit to the total DFT-SAPT
energy,
\begin{align}
\etot = \erep + \eelst + \eind + \edhf + \edisp,
\end{align}
for either the \isaff or the \saptff. Rather, \vtot was calculated according to \eqref{eq:bothff}.

Data points for each fit were weighted using a Fermi-Dirac functional form
given by
\begin{align}
\label{eq:weighting-function}
w_i = \frac{1}{\exp((-E_i - \mu_{\text{eff}})/kT) + 1},
\end{align}
where $E_i$ is the reference energy and $\mu_{\text{eff}}$ and $kT$ were treated as
adjustable parameters. The parameter $kT$, which sets the energy scale for the
weighting function, was taken to be $kT = \lambda |E_{\text{min}}|$; here $E_{\text{min}}$
is an estimate of the global minimum well depth. Unless otherwise stated, we have used
$\lambda = 2.0$ and $\mu_{\text{eff}} = 0.0$.   
These defaults were chosen to minimize overall average attractive RMSE for all 91 dimer
sets. 
Increases or decreases in the $\lambda$ factor correspond to the weighting of
more or fewer repulsive configurations, respectively.  

In the case of Lennard-Jones, the standard Lennard-Jones functional form was
used for the van der Waals terms, with Coulomb and polarization terms modeled
exactly as for the \isaff:
\begin{align}
\vtot^{\text{LJ}} &= \sum\limits_{ij} \frac{A_{ij}}{r^{12}_{ij}} - \frac{C_{ij,6}}{r^{6}_{ij}} + \vdrude + \vmultipole
\end{align}
Lorentz-Berthelot combination rules were used to obtain heteroatomic $A_{ij}$
and $C_{ij}$ parameters. Unlike with the \isaff and Born-Mayer models, $\vtot^{\text{LJ}}$ was fit to
the total \sapt energy, with $A_{ij}$ and $C_{ij,6}$ as fitting parameters. The
weighting function from \eqref{eq:weighting-function} was used in fitting.


\end{subsection}
\begin{subsection}{Potential Energy Surface Scans}

In order to visually assess fit quality, representative one-dimensional scans
of the potential energy surface were calculated for several dimer pairs along
low-energy dimer orientations. For each dimer pair, the minimum energy
configuration of the 1000 random dimer points was selected as a starting
configuration, and additional dimer configurations (not necessarily included in the
original 1000 points) were generated by scanning along some bond vector. In the case of the
ethane dimer, two carbon atoms (one on each monomer) were used; for acetone,
the carbonyl carbon on each monomer defined the bond vector.

\end{subsection}

\begin{subsection}{Molecular Simulations}

All bulk simulations were run using OpenMM release version 7.0.
\cite{Eastman2013}
Enthalpies of vaporization were computed from 
\begin{align*}
\Delta H_{\text{vap}} = (E_{\text{pot}}(g) + RT) - E_{\text{pot}}(l)
\end{align*}
where $E_{\text{pot}}(g)$ and $E_{\text{pot}}(l)$ were determined from NVT
simulations at the experimental gas and liquid densities, respectively.
Calculated liquid
densities were determined from NPT simulations. In all cases, the OPLS/AA
force field was used for the intramolecular potential.
\cite{Jorgensen1996}
All simulations used a Langevin integrator with a 0.5 fs time step and a 1
ps$^{-1}$ friction coefficient; NPT simulations used a Monte Carlo barostat
with a trial volume step every 5\super{th} move. Periodic boundary conditions,
particle-mesh Ewald, and a non-bonding cutoff of 1.2nm with added long-range
corrections were used to simulate a
unit cell of 222 molecules. After an equilibration period
of at least 600ps, simulation data was gathered from production runs lasting
at least 200ns.

\end{subsection}

%% file: results.tex

The Slater-ISA methodology for short-range intermolecular interactions has
been derived from a simple but rigorous physical model of overlapping
monomer electron densities. In practice, this approach differs from the
conventional Born-Mayer approach in both the choice of the short-range
functional form (with the latter omitting the polynomial pre-factor) and the
source of the exponents (with the former derived from ISA analysis of the
monomer density). Our principal goal is to examine the influence of these
modifications on the accuracy and transferability of the resulting force
fields.

We initially benchmark the \isaff against a conventional Born-Mayer potential,
\saptff.
The latter approach has been used extensively in prior
work,\cite{Schmidt2015,McDaniel2013} and both approaches use identical numbers of
fitted parameters.  Following prior work, combination rules for the \saptff are as in
\citen{McDaniel2013}. (We have tested the effect of using a geometric mean
for the \saptff; results do not differ qualitatively from those presented
below.) Owing to its popularity, we also compare the \isaff to a Lennard-Jones
functional form (\ljff).

We first assess the accuracy of the \isaff, \saptff, and \ljff against benchmark
ab initio intermolecular interaction energies and experimental 2\super{nd}
virial coefficients, enthalpies of vaporization, and liquid densities. 
We next examine parameter transferability, assessing the
extent to which parameters from pure homo-monomeric systems can be re-used
(without further optimization) to describe mixed interactions. To assess 
parameter robustness, we also study the sensitivity of each methodology to
changes in the weighting function (\eqref{eq:weighting-function}). Finally, we
explore the application of \isa-derived exponents within the Born-Mayer
functional form as a straightforward method for simplifying the
parameterization (and potentially increasing the accuracy) of a wide variety of
standard ab initio and empirically-parameterized force fields.

\begin{subsection}{Accuracy: Comparison with DFT-SAPT}

For each of the 91 molecule pairs described in the \nameref{sec:methods} section, 
parameters for the \isaff, \saptff, and \ljff
were fit to reproduce \sapt interaction energies calculated for a set of 1000 dimer
configurations. These 91,000 total configurations and corresponding DFT-SAPT
energies are collectively referred to as the `91 dimer test set'.  As a
primary indication of accuracy, root-mean-square errors (RMSE) and mean signed
errors (MSE), both with respect to DFT-SAPT, were computed for each
methodology and for each dimer pair. Because these RMSE and MSE are dominated by
repulsive contributions, and owing to the thermodynamic importance of
attractive configurations, so-called `attractive RMSE/MSE' were also computed by
excluding net repulsive configurations (as measured by the DFT-SAPT total energy). 
The overall RMSE/MSE for all 91 dimers were then
averaged to produce one `characteristic RMSE/MSE' for the entire test set.  Since
these errors varied considerably in magnitude depending on the dimer in
question, this overall average was taken in the geometric mean sense. (Results
with an arithmetic mean do not differ qualitatively). Note that when computing
the characteristic MSE, only the magnitude of each MSE, \mse, was considered.

Characteristic RMSE and \mse across the 91 dimer test set are shown in \figref{fig:rmse}
and \tabref{tab:rmse}.  Overall, the \isaff exhibits smaller errors compared
to the \saptff. On average, the characteristic total energy RMSE for the \isaff
decrease by 33\% relative to the \saptff.  Even excluding repulsive configurations
(dominated by short-range interactions), errors for the \isaff are lower by 11\%
compared to the \saptff, demonstrating modest gains in accuracy even over the most
energetically-relevant regions of the potential.  
A more detailed analysis of each of the 91 pairs of molecules shows that in an
overwhelming 93\% of such cases, force fields
derived from the Slater-ISA method have smaller RMSEs compared to their
Born-Mayer-IP counterparts (70\% if only attractive configurations are
considered). Regardless of the metric used, the \isaff produces force
fields with higher fidelity to the underlying benchmark interaction energies.

    \begin{figure}
    \includegraphics[width=0.9\textwidth]{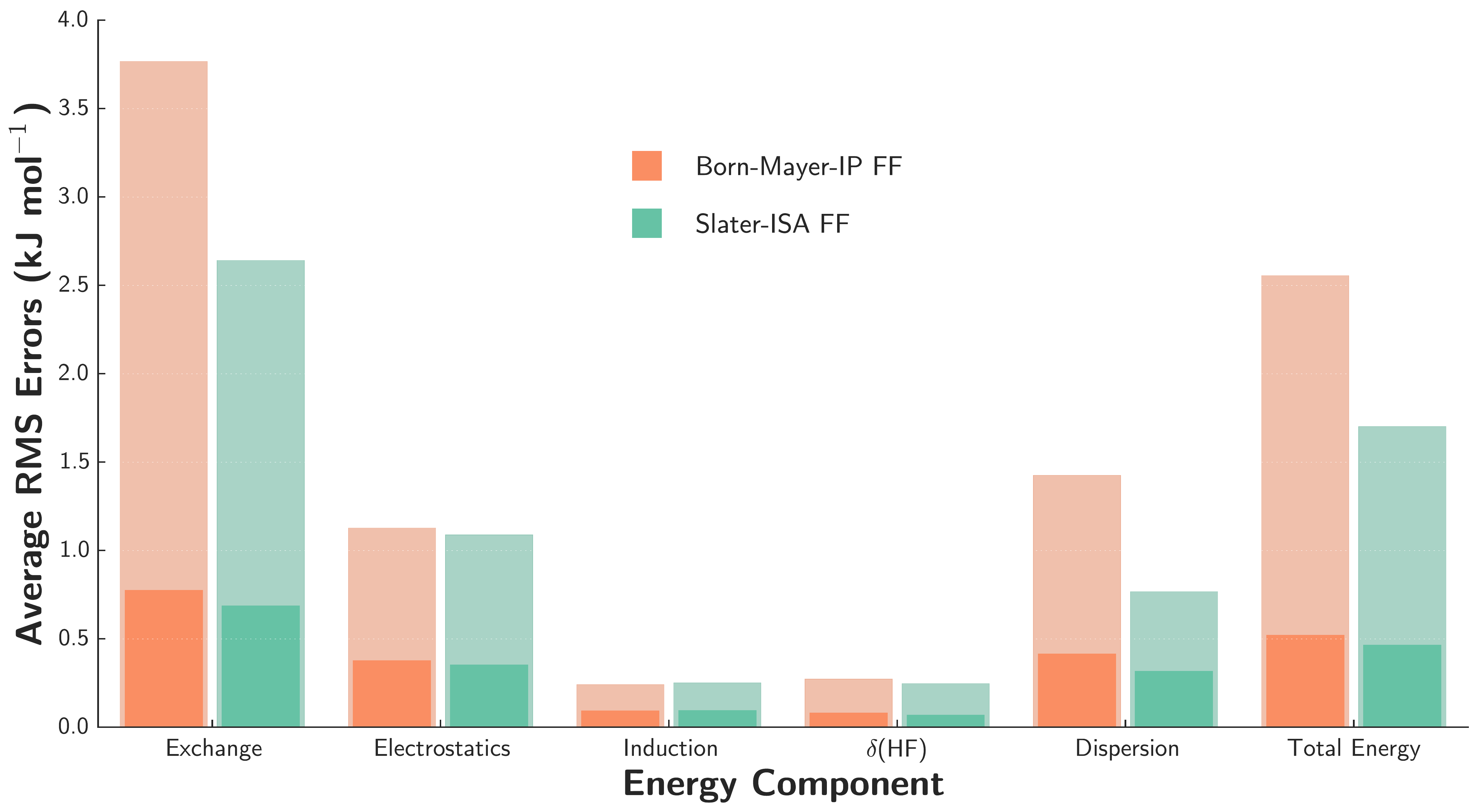}
    \caption{
    Characteristic RMSE (as described in the main text) for the \saptff (orange) and the \isaff (green) over the 91
    dimer test set. The translucent bars represent total RMSE
    for each energy component, while the smaller solid bars represent `Attractive'
    RMSE, in which repulsive points have been excluded.
            }
    \label{fig:rmse}
    \end{figure}

\begin{landscape}
\begin{table}
\small
\centering
\renewcommand\arraystretch{1.1}
\begin{tabular}{@{}rcccccccc@{}}
\hline
\toprule
& \phantom{} &
  \multicolumn{3}{c}{Dimer-Specific Fits} &
  \phantom{ab} &
  \multicolumn{3}{c}{Transferable Fits} \\
\cmidrule{3-5} \cmidrule{7-9}

Component & & \isaff & \saptff & \ljff & & \isaff & \saptff & \ljff \\
     & & \multicolumn{1}{c}{(kJ mol$^{-1}$)} & \multicolumn{1}{c}{(kJ mol$^{-1}$)} &  \multicolumn{1}{c}{(kJ mol$^{-1}$)}
     & & \multicolumn{1}{c}{(kJ mol$^{-1}$)}& \multicolumn{1}{c}{(kJ mol$^{-1}$)} &  \multicolumn{1}{c}{(kJ mol$^{-1}$)}\\
\midrule
Exchange        & &    2.641 (0.686)   &    3.766 (0.775)   &     ---          & &    2.718 (0.720)  &    4.033 (0.836)   &     ---        \\
Electrostatics  & &    1.087 (0.351)   &    1.126 (0.377)   &     ---          & &    1.134 (0.351)  &    1.231 (0.378)   &     ---        \\
Induction       & &    0.251 (0.095)   &    0.241 (0.093)   &     ---          & &    0.278 (0.101)  &    0.265 (0.098)   &     ---        \\
\dhf            & &    0.246 (0.068)   &    0.272 (0.079)   &     ---          & &    0.274 (0.076)  &    0.304 (0.081)   &     ---        \\
Dispersion      & &    0.766 (0.317)   &    1.425 (0.414)   &     ---          & &    0.766 (0.317)  &    1.425 (0.414)   &     ---        \\
\addlinespace
\textbf{
Total Energy}  \\
\emph{RMSE}     & &    1.701 (0.464)   &    2.554 (0.520)   &   1.984 (0.603)  & &    1.650 (0.456)  &    2.698 (0.555)   &  2.054 (0.640) \\
\emph{\mse}
                & &    0.216 (0.057)   &    0.539 (0.127)   &    0.322 (0.345) & &    0.175 (0.051)  &    0.569 (0.112)  &    0.311 (0.368) \\
\bottomrule
\hline
\end{tabular}
\caption{
    Comparison of characteristic RMSE (as described in the main text) over the 91 dimer test
set for the \isaff, \saptff and \ljff.
For the total energy, both characteristic RMSE and MSE have been shown, with
only the magnitude of the MSE, \mse, displayed.
    `Attractive' RMSE, representing the characteristic RMSE for
    the subset of points whose energies are net attractive ($\etot <
    0$), are shown in parentheses to the right of the total RMS
    errors; `attractive' \mse are likewise displayed for the total
    energy.
As discussed in \secref{ss:transferability}, the `Dimer-Specific Fits'
refer to force fields whose parameters have been optimized for each of the 91
dimers separately, whereas the `Transferable Fits' refer
to force fields whose parameters have been optimized for the 13 homodimers and
then applied (without further optimization) to the remaining 78 mixed systems.
    Unless otherwise stated, a default weighting function of $\lambda=2.0$
    (see \eqref{eq:weighting-function}) has been used for all force fields in
    this work.
	}
\label{tab:rmse}
\end{table}
\normalsize
\end{landscape}

It is also instructive to consider each energy component individually.  As
might be expected, improvements in the description of \erep are pronounced,
with the characteristic RMSE from the \isaff being 30\% smaller than that from the \saptff.
Examining each dimer pair separately (see Supporting Info for homo-monomeric
fits, representative of the entire test set) we also find that, in general, the \isaff
is far better at reproducing \emph{trends} in the exchange energy compared to
the \saptff. This qualitative result is also reflected in the smaller \mse
values for the \isaff as compared to the \saptff.
Nevertheless, there remains a fair amount of scatter in the exchange
energies for several dimer pairs, particularly for molecules with exposed lone
pairs or delocalized $\pi$ systems. We hypothesize that this scatter is due to
a breakdown of the isotropic approximation made in the \nameref{sec:theory}
section, a conclusion supported by observations on the pyridine dimer system
recently made by some of us.\cite{Misquitta2015b}
It it therefore quite possible that the observed 30\% RMSE reduction 
underestimates the true error reduction that might be observed
if such anisotropy were accounted for.

From  \figref{fig:rmse}, we see that the dispersion energy model from the 
\isaff is also a substantial improvement; for dispersion, characteristic RMSE
are 46\% smaller for the \isaff compared to the Born-Mayer model.
This should not be a counter-intuitive result: while both potentials use identical
dispersion coefficients, they differ in the damping model used.
In the \saptff, the standard Tang--Toennies damping model is employed, and 
the damping parameters only depend on free atom ionization potentials; 
in the \isaff, on the other hand, the damping parameters are obtained from the
ISA shape functions, and thus take 
molecular environment effects into account.
Even when only considering attractive dimer configurations (solid bar in
\figref{fig:rmse}), errors in the dispersion energy component are reduced by 23\%,
demonstrating the importance of the damping function across the potential surface. 
From these results, and in agreement with related literature
studies,\cite{Sebetci2010} we conclude that use of
the standard Tang-Toennies damping function based on atomic ionization
potentials
\cite{Tang1984, Misquitta2008a, Price2010, Totton2010a, McDaniel2013, Hermida-Ramon2000, Nyeland1990} 
lacks quantitative predictive power compared to the Slater-ISA model.
Note that neither the \isaff nor the \saptff are directly fitted to the DFT-SAPT
dispersion energies (all parameters are determined from monomer properties),
making this accuracy particularly striking.  We hypothesize that the effect of
the Slater-ISA approach is greater for dispersion than for first-order exchange
because here (in contrast to the exchange energy) there are no fitted
parameters to compensate for deficiencies in the exponents or functional form
of the \saptff.

In contrast to the exchange and dispersion energies, the \isaff and the \saptff show nearly identical errors for the
electrostatic and the induction (2\super{nd} order induction plus \dhf) energies.
In these cases, the two models differ only in the parameters and functional form used to represent
the exponentially-dependent short-range terms of these energy components, 
namely the penetration component for the electrostatic term and the 
penetration/charge-transfer term for the induction.
The lack of improvement between the Slater-ISA and Born-Mayer-IP models may
imply that we are not 
able to capture the physics of these particular short-range interactions with
either the Slater-functional of Born-Mayer functional forms.
Alternatively, the assumption that the short-range components of the electrostatic 
and induction energies are proportional to the exchange-repulsion may need to be 
re-examined. As discussed in \secref{sec:heteroatomic_vsr},
this proportionality is known to be approximately valid, but as yet there does 
not seem to be a deeper theoretical understanding of these short-range terms
that may lead to a better model.
Nevertheless, absolute errors in the electrostatic and induction components are
relatively small for both models.  Thus overall, the \isaff functional form
is promising for treating a wide variety of short-range effects.

\begin{table}
\small
\centering
\renewcommand\arraystretch{1.1}
\begin{tabular}{@{}rcccccc@{}}
\hline
\toprule
& \phantom{} &
  \multicolumn{2}{c}{\ljff Dimer-Specific Fits} &
  \phantom{ab} &
  \multicolumn{2}{c}{\ljff Transferable Fits} \\
\cmidrule{3-4} \cmidrule{6-7}


          & &           $\lambda=2.0$ &           $\lambda=0.1$      
          & &           $\lambda=2.0$ &           $\lambda=0.1$   \\ 
     & & \multicolumn{1}{c}{(kJ mol$^{-1}$)} & \multicolumn{1}{c}{(kJ mol$^{-1}$)} 
     & & \multicolumn{1}{c}{(kJ mol$^{-1}$)}& \multicolumn{1}{c}{(kJ mol$^{-1}$)}  \\ 
\midrule
\emph{RMSE}             & &  1.984 (0.603)   &  6.058 (0.413)  && 2.054 (0.640)  &  5.760 (0.457)    \\
\emph{\mse}
                 & &  0.322 (0.345)   &  1.610 (0.041)  && 0.311 (0.368)  &  1.410 (0.060)    \\
\bottomrule
\hline
\end{tabular}
\caption{
    Comparison of characteristic RMSE and \mse over the 91 dimer test
set for the various Lennard-Jones models. The LJ models are not parameterized
on a component-by-component basis, thus RMSE/\mse values are only
shown for the total FF energies.
    `Attractive' errors, representing the characteristic RMSE/\mse for
    the subset of points whose energies are net attractive ($\etot <
    0$), are shown in parentheses to the right of the total 
    errors. `Dimer-Specific Fits' and `Transferable Fits' are as in \tabref{tab:rmse}.
	}
\label{tab:lj_rmse}
\end{table}
\normalsize

The comparison between the \isaff and the \ljff is slightly more complicated,
owing to the differences in long-range potential and fitting methodology (see
\secref{sec:FF-forms}). As such, we compare the \isaff to several versions of the
\ljff (for which characteristic RMSE and \mse are shown in 
\tabref{tab:lj_rmse}). Using the same weighting function and
constraining the Coulombic and polarization terms to be identical to the
\isaff, we see that the resulting Lennard-Jones force field (\ljff,
$\lambda=2.0$) is significantly worse than the \isaff, both in terms of total
RMSE and attractive RMSE. Furthermore, by comparing the \mse
of both force fields, we see that errors in
\ljff are much more \emph{systematic} than in the \isaff: in order to
reproduce the repulsive wall correctly, the Lennard-Jones potential
generally underestimates the well-depth by a considerable fraction (see
the Supporting Information for ethane as a typical example). 

Given the failure of the \ljff ($\lambda=2.0$) force field to reproduce the
energetically important region of the PES, we also compared the \isaff to a
`best-case' scenario Lennard-Jones force field which correctly
reproduces the minimum energy region at the expense of the repulsive wall. These
\ljff ($\lambda=0.1$) fits have total RMSE errors nearly 4 times that of the
\isaff; indeed, the \ljff ($\lambda=0.1$) reproduces the repulsive wall only
qualitatively. Insofar as the repulsive wall is concerned, the \isaff is far
superior to the Lennard-Jones short-range model. Nevertheless (and much more
importantly for molecular simulation), the attractive region of the potential
is reproduced surprisingly well by \ljff. Characteristic attractive RMSE for
the \ljff ($\lambda=0.1$) are slightly lower than those for \isaff, although
the former has one additional free parameter per atom type and is also fit directly
to reproduce the total energy. Likewise, attractive \mse between the
\isaff and the \ljff ($\lambda=0.1$) are comparable. As we show in the
Supporting Information,
however, and as is well known in the literature, weighting the Lennard-Jones potential in this
manner does not necessarily capture important information from the long-range attractive tail or repulsive wall of the PES, such
that the \ljff ($\lambda=0.1$) is not always expected to yield good
property predictions. This latter point will be demonstrated in 
\secref{sec:accuracy_experiment}.

In order to compare the performance of the \isaff against popular standard
force fields, we also developed a `best case scenario' non-polarizable point
charge Lennard-Jones model, results for which are shown in the Supporting
Information. Unsurprisingly, this force field is worse (in an
RMSE and \mse sense) than all
other force fields studied in this work, thus demonstrating how important
accurate models for long-range electrostatics and polarization are to the
overall accuracy of ab initio force fields.

\begin{subsubsection}{Argon Dimer}

We now turn to several specific case studies. The Ar dimer provides an interesting test
case to examine directly the impact of the polynomial pre-factor included in
the \isaff functional form. 
Since Ar is an atomic species, we should have $\Bisa{\text{Ar}} = \Bip{\text{Ar}}$.
For numerical reasons, the \isaff and \saptff exponents differ by 0.03 a.u.;
however, this difference is insignificant, and the two FFs differ mainly in the
polynomial pre-factor.
\figref{fig:ar-pes} shows the potential energy
surface (PES) for the argon dimer computed using the \isaff and the \saptff.
Here the default weighting scheme has been used so as to best reproduce the energetically
attractive region. Note that, while both potentials reproduce the minimum energy
configurations correctly, the \saptff considerably overestimates the exchange energy (and thus
the total energy) along the repulsive wall.
The \isaff, on the other hand, maintains excellent accuracy in this region of the
potential. This result is particularly notable because the repulsive wall is not heavily weighted in the fit.
(A point 10 kJ mol$^{-1}$ along the repulsive wall, for instance, is weighted only 3\% as
heavily as a point near the bottom of the well).
A similar, though smaller, increase in accuracy is seen in the fit to the
DFT-SAPT dispersion energies, where the \isaff is better able to model the
energies for shorter interatomic separations.  This increased accuracy is
entirely attributable to the functional form employed, as the dispersion
parameters are identical between the two FFs.

Consistent with prior literature,\cite{Ihm1990, McDaniel2012} these results
suggest that neglect of the polynomial pre-factor $P$ (as in standard
Born-Mayer potentials) is \emph{by itself} a poor approximation. However, as
we show below, the Born-Mayer form can still be used as an accurate model
in conjunction with appropriately scaled atomic exponents. Nonetheless, the
more physically-motivated Slater form provides increased accuracy over a wider
range of separations without recourse to empirical scaling.

    \begin{figure}
    \includegraphics[width=0.9\textwidth]{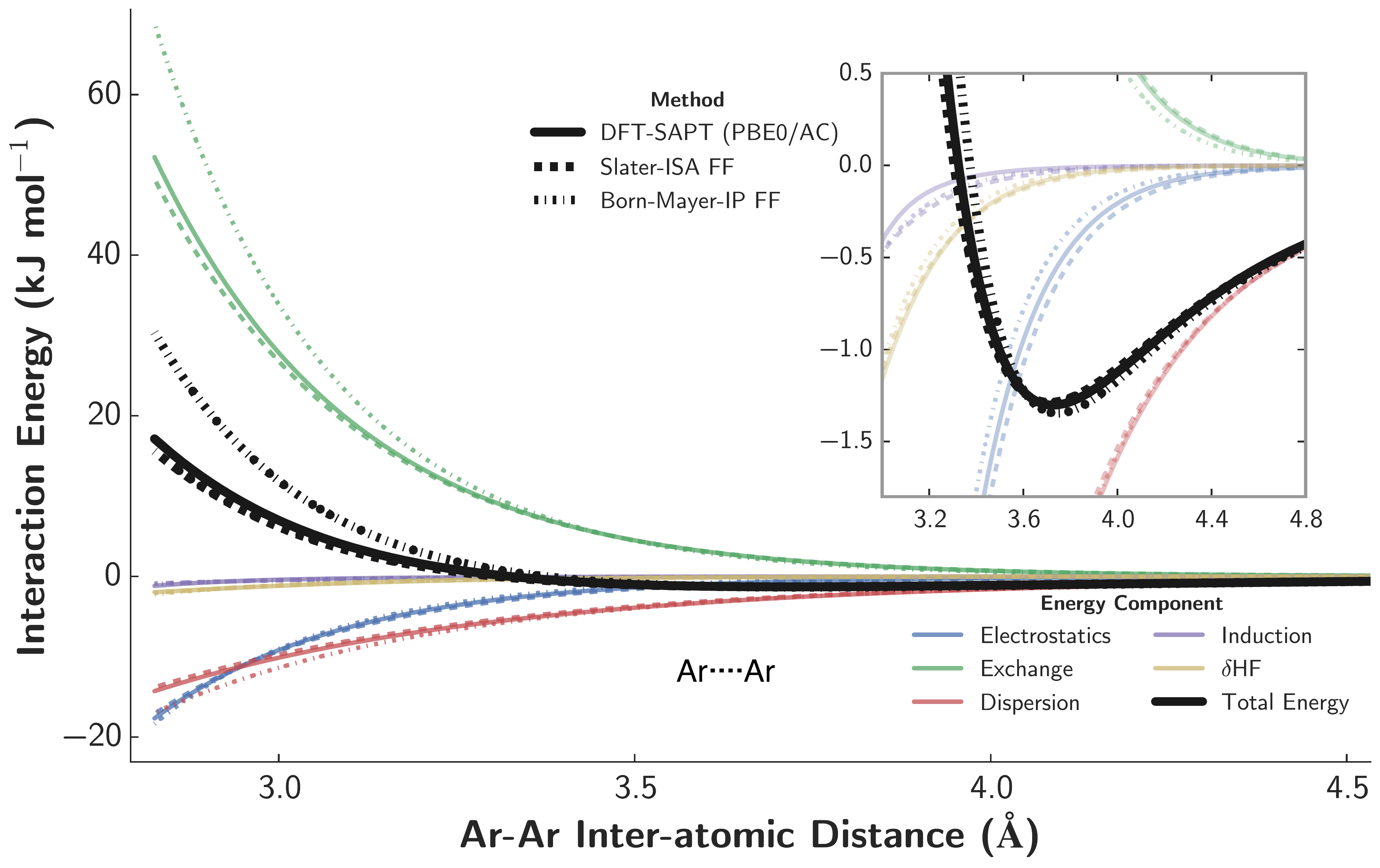}
    \caption{
    Potential energy surface for the argon dimer. 
    Interaction energies for the \isaff (dashed curves) and the \saptff (dash-dotted
    curves) are shown alongside benchmark \sapt energies (solid curves). The
    energy decomposition for DFT-SAPT and for each force field is shown for reference.
    }
    \label{fig:ar-pes}
    \end{figure}

Results for \ljff are shown in the Supporting Information; consistent with expectations for the
Lennard-Jones model, the repulsive wall is overestimated by the
$1/r_{ij}^{12}$ short-range functional form, and the magnitude of the
attractive tail region is similarly overestimated by the effective $C_{ij,6}$
dispersion parameter. 
Note that this $C_{ij,6}$ coefficient has been fit to the total energy, and thus differs
from the asymptotically-correct $C_{ij,6}$ parameter used for both the \isaff
and the \saptff. 
An alternative parameterization strategy would have been to use the asymptotically-correct $C_{ij,6}$
parameter in the \ljff, but this would have worsened predictions
along both the repulsive wall and the
minimum energy configurations.

\end{subsubsection}
\begin{subsubsection}{Ethane Dimer}

We next discuss the ethane dimer and show both a scatter plot
of the 1000 dimer interactions (\figref{fig:ethane-scatter}) and a cut through the
potential energy surface near the minimum (\figref{fig:ethane-pes}) as
indications of force field quality.

    \begin{figure}
    \includegraphics[width=0.9\textwidth]{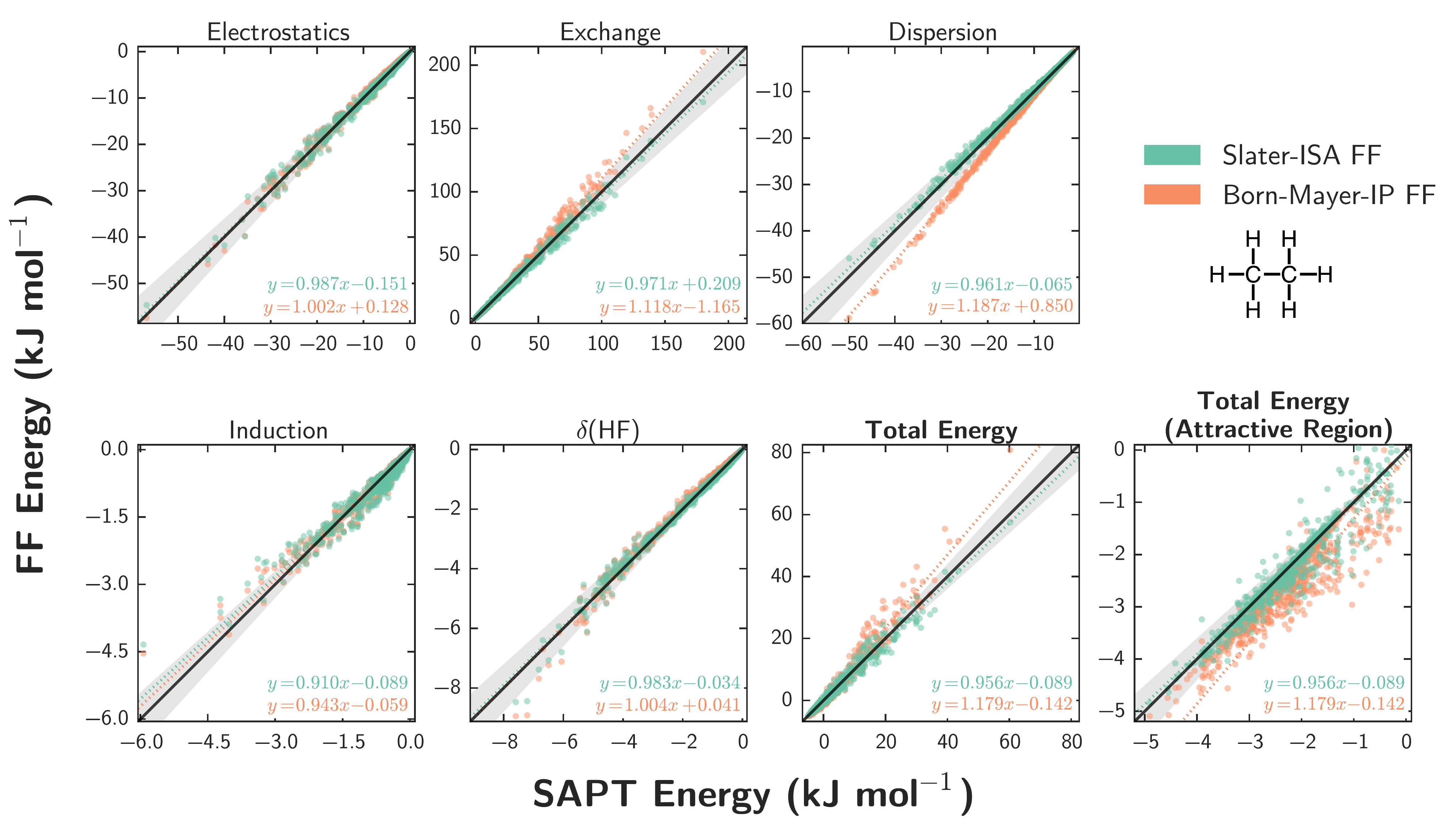}
    \caption{
    Force field fits for the ethane dimer using the Slater-ISA (green) and
    Born-Mayer-IP (orange) FFs.
    Fits for each energy component are displayed along with two views of the total interaction energy.
    The diagonal line (black) indicates perfect agreement between reference energies
    and each force field, while shaded grey areas represent points within $\pm
    10\%$ agreement of the benchmark. To guide the eye, a line of best fit (dotted
    line) has been computed for each force field and for each energy component.
     }
    \label{fig:ethane-scatter}
    \end{figure}

As with the argon dimer, for the ethane dimer the \isaff produces more
accurate exchange and dispersion energies compared to the \saptff. Here, the
effects of the \isaff for dispersion are even more pronounced, likely because
the conventional damping of the \saptff is systematically in error due to
differences in both the form of the damping function and exponents. As for the
total interaction energy, we again find that the \saptff exhibits large errors
for repulsive contributions, while the \isaff naturally reproduces
interactions for both attractive and strongly repulsive configurations.  Even
in the attractive regime, the \saptff is systematically too attractive. These
systematic errors are the result of imperfect error cancellation
between the exchange and dispersion components of the fit, and are discussed in
more detail in Section \ref{sec:results-robustness}. 

    \begin{figure}
    \includegraphics[width=0.9\textwidth]{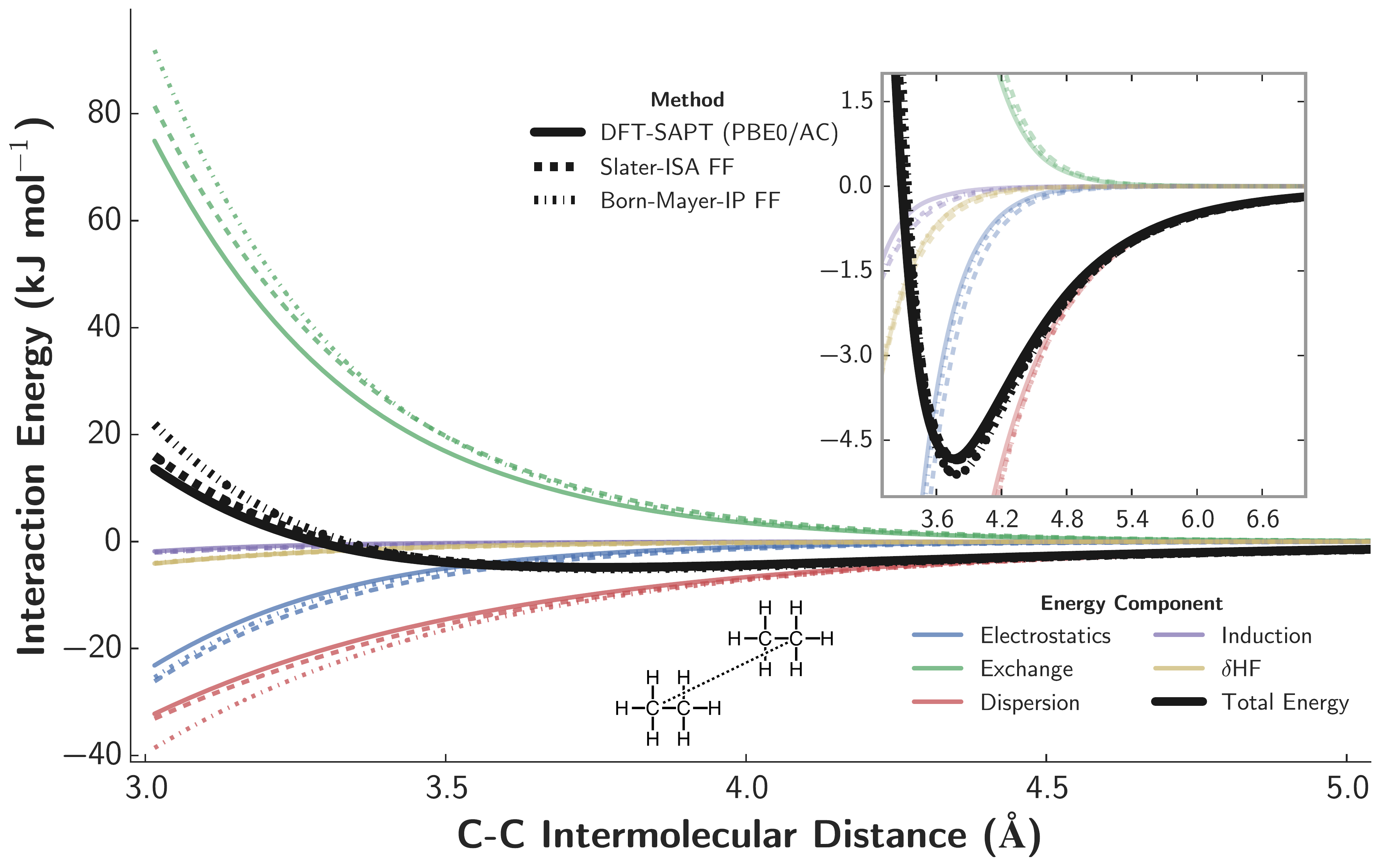}
    \caption{
    A representative potential energy scan near a local minimum for the ethane dimer. 
    Interaction energies for the \isaff (dashed curves) and the \saptff (dash-dotted
    curves) are shown alongside benchmark \sapt energies (solid curves). The
    energy decomposition for DFT-SAPT and for each force field is shown for reference.
     The ethane dimer configuration in this scan corresponds to the most
    energetically attractive dimer included in the training set; other
    points along this scan are not included in the training set.
    }
    \label{fig:ethane-pes}
    \end{figure}

Examining a specific cut across the ethane-ethane PES
(\figref{fig:ethane-pes}) visually confirms these results.  Both potentials do
an excellent job of reproducing the benchmark DFT-SAPT energies in the minimum
energy region, though the \saptff is slightly too attractive. (Other cuts of
the PES would show the Born-Mayer-IP predictions to be significantly more in error,
consistent with the scatter plots). Along the repulsive wall, however, the
\saptff predictions worsen in comparison to those from the \isaff. Finally,
the PES shows an increased reliance on error cancellation between the
various energy components for the \saptff compared to the \isaff.

As shown in the Supporting Information, the Lennard-Jones force field models
are incapable of reproducing the entirety of the ethane PES; depending on the
weighting function, either the repulsive wall or the attractive well can be
reproduced, however no set of parameters can predict both regions
simultaneously.

\end{subsubsection}
\begin{subsubsection}{Acetone Dimer}

The acetone dimer provides a final interesting example involving a moderately sized
organic molecule.
From both the scatter plots (\figref{fig:acetone-scatter}) and the PES cross section
(\figref{fig:acetone-pes}), it is evident that both the Slater-ISA and
Born-Mayer-IP force fields do an
excellent job of reproducing DFT-SAPT energies for the low energy dimers.
Along the repulsive wall, however, the \saptff shows larger systematic
errors in each energy component, and seems to rely on error cancellation
to achieve good agreement in the total energy.
This reliance on error cancellation has two negative effects: 
Firstly, the additional scatter in the total energy of the \saptff
fit, especially prominent for attractive configurations, indicates that this
error cancellation is imperfect in certain cases. MSE
for the \isaff ($-0.0115$ kJ mol$^{-1}$) are an order of magnitude lower than for
the \saptff ($0.182$ kJ mol$^{-1}$) in the attractive region of the potential.  Secondly,
as we shall later explore, reliance on error cancellation likely
contributes to the somewhat decreased transferability of the \saptff as
compared to the \isaff. 

As shown in the Supporting Information, the
\ljff predictions for acetone are reasonably good in both the tail and minimum
energy regions of the potential, however the \ljff grossly overpredicts the
\sapt energies along the repulsive wall.

    \begin{figure}
    \includegraphics[width=0.9\textwidth]{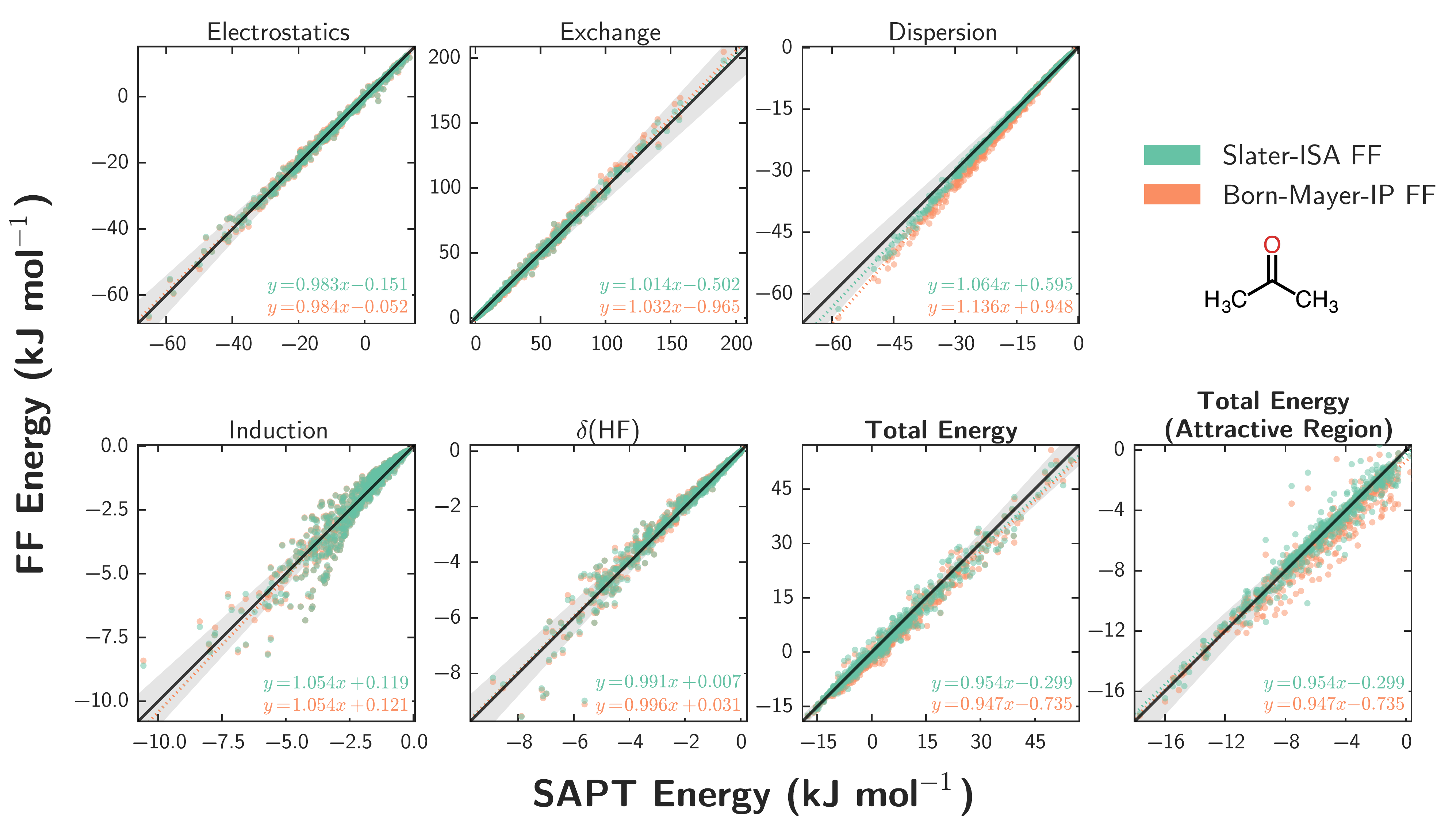}
    \caption{
    Force field fits for the acetone dimer using the Slater-ISA (green) and
    Born-Mayer-IP (orange) FFs, as in \figref{fig:ethane-scatter}.
            }
    \label{fig:acetone-scatter}
    \end{figure}

    \begin{figure}
    \includegraphics[width=0.9\textwidth]{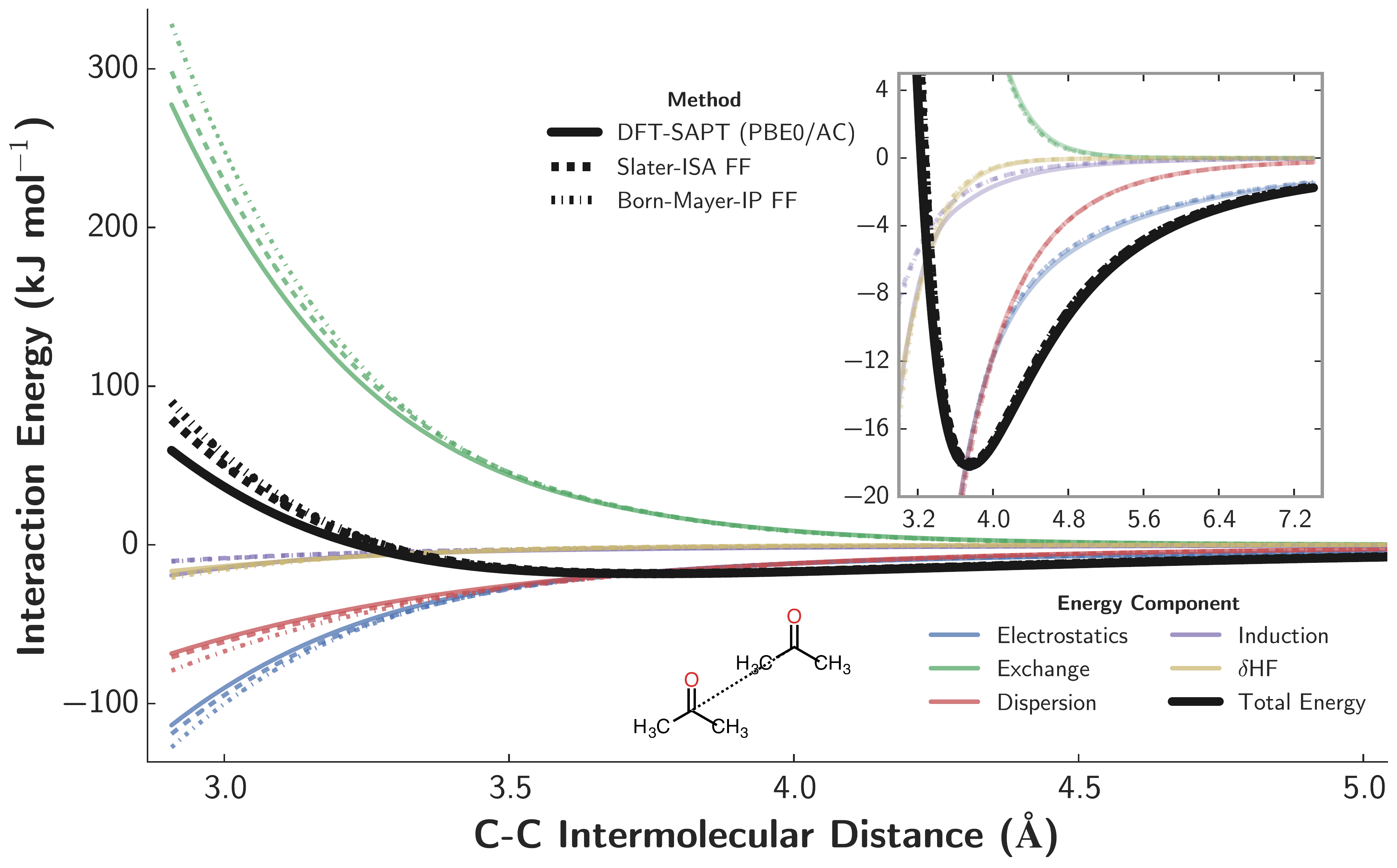}
    \caption{
      A representative potential energy scan near a local minimum for the
      acetone dimer.  Interaction energies for the \isaff (dashed curves) and
      the \saptff
      (dash-dotted curves) are shown alongside benchmark \sapt energies (solid
      curves). The energy decomposition for DFT-SAPT and for each force field is
      shown for reference.  The intermolecular distance is taken to be the
      internuclear distance between the two carbonyl carbons on each acetone
      monomer.  The configuration in this scan corresponds to the
      most attractive dimer configuration included in the training set for the acetone dimer;
      other points along this scan have not explicitly been included in the training
      set.
      }
    \label{fig:acetone-pes}
    \end{figure}

\end{subsubsection}
\end{subsection}
\begin{subsection}{Accuracy: Comparison with experiment}
\label{sec:accuracy_experiment}

    \begin{figure}
    \includegraphics[width=0.9\textwidth]{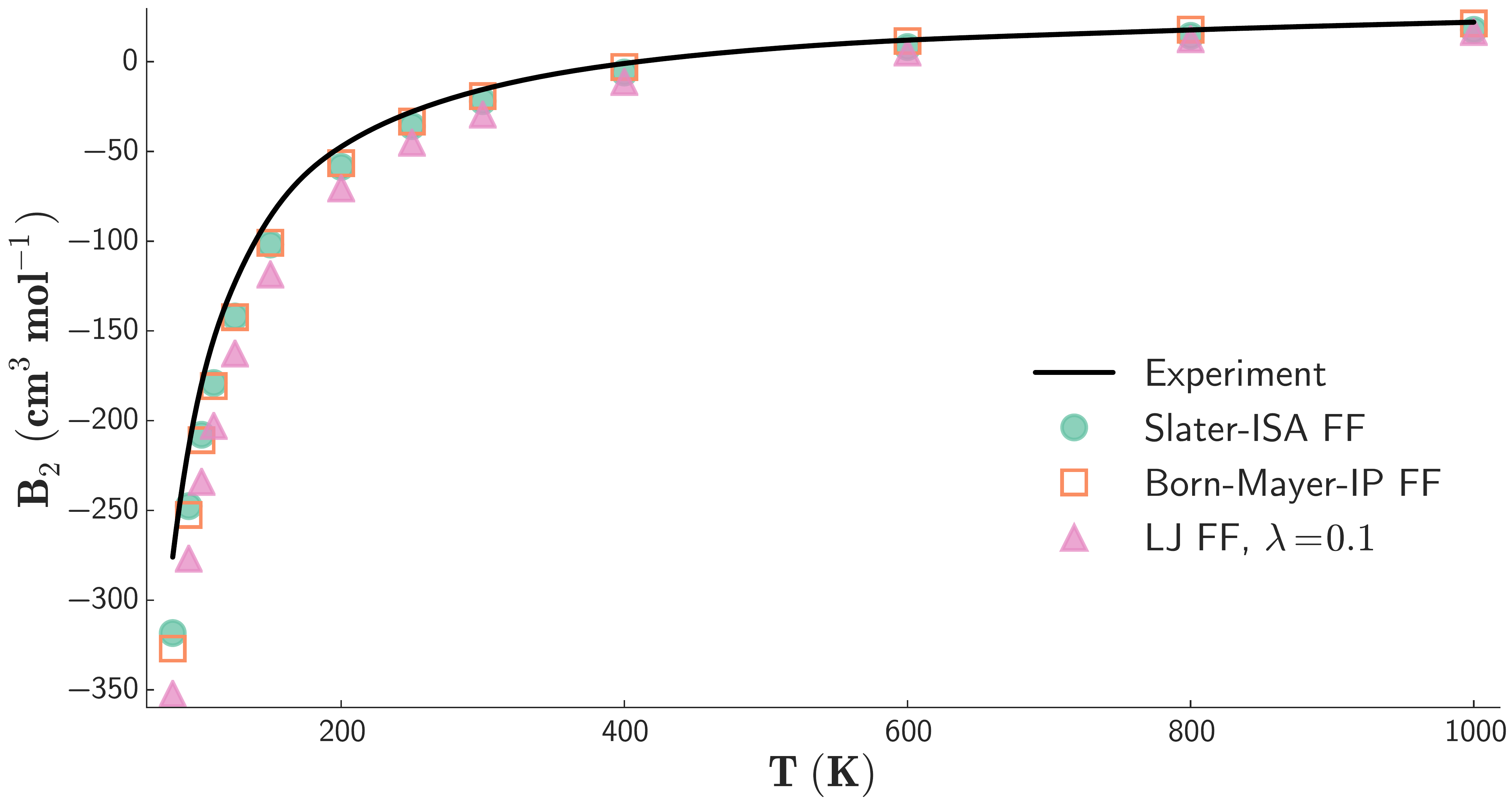}
   \caption{
    Second virial coefficients for argon. The Slater-ISA and the Born-Mayer-IP
    FFs are shown as green
    circles and orange squares, respectively; the black line corresponds to
    experiments from \citen{Dymond1980}.
           }
    \label{fig:ar-virial}
    \end{figure}

We have benchmarked the above force fields against experimental 
second virial coefficients and, in the case of ethane, enthalpies of
vaporization and liquid densities.
The classical 2\super{nd} virial coefficients were calculated for both argon and
ethane using rigid monomer geometries, following the procedure described in
\citen{McDaniel2013}. 
Enthalpies of
vaporization and liquid densities were calculated using the OpenMM molecular
simulation package
\cite{Eastman2013}
as described in \secref{sec:methods}.
Higher-order multipole moments
--- which were negligible for these molecules --- were neglected, and so
only rank $0$ terms were used in these calculations. 
Results are shown in Figures \ref{fig:ar-virial} and \ref{fig:ethane-virial}
as well as \tabref{tab:deltah}.

For argon, since both \isaff and \saptff accurately reproduce the energetics 
of low-energy configurations, it
is unsurprising that both force fields yield accurate virial
coefficients over a wide range of temperatures.  Errors in computed $B_2$
coefficients (for both potentials) are likely attributable to small errors in
the \sapt potential itself,
\cite{Podeszwa2005a}
and, to a much lesser extent, the neglect of nuclear
quantum effects at lower temperatures.
\cite{Vogel2010}
Despite the good (in an RMSE sense) fit quality of the \ljff ($\lambda=0.1$),
this force field overpredicts the magnitude of the 2\super{nd} virial for argon, likely as a
result of the effective dispersion coefficient, which overestimates the
attraction in the tail
region of the PES (see Supporting Information). Although it is certainly
possible to parameterize a Lennard-Jones model \emph{empirically} for argon,
such a force field would rely on a subtle cancellation of errors between the
minimum energy- and tail-regions of the PES. As the proper balance is
impossible to predict a priori, this result highlights one of the difficulties
of using the less physical LJ model in the development of ab-initio force
fields.

%
    \begin{figure}
    \includegraphics[width=0.9\textwidth]{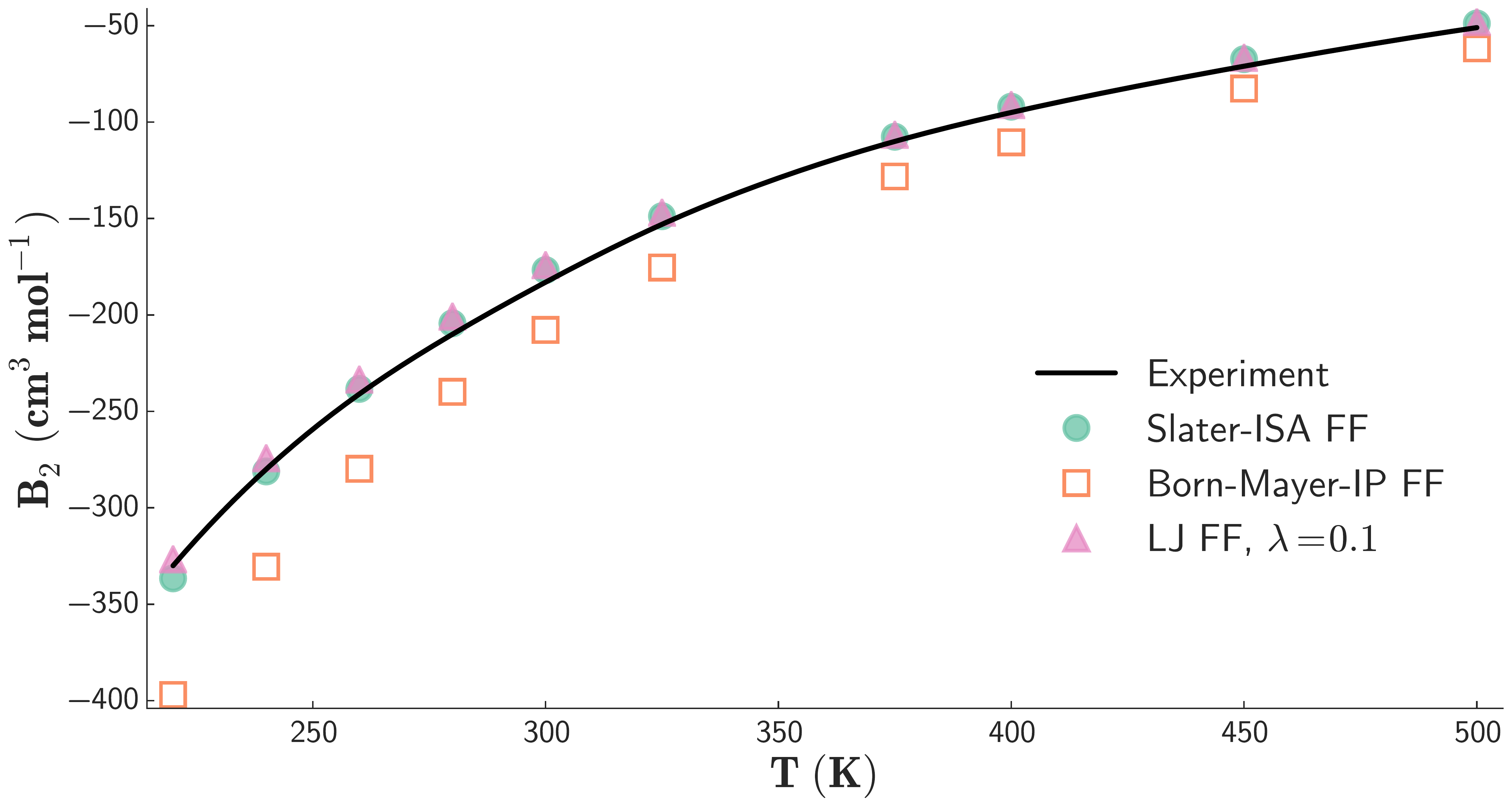}
   \caption{
    Second virial coefficients for ethane. 
    The Slater-ISA and Born-Mayer-IP FFs are shown as green
    circles and orange squares, respectively; the black line corresponds to
    experiments from \citen{Dymond1980}.
    }
    \label{fig:ethane-virial}
    \end{figure}
%

In the case of ethane, the \isaff is in excellent agreement with experiment,
whereas the \saptff underpredicts $B_2$ by as much as 20\%. These results are
indicative, not only of the more accurate functional form and parameterization
of \isaff, but also of the high accuracy of the underlying \sapt benchmark
energies. In this case, \ljff also correctly predicts the virial.  Using
weighting functions for each model that are optimal for the 91 dimer test set
as a whole ($\lambda = 2.0$ for the \isaff and the \saptff,
$\lambda = 0.1$ for the \ljff), all force fields produce similar results for
$\Delta H_{\text{vap}}$ and $\rho$ (\tabref{tab:deltah}). These values are
slightly overestimated by all force fields (especially in the case of the
\saptff), which is to be expected given our neglect of many-body effects.
\citeauthor{McDaniel2014} have calculated the 3-body correction for the
\saptff; using this value as a global 3-body correction for all force fields,
we see that both the Slater-ISA and the Lennard-Jones force fields compare
very favorably to experiment, with the \isaff perhaps slightly more accurate. 

\end{subsection}

\begin{subsection}{Transferability}
\label{ss:transferability}

The transferability of interaction potentials is a crucial aspect of practical
molecular simulations. Here we examine `parameter transferability', by which
we mean the extent to which parameters from two homo-monomeric systems can
combined to predict the intermolecular interactions of the resulting mixed
hetero-monomeric system. 
As a measure of parameter transferability, we compared characteristic RMSE
and \mse relative to
the benchmark data for two different parameterization schemes.  For the
`Dimer-Specific Fits', \A parameters were obtained for each of the 91 dimer
pairs individually; these results are identical to those discussed in the
previous two subsections. In contrast, for the `Transferable Fits', the \A
parameters were fit to the 13 homonomeric dimer pairs and were re-used
(without any further optimization) to calculate energies for the 78 mixed
systems using the combination rules listed in \secref{sec:FF-forms}.
Results for each parameterization scheme are shown in \tabref{tab:rmse}. 
From the RMSE and \mse from the competing schemes, 
we see excellent parameter transferability for all force fields studied. 
For the \isaff, characteristic RMSE and \mse for each component increase by a very small
fraction upon constraining the fit; due to small error cancellation, errors in the
total energy actually \emph{decrease} somewhat with these constraints. 
(This is possible since the total energy is not directly fit.)
The \saptff also displays a significant degree of transferability, though
errors in the total energy increase slightly upon constraining the fit.
As in prior work, the observed parameter transferability for both force fields can be
attributed to our use of a term-by-term parameterization scheme
(\secref{sec:FF-forms}),
which serves to minimize error cancellation between energy components and generate a
more physically-meaningful (and thus transferable) set of parameters.
\cite{McDaniel2013,McDaniel2012a}
Finally, note that for four of the five interaction energy components the relative
change in RMSE on constraining the fit is smaller for the \isaff than
the \saptff. The \dhf term is the exception, but even here the relative 
change in errors from the two methods are comparable. This suggests that the
\isaff may be the more transferable of the force fields studied.
Nevertheless, the Lennard-Jones model is surprisingly transferable,
likely in part due to the same accurate and transferable `long-range'
electrostatics and polarization as the \isaff. The non-polarizable, point-charge
Lennard-Jones model (results for which are shown in the Supporting
Information) displays the least transferability (in both an RMSE and \mse
sense) of all force fields studied.

Although we do not examine it here, we expect that the previously demonstrated success\cite{McDaniel2013, McDaniel2012,
McDaniel2012a, McDaniel2014} of the \saptff with respect to `environment transferability'
--- the extent to which a single set of parameters can model a variety of phases and
molecular environments --- and `atom type transferability' --- the extent to which
atoms in chemically similar environments can accurately be grouped together
into `types' and treated using one parameter set --- 
would also apply to, or even be improved by, \isaff. These issues are under
investigation in our groups.

\end{subsection}
\begin{subsection}{Robustness}
\label{sec:results-robustness}

One of the practical challenges of ab initio force field development is the
robustness of the resulting force field quality with respect to the choice of an
appropriate training set and/or weighting function.  To this end, the default
weighting function (\eqref{eq:weighting-function}, $\lambda = 2.0$) was varied
to produce unconstrained fits that were skewed either towards attractive
($\lambda = 0.5$) or repulsive ($\lambda = 5.0$) configurations, and pairwise
differences in force field total energies were computed between each weighting
scheme. Characteristic root-mean-square pairwise differences (RMSD) between
each weighting function are shown in
\tabref{tab:rmsd-weightings}; as before, `attractive RMSD' were
calculated by excluding repulsive points from consideration. Note that, on
average, the default $\lambda = 2.0$ weighting scheme is optimal (in an RMSE
sense) for both the Slater-ISA and Born-Mayer-IP FFs.

Overall, both the \saptff and the \ljff display significant weighting function
sensitivity. This sensitivity is not surprising; as both force fields are
unable to reproduce the entirety of the potential energy surface, changing the
weighting scheme (or equivalently, the balance of configurations in the
training set) alters the parameters in the \saptff or the \ljff models quite 
substantially. Even excluding repulsive configurations, RMSD
of $\sim0.5$ kJ mol$^{-1}$ are typical for the \saptff. RMSD are
somewhat smaller for the \ljff ($\sim0.3$ kJ mol$^{-1}$), however
qualitatively we see that differences in computed force field energies are systematic: smaller weighting
functions capture the minimum energy region of the potential while
overestimating the magnitudes of both the repulsive and tail regions of the
potential, whereas larger weighting functions tend to underestimate the
minimum energy region in order to correctly reproduce the repulsive wall.
Consequently, the Lennard-Jones model shows weighting-function sensitivity in
a manner that is not entirely captured by the RMSD, but is instead
reflected in the greater sensitivity of the \ljff (as compared to the \saptff)
in the prediction of experimental properties (\emph{vide infra}).
 
Note that for practical force field development (as opposed to minimization of
overall RMSE), the default weighting scheme for the \saptff
and the \ljff is suboptimal for many dimers in the test set.  Because both the
\saptff and the \ljff must inherently compromise between accuracy near the
minimum and along the repulsive wall, the weighting function requires
system-specific fine-tuning in order to achieve proper balance. This
empiricism creates significant challenges in the development of ab initio force
fields.

\begin{table}[t]
\small
\centering
\renewcommand\arraystretch{1.1}
\begin{tabular}{@{}rcccccc@{}}
\hline
\toprule
\multirow{2}{*}{Characteristic RMSD} & \phantom{ab} &
  {$\lambda = 0.5$ vs 2.0} & \phantom{ab} &
  {$\lambda = 0.5$ vs 5.0} & \phantom{ab} &
  {$\lambda = 2.0$ vs 5.0} \\
  & \phantom{ab} &
  (kJ mol$^{-1}$) & \phantom{ab} &
  (kJ mol$^{-1}$) & \phantom{ab} &
  (kJ mol$^{-1}$) \\
\midrule
\isaff    & & 0.742 (0.207) & & 0.990 (0.273) & & 0.306 (0.086) \\
\saptff   & & 1.866 (0.409) & & 2.632 (0.550) & & 0.797 (0.153) \\
\ljff  & & 1.301 (0.216) & & 1.605 (0.309) & & 0.324 (0.099) \\
\bmsisaff & & 0.611 (0.178) & & 0.810 (0.236) & & 0.293 (0.081) \\ 

\bottomrule
\hline
\end{tabular}
\caption{
    Characteristic RMS pairwise differences (RMSD) in force field total energies 
    for different weighting functions with $\lambda$ values as defined in
    \eqref{eq:weighting-function}; values shown are the (arithmetic mean,
    rather than geometric)
    RMSD across the 91 dimer test set.
    Characteristic `Attractive' RMSD (as defined
    in \tabref{tab:rmse}) are shown in parentheses to the right of each overall RMSD.
	}
\label{tab:rmsd-weightings}
\end{table}
\normalsize

By contrast, we find the \isaff to be robust with respect to the choice of
weighting function due to its more balanced treatment of repulsive and
attractive regions of the potential energy surface.
Average RMSD for the \isaff are between two to three \emph{times} smaller compared
to the \saptff, and the \isaff is relatively insensitive to the choice of
weighting function. 
These conclusions hold for both attractive and overall RMSD.
As a result, the Slater-ISA model largely eliminates the need for empirical
fine-tuning of the weighting function, which in turn greatly simplifies the
parameterization process and allows for a more robust prediction of chemical
and physical properties. 

For the ethane dimer, \figref{fig:ethane-weighting} shows overall
force field energies for both the Slater-ISA and Born-Mayer-IP FFs for three
weighting functions. Results for the Lennard-Jones models are shown in the SI,
and are qualitatively similar to the \saptff results.
The \saptff fits vary qualitatively with $\lambda$, leading to a relatively large
uncertainty in calculated $B_2$ coefficients, enthalpies of vaporization, and
liquid densities (see \tabref{tab:deltah}). 
By skewing the fits towards attractive
configurations ($\lambda = 0.5$), the majority of attractive configurations
are predicted without systematic error, though points along the repulsive wall
(including those with net negative energies) are systematically too repulsive. 
Using a scheme which more heavily weights repulsive configurations, the \saptff
regains semi-quantitative accuracy for repulsive configurations, albeit at the expense
of a systematic increase in errors for the attractive dimer configurations.
Finally, we reiterate that the optimal weighting function for the
ethane dimer (here $\lambda = 0.5$ best reproduces the 2\super{nd} virial for
the \saptff) is
by no means universal for the molecules in the 91 dimer test set.

\begin{table}
\small
\centering
\renewcommand\arraystretch{1.1}
\begin{tabular}{@{}rcccccc@{}}
\hline
\toprule
& \phantom{} &
  \multicolumn{4}{c}{Weighting Function} \\
\cmidrule{3-6} 

Force Field && $\lambda = 0.1$ &  $\lambda = 0.5$ &  $\lambda = 2.0$ &
$\lambda = 5.0$ & Experiment\\

\midrule
\addlinespace
\multicolumn{7}{c}{\textbf{$\boldsymbol{\Delta H_{\text{vap}}}$ (kJ mol\super{-1});
$\boldsymbol{\rho = 0.546}$ g L\super{-1}, $\boldsymbol{T=184}$ K}} \\
\isaff  && 15.3 (14.7)  & 15.3 (14.6) & 15.3 (14.7) & 15.2 (14.6) & \multirow{3}{*}{14.7}\\
\saptff && 14.3 (13.7)  & 15.1 (14.5) & 16.6 (15.9) & 18.6 (18.0) & \\
\ljff   && 15.5 (14.9)  & 14.6 (13.9) & 11.4 (10.7) & 10.1 ( 9.5) & \\
\addlinespace
\multicolumn{7}{c}{\textbf{$\boldsymbol{\rho}$ (g L\super{-1}); 
$\boldsymbol{P = 1}$ atm, $\boldsymbol{T=184}$ K}} \\
\isaff  && 0.600 (0.566) & 0.602 (0.568) & 0.600 (0.566) & 0.593 (0.559) & \multirow{3}{*}{0.546} \\
\saptff && 0.521 (0.487) & 0.567 (0.533) & 0.632 (0.598) & 0.678 (0.644) & \\
\ljff   && 0.607 (0.573) & 0.610 (0.576) & 0.555 (0.521) & 0.494 (0.460) \\
\bottomrule
\hline
\end{tabular}
\caption{
    Enthalpies of vaporization and liquid densities for ethane as a function
    of force field and weighting function. Values in parentheses include an
    estimation of the 3-body correction (0.628 kJ mol\super{-1} and 0.034
    g mL\super{-1} for the enthalpy of vaporization and liquid density,
    respectively) as computed in \citen{McDaniel2014}. Experimental data taken
    from \citen{Witt1937} and \citen{Riddick1986}.
	}
\label{tab:deltah}
\end{table}
\normalsize

    \begin{figure}
        \includegraphics[width=0.9\textwidth]{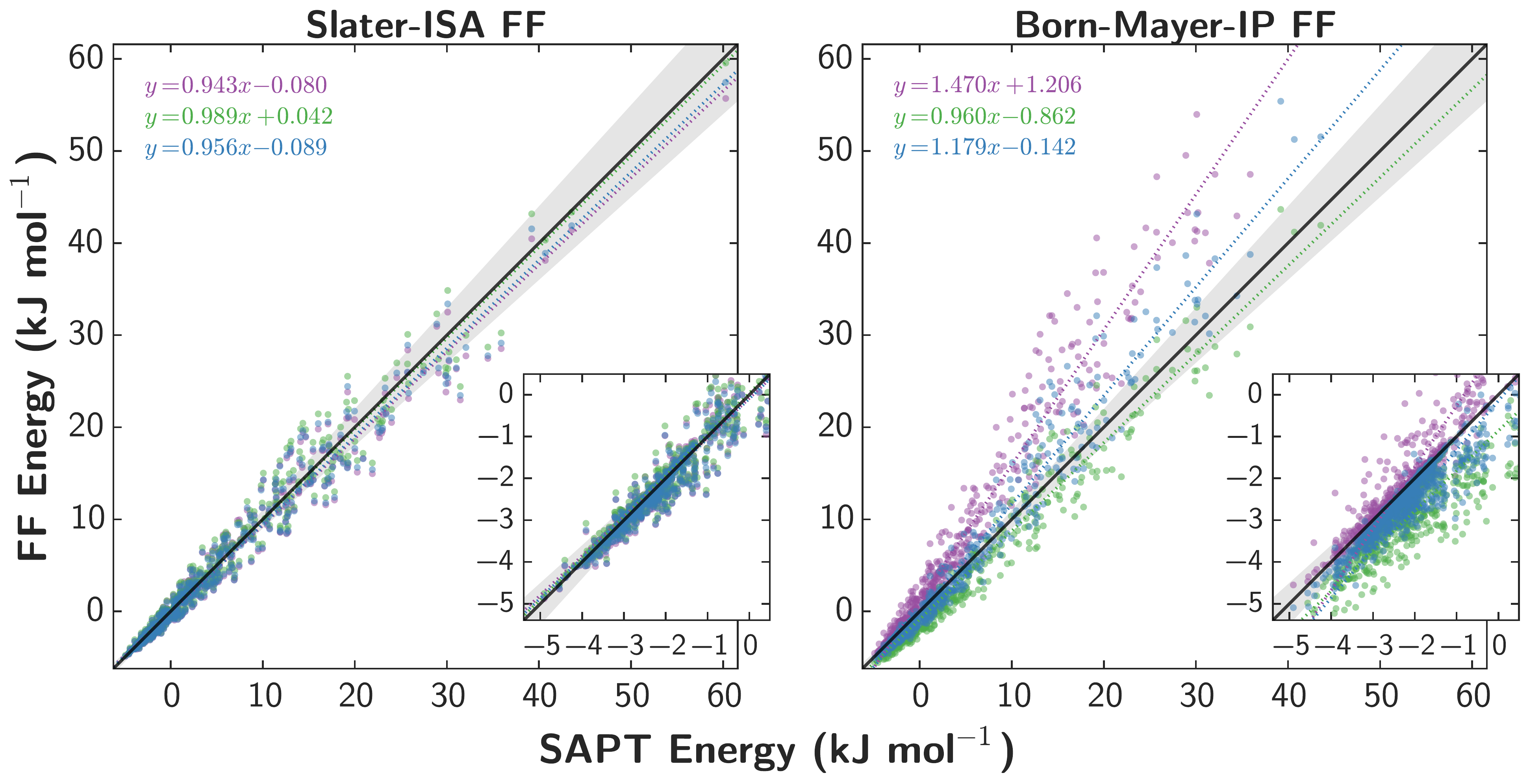}
      \vfill
      \vfill
        \includegraphics[width=0.9\textwidth]{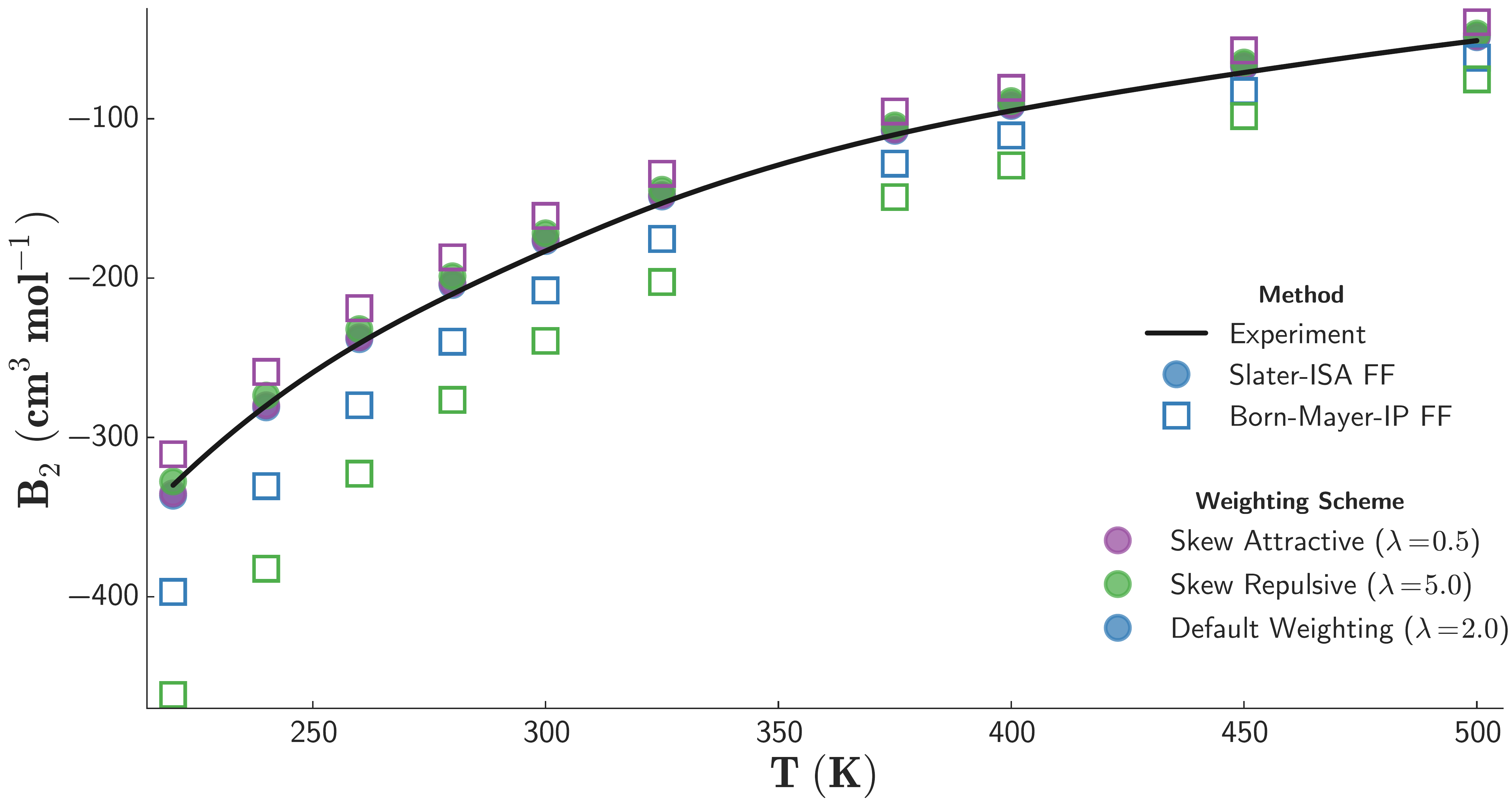}
      \caption[foo]{
        Comparison of the \isaff and the \saptff in terms of sensitivity to the
        weighting function employed in parameter optimization for the ethane
        dimer. Three weighting functions, $\lambda = 0.5$ (purple), $\lambda = 2.0$
        (blue), and $\lambda = 5.0$ (green) are shown, with higher $\lambda$ values
        indicating more weighting of repulsive configurations.
        
        (top) Total interaction energies for the \isaff (left) and the \saptff (right)
        indicating the accuracy of each force field with respect to \sapt benchmark
        energies.  The diagonal line (black) indicates perfect agreement between
        reference energies and each force field, while shaded grey areas represent
        points within $\pm 10\%$ agreement of the benchmark.  To guide the eye, a line
        of best fit (dotted line) has been computed for each force field and for each
        weighting function.
        
        (bottom) Computed 2$^\text{nd}$ virial coefficients for ethane. Data for
        the \isaff and the \saptff are depicted using shaded circles and open squares,
        respectively; colors for the different weighting functions are as above.
        Experimental data from \citen{Dymond1980} (black line) is also shown.
    }
    \label{fig:ethane-weighting}

    \end{figure}

The \isaff fits for the ethane dimer, on the other hand, are nearly completely
insensitive to the weighting function, leading to little intrinsic uncertainty
in the determination of parameters or in the computation of macroscopic
properties. Some other dimers, particularly those where
atomic anisotropy would be anticipated (e.g., water), exhibited slightly larger
sensitivity to the weighting function. Nevertheless, the
vast majority of dimers in the test set are qualitatively insensitive to the choice of
weighting function, and can be optimized with the default $\lambda = 2.0$
weighting function without yielding undue systematic error in the attractive
region of the potential, thus proving the enhanced robustness of the \isaff
model relative to conventional force fields.

\end{subsection}
\begin{subsection}{Next-Generation Born-Mayer Models: \bmsisaff}

We hypothesize that the increased accuracy, transferability, and robustness of
the \isaff is a direct result of its more physically-motivated functional form and
its use of ISA-derived atomic exponents that directly account for the influence
of the molecular environment. Nonetheless, we recognize that the standard
Born-Mayer functional form remains extremely common, both in simulation software and in
existing force fields. It is therefore fruitful to explore the extent to which the \isa
exponents themselves could be used in conjunction with a Born-Mayer functional
form. These results are shown in \tabref{tab:bmsisaff_rmse}.

\begin{landscape}
\begin{table}
\small
\centering
\renewcommand\arraystretch{1.1}
\begin{tabular}{@{}rcccccccc@{}}
\hline
\toprule
& \phantom{ab} &
  \multicolumn{3}{c}{Dimer-Specific Fits} &
  \phantom{ab} &
  \multicolumn{3}{c}{Transferable Fits} \\
\cmidrule{3-5} \cmidrule{7-9}

Component & & \isaff & Born-Mayer-ISA & Born-Mayer-sISA & & \isaff & Born-Mayer-ISA & Born-Mayer-sISA  \\
     & & \multicolumn{1}{c}{(kJ mol$^{-1}$)} & \multicolumn{1}{c}{(kJ mol$^{-1}$)} &  \multicolumn{1}{c}{(kJ mol$^{-1}$)}
     & & \multicolumn{1}{c}{(kJ mol$^{-1}$)}& \multicolumn{1}{c}{(kJ mol$^{-1}$)} &  \multicolumn{1}{c}{(kJ mol$^{-1}$)} \\
\midrule
Exchange        & &    2.641 (0.686)   &  7.030 (1.203)     &  2.677 (0.686)  & &    2.718 (0.720)  &  6.968 (1.228)  & 2.764 (0.706) \\
Electrostatics  & &    1.087 (0.351)   &  1.406 (0.589)     &  1.083 (0.352)  & &    1.134 (0.351)  &  1.461 (0.598)  & 1.141 (0.352) \\
Induction       & &    0.251 (0.095)   &  0.229 (0.097)     &  0.250 (0.096)  & &    0.278 (0.101)  &  0.257 (0.101)  & 0.275 (0.101) \\
\dhf            & &    0.246 (0.068)   &  0.327 (0.120)     &  0.248 (0.068)  & &    0.274 (0.076)  &  0.353 (0.122)  & 0.274 (0.076) \\
Dispersion      & &    0.766 (0.317)   &  3.584 (0.890)     &  0.856 (0.336)  & &    0.766 (0.317)  &  3.584 (0.890)  & 0.856 (0.336) \\
\addlinespace                                                                                                           
\textbf{                                                                                                                    
Total Energy}   \\
\emph{RMSE}
                & &    1.701 (0.464)   &  4.934 (1.054)     &  1.751 (0.453)  & &    1.650 (0.456)  &  4.555 (1.035)  & 1.713 (0.446) \\
\emph{\mse}
                & &    0.216 (0.057)   &  1.127 (0.505)     &  0.258 (0.063)  & &    0.175 (0.051)  &  0.882 (0.516)  & 0.245 (0.057) \\
\bottomrule

\hline
\end{tabular}
\caption{
    Comparison of characteristic RMSE (as described in the main text) over the 91 dimer test
set for the 
    Born-Mayer-sISA approximation compared with other methods.
    For the total energy, both RMSE and absolute mean signed errors
    (MSE) have been shown.
    `Attractive' RMSE, representing the characteristic RMSE for
    the subset of points whose energies are net attractive ($\etot <
    0$), are shown in parentheses to the right of the total RMS
    errors; `attractive' \mse are likewise displayed for the total
    energy.
    \isaff, Born-Mayer-ISA, and \bmsisaff are as described in the main text,
and the `Dimer-Specific' and `Transferable' fits are as described in
\tabref{tab:rmse}.
	}
\label{tab:bmsisaff_rmse}
\end{table}
\normalsize
\end{landscape}

As expected, direct insertion of the \isa exponents into the
Born-Mayer functional form (Born-Mayer-ISA) does not yield promising results.
Indeed, the Born-Mayer-ISA FF has significantly worse RMSE and \mse than
the \saptff. 
We reiterate that the $P = 1$ approximation from \eqref{eq:isaff_sr}, yielding the
conventional Born-Mayer form, is by itself a crude model.
Rather, it becomes necessary to accompany this approximation by a
corresponding exponent scale factor, $\xi$:
\begin{align}
\label{eq:bmsisaff_bi}
B_i &= \xi \Bisa{i}.
\end{align}
Following literature precedent, 
\cite{Ihm1990, McDaniel2012}
we hypothesized that $\xi$ could be treated as a
universal constant. To test this conjecture, we computed reference density overlaps for a
variety of isolated atom pairs (details in the Supporting Information), and
fitted each of these overlaps to a Born-Mayer function of the form 
$S_{ij} \approx K_{ij}\exp(-\xi \Bisa{ij} \R)$, where $K_{ij} =
\frac{K}{B^{3}_{ij}}$ in line with \eqref{eq:isaff_aij}. To very good
approximation, both $K$ and $\xi$ can be treated as universal constants;
that is, neither $K$ nor $\xi$ is sensitive to the value of $\Bisa{}$.
However, fitted values of $K$ and $\xi$ do depend strongly on the range of \R values
used in the optimization, yielding estimates ranging from 0.74 to 0.88.

As an alternative, we optimized $\xi$ directly by minimizing RMSE
against the 91 dimer test set. Results from various choices
of $\xi$ can be found in the Supporting Information.  In agreement
with prior literature and our `first-principles' analysis of overlaps, we find $\xi
= 0.84$ to be optimal for minimizing characteristic overall and attractive RMSE,
though in practice the errors are insensitive to $\xi \in [0.82,0.86]$.
We henceforth use $\xi=0.84$ and refer to to this force field methodology
(Born-Mayer functional form, ISA-derived exponents with
scale factor $\xi=0.84$) as the \bmsisaff. 
Parameters and homo-monomeric fits for the \bmsisaff
can be found in the Supporting Information.
 
From \tabref{tab:bmsisaff_rmse} we see that the \bmsisaff is comparable in
quality to our original \isaff methodology. 
For all attractive configurations, the \bmsisaff is equally
accurate and transferable (\tabref{tab:bmsisaff_rmse}). Furthermore, as shown in
\tabref{tab:rmsd-weightings}, \bmsisaff displays similar parameter robustness
to \isaff. These results suggest that many of the advantages of the \isaff
procedure can be captured simply by using the (scaled) ISA exponents.
Note, however, that the optimal scale factor likely exhibits some system dependence,
and furthermore that the enhanced Slater functional form may be important
where an accurate description of highly repulsive configurations is crucial.

We also examined the \isaff and the \bmsisaff against force
fields where $B_i$ values were instead treated as soft constraints, rather than fixed parameters.
Using entirely unconstrained exponents yields unphysical parameters and a 
severe degradation in force field transferability. Using exponents from the \isaff and
the \bmsisaff as Bayesian priors (in the sense used in
\citens{Misquitta2015a,Misquitta2015b}), 
we generated two new force fields with optimized exponents, denoted Slater-OPT and Born-Mayer-OPT,
respectively. Characteristic RMSE and \mse for these force fields can be found in the Supporting Information.
We find that both methods yield only very minimal improvement, suggesting that the
first-principles ISA exponents are already nearly optimal. Comparing the Born-Mayer-OPT
exponents to those from Slater-ISA, we find a nearly identical average scale factor
of $\gamma = 0.83 \pm 0.07$.  Given that these optimal exponents can now be generated
directly from first principles calculations of the molecular densities via the \isa 
approach of \citeauthor{Misquitta2014}, we anticipate that the \isa densities and
resulting ISA exponents will be extremely useful in next-generation force field
development in order to greatly simplify force field parameterization.

\end{subsection}

%% file: conclusions.tex
We have presented a new methodology for describing short-range intermolecular
interactions based upon a simple model of atom-in-molecule electron density
overlap. The resulting \isaff is a simple extension of the conventional
Born-Mayer functional form, supplemented with atomic exponents determined from
an ISA analysis of the molecular electron density. 
In contrast to simple Born-Mayer or Lennard-Jones models, the \isaff is capable of
reproducing ab initio interaction energies over a wide range of inter-atomic
distances, and displays
extremely low sensitivity to the details of parameterization. Furthermore, the
\isaff exhibits excellent parameter transferability. We thus recommend
\isaff for use in the development of future ab initio (and possibly
empirically-parameterized) potentials, particularly where accuracy across wide
regions of the potential surface is paramount. 

More generally, we find that analysis of the ISA densities provides an
excellent first-principles procedure for the determination of atomic-density
decay exponents.  This analysis improves upon existing approaches (which rely
upon exponents derived from atomic radii or ionization
potentials)\cite{Rappe1992, Mayo1990, Lim2009, VanDuin2001} and explicitly
incorporates the influence of the molecular environment.  These exponents can
be used within \isaff without further parameterization.  Alternatively, in
conjunction with an appropriate scale factor, the exponents can be used to
enhance the accuracy of standard Born-Mayer potentials and/or Tang-Toennies
damping functions. The resulting \bmsisaff retains many of the advantages of
\isaff, but also maintains compatibility with existing force fields and
simulations packages that do not support the Slater functional form.  
Given that the \isa exponents appear to be essentially
optimal with respect to additional empirical optimization, we strongly
recommend use of these first-principles exponents in order to simplify 
(both ab initio and empirical) future force field development involving Born-Mayer
or related functional forms.\cite{Gordon2006}

Overall, \isaff enables a significantly increase in force field accuracy,
particularly in describing short intermolecular contacts. Nevertheless, the
neglect of atomic anisotropy remains, in some cases, a severe approximation.
\cite{Eramian2013, Badenhoop1997, Kim2014b}
Indeed, it has been shown by many
authors\cite{stone2013theory,Day2003,Totton2010a,Wheatley1990} that
quantitatively accurate \A parameters (and to a lesser extent, \B parameters)
require incorporation of angular dependence for the generation of
highly-accurate force fields. This anisotropy becomes crucial when describing
systems containing lone pairs, hydrogen bonds, and/or $\pi$-interactions.
Promisingly, \isa densities naturally describe such anisotropy, 
\cite{Wheatley2012,Misquitta2015a,Misquitta2015b}
and a straightforward method for its inclusion (where essential) in
ab initio force fields is the subject of ongoing work.

%% file: isa_ff.bbl
\providecommand{\latin}[1]{#1}
\providecommand*\mcitethebibliography{\thebibliography}
\csname @ifundefined\endcsname{endmcitethebibliography}
  {\let\endmcitethebibliography\endthebibliography}{}
\begin{mcitethebibliography}{132}
\providecommand*\natexlab[1]{#1}
\providecommand*\mciteSetBstSublistMode[1]{}
\providecommand*\mciteSetBstMaxWidthForm[2]{}
\providecommand*\mciteBstWouldAddEndPuncttrue
  {\def\EndOfBibitem{\unskip.}}
\providecommand*\mciteBstWouldAddEndPunctfalse
  {\let\EndOfBibitem\relax}
\providecommand*\mciteSetBstMidEndSepPunct[3]{}
\providecommand*\mciteSetBstSublistLabelBeginEnd[3]{}
\providecommand*\EndOfBibitem{}
\mciteSetBstSublistMode{f}
\mciteSetBstMaxWidthForm{subitem}{(\alph{mcitesubitemcount})}
\mciteSetBstSublistLabelBeginEnd
  {\mcitemaxwidthsubitemform\space}
  {\relax}
  {\relax}

\bibitem[Stone(2013)]{stone2013theory}
Stone,~A.~J. \emph{{The Theory of Intermolecular Forces}}, 2nd ed.; OUP Oxford,
  2013\relax
\mciteBstWouldAddEndPuncttrue
\mciteSetBstMidEndSepPunct{\mcitedefaultmidpunct}
{\mcitedefaultendpunct}{\mcitedefaultseppunct}\relax
\EndOfBibitem
\bibitem[Margenau and Kestner(1969)Margenau, and Kestner]{margenau1969theory}
Margenau,~H.; Kestner,~N.~R. \emph{{Theory of Intermolecular Forces}};
  International series of monographs in natural philosophy; Pergamon Press:
  Oxford, 1969\relax
\mciteBstWouldAddEndPuncttrue
\mciteSetBstMidEndSepPunct{\mcitedefaultmidpunct}
{\mcitedefaultendpunct}{\mcitedefaultseppunct}\relax
\EndOfBibitem
\bibitem[Riley \latin{et~al.}(2010)Riley, Piton{\v{c}}{\'{a}}k, Jurec{\v{c}}ka,
  and Hobza]{Riley2010}
Riley,~K.~E.; Piton{\v{c}}{\'{a}}k,~M.; Jurec{\v{c}}ka,~P.; Hobza,~P.
  \emph{Chem. Rev.} \textbf{2010}, \emph{110}, 5023--5063\relax
\mciteBstWouldAddEndPuncttrue
\mciteSetBstMidEndSepPunct{\mcitedefaultmidpunct}
{\mcitedefaultendpunct}{\mcitedefaultseppunct}\relax
\EndOfBibitem
\bibitem[Stone and Misquitta(2007)Stone, and Misquitta]{Stone2007}
Stone,~A.~J.; Misquitta,~A.~J. \emph{Int. Rev. Phys. Chem.}; 2007; Vol.~26; pp
  193--222\relax
\mciteBstWouldAddEndPuncttrue
\mciteSetBstMidEndSepPunct{\mcitedefaultmidpunct}
{\mcitedefaultendpunct}{\mcitedefaultseppunct}\relax
\EndOfBibitem
\bibitem[Dykstra and Lisy(2000)Dykstra, and Lisy]{Dykstra2000}
Dykstra,~C.~E.; Lisy,~J.~M. \emph{J. Mol. Struct. THEOCHEM} \textbf{2000},
  \emph{500}, 375--390\relax
\mciteBstWouldAddEndPuncttrue
\mciteSetBstMidEndSepPunct{\mcitedefaultmidpunct}
{\mcitedefaultendpunct}{\mcitedefaultseppunct}\relax
\EndOfBibitem
\bibitem[Stone and Tough(1984)Stone, and Tough]{Stone1984}
Stone,~A.~J.; Tough,~R. \emph{Chem. Phys. Lett.} \textbf{1984}, \emph{110},
  123--129\relax
\mciteBstWouldAddEndPuncttrue
\mciteSetBstMidEndSepPunct{\mcitedefaultmidpunct}
{\mcitedefaultendpunct}{\mcitedefaultseppunct}\relax
\EndOfBibitem
\bibitem[Williams and Stone(2003)Williams, and Stone]{Williams2003}
Williams,~G.~J.; Stone,~A.~J. \emph{J. Chem. Phys.} \textbf{2003}, \emph{119},
  4620--4628\relax
\mciteBstWouldAddEndPuncttrue
\mciteSetBstMidEndSepPunct{\mcitedefaultmidpunct}
{\mcitedefaultendpunct}{\mcitedefaultseppunct}\relax
\EndOfBibitem
\bibitem[Misquitta and Stone(2006)Misquitta, and Stone]{Misquitta2006}
Misquitta,~A.~J.; Stone,~A.~J. \emph{J. Chem. Phys.} \textbf{2006}, \emph{124},
  024111\relax
\mciteBstWouldAddEndPuncttrue
\mciteSetBstMidEndSepPunct{\mcitedefaultmidpunct}
{\mcitedefaultendpunct}{\mcitedefaultseppunct}\relax
\EndOfBibitem
\bibitem[Dehez \latin{et~al.}(2001)Dehez, Chipot, Millot, and
  {\'{A}}ngy{\'{a}}n]{Dehez2001}
Dehez,~F.; Chipot,~C.; Millot,~C.; {\'{A}}ngy{\'{a}}n,~J.~G. \emph{Chem. Phys.
  Lett.} \textbf{2001}, \emph{338}, 180--188\relax
\mciteBstWouldAddEndPuncttrue
\mciteSetBstMidEndSepPunct{\mcitedefaultmidpunct}
{\mcitedefaultendpunct}{\mcitedefaultseppunct}\relax
\EndOfBibitem
\bibitem[Stone(2005)]{Stone2005}
Stone,~A.~J. \emph{J. Chem. Theory Comput.} \textbf{2005}, \emph{1},
  1128--1132\relax
\mciteBstWouldAddEndPuncttrue
\mciteSetBstMidEndSepPunct{\mcitedefaultmidpunct}
{\mcitedefaultendpunct}{\mcitedefaultseppunct}\relax
\EndOfBibitem
\bibitem[Misquitta \latin{et~al.}(2014)Misquitta, Stone, and
  Fazeli]{Misquitta2014}
Misquitta,~A.~J.; Stone,~A.~J.; Fazeli,~F. \emph{J. Chem. Theory Comput.}
  \textbf{2014}, \emph{10}, 5405--5418\relax
\mciteBstWouldAddEndPuncttrue
\mciteSetBstMidEndSepPunct{\mcitedefaultmidpunct}
{\mcitedefaultendpunct}{\mcitedefaultseppunct}\relax
\EndOfBibitem
\bibitem[Jeziorski \latin{et~al.}(1994)Jeziorski, Moszynski, and
  Szalewicz]{Jeziorski1994}
Jeziorski,~B.; Moszynski,~R.; Szalewicz,~K. \emph{Chem. Rev.} \textbf{1994},
  \emph{94}, 1887--1930\relax
\mciteBstWouldAddEndPuncttrue
\mciteSetBstMidEndSepPunct{\mcitedefaultmidpunct}
{\mcitedefaultendpunct}{\mcitedefaultseppunct}\relax
\EndOfBibitem
\bibitem[Szalewicz(2012)]{Szalewicz2012}
Szalewicz,~K. \emph{Wiley Interdiscip. Rev. Comput. Mol. Sci.} \textbf{2012},
  \emph{2}, 254--272\relax
\mciteBstWouldAddEndPuncttrue
\mciteSetBstMidEndSepPunct{\mcitedefaultmidpunct}
{\mcitedefaultendpunct}{\mcitedefaultseppunct}\relax
\EndOfBibitem
\bibitem[Raghavachari \latin{et~al.}(1989)Raghavachari, Trucks, Pople, and
  Head-Gordon]{Raghavachari1989}
Raghavachari,~K.; Trucks,~G.~W.; Pople,~J.~A.; Head-Gordon,~M. \emph{Chem.
  Phys. Lett.} \textbf{1989}, \emph{157}, 479--483\relax
\mciteBstWouldAddEndPuncttrue
\mciteSetBstMidEndSepPunct{\mcitedefaultmidpunct}
{\mcitedefaultendpunct}{\mcitedefaultseppunct}\relax
\EndOfBibitem
\bibitem[Grimme and Djukic(2011)Grimme, and Djukic]{Grimme2011}
Grimme,~S.; Djukic,~J.~P. \emph{Inorg. Chem.} \textbf{2011}, \emph{50},
  2619--2628\relax
\mciteBstWouldAddEndPuncttrue
\mciteSetBstMidEndSepPunct{\mcitedefaultmidpunct}
{\mcitedefaultendpunct}{\mcitedefaultseppunct}\relax
\EndOfBibitem
\bibitem[Lennard-Jones(1931)]{Lennard-Jones1931}
Lennard-Jones,~J. \emph{Proc. Phys. Soc.} \textbf{1931}, \emph{43},
  461--482\relax
\mciteBstWouldAddEndPuncttrue
\mciteSetBstMidEndSepPunct{\mcitedefaultmidpunct}
{\mcitedefaultendpunct}{\mcitedefaultseppunct}\relax
\EndOfBibitem
\bibitem[Born and Mayer(1932)Born, and Mayer]{Born1932}
Born,~M.; Mayer,~J.~E. \emph{Zeitschrift f{\"{u}}r Phys.} \textbf{1932},
  \emph{75}, 1--18\relax
\mciteBstWouldAddEndPuncttrue
\mciteSetBstMidEndSepPunct{\mcitedefaultmidpunct}
{\mcitedefaultendpunct}{\mcitedefaultseppunct}\relax
\EndOfBibitem
\bibitem[Buckingham(1938)]{Buckingham1938}
Buckingham,~R.~A. \emph{Proc. R. Soc. A Math. Phys. Eng. Sci.} \textbf{1938},
  \emph{168}, 264--283\relax
\mciteBstWouldAddEndPuncttrue
\mciteSetBstMidEndSepPunct{\mcitedefaultmidpunct}
{\mcitedefaultendpunct}{\mcitedefaultseppunct}\relax
\EndOfBibitem
\bibitem[Nezbeda(2005)]{Nezbeda2005}
Nezbeda,~I. \emph{Mol. Phys.} \textbf{2005}, \emph{103}, 59--76\relax
\mciteBstWouldAddEndPuncttrue
\mciteSetBstMidEndSepPunct{\mcitedefaultmidpunct}
{\mcitedefaultendpunct}{\mcitedefaultseppunct}\relax
\EndOfBibitem
\bibitem[Galliero and Boned(2008)Galliero, and Boned]{Galliero2008}
Galliero,~G.; Boned,~C. \emph{J. Chem. Phys.} \textbf{2008}, \emph{129}\relax
\mciteBstWouldAddEndPuncttrue
\mciteSetBstMidEndSepPunct{\mcitedefaultmidpunct}
{\mcitedefaultendpunct}{\mcitedefaultseppunct}\relax
\EndOfBibitem
\bibitem[Gordon(2006)]{Gordon2006}
Gordon,~P.~A. \emph{J. Chem. Phys.} \textbf{2006}, \emph{125}\relax
\mciteBstWouldAddEndPuncttrue
\mciteSetBstMidEndSepPunct{\mcitedefaultmidpunct}
{\mcitedefaultendpunct}{\mcitedefaultseppunct}\relax
\EndOfBibitem
\bibitem[Ruckenstein and Liu(1997)Ruckenstein, and Liu]{Ruckenstein1997}
Ruckenstein,~E.; Liu,~H. \emph{Society} \textbf{1997}, 3927--3936\relax
\mciteBstWouldAddEndPuncttrue
\mciteSetBstMidEndSepPunct{\mcitedefaultmidpunct}
{\mcitedefaultendpunct}{\mcitedefaultseppunct}\relax
\EndOfBibitem
\bibitem[Galliero \latin{et~al.}(2007)Galliero, Boned, Baylaucq, and
  Montel]{Galliero2007}
Galliero,~G.; Boned,~C.; Baylaucq,~A.; Montel,~F. \emph{Chem. Phys.}
  \textbf{2007}, \emph{333}, 219--228\relax
\mciteBstWouldAddEndPuncttrue
\mciteSetBstMidEndSepPunct{\mcitedefaultmidpunct}
{\mcitedefaultendpunct}{\mcitedefaultseppunct}\relax
\EndOfBibitem
\bibitem[Wu and Sadus(2000)Wu, and Sadus]{Wu2000}
Wu,~G.-W.; Sadus,~R.~J. \emph{Fluid Phase Equilib.} \textbf{2000}, \emph{170},
  269--284\relax
\mciteBstWouldAddEndPuncttrue
\mciteSetBstMidEndSepPunct{\mcitedefaultmidpunct}
{\mcitedefaultendpunct}{\mcitedefaultseppunct}\relax
\EndOfBibitem
\bibitem[Errington and Panagiotopoulos(1998)Errington, and
  Panagiotopoulos]{Errington1998}
Errington,~J.~R.; Panagiotopoulos,~A.~Z. \emph{J. Chem. Phys.} \textbf{1998},
  \emph{109}, 1093--1100\relax
\mciteBstWouldAddEndPuncttrue
\mciteSetBstMidEndSepPunct{\mcitedefaultmidpunct}
{\mcitedefaultendpunct}{\mcitedefaultseppunct}\relax
\EndOfBibitem
\bibitem[McGrath \latin{et~al.}(2010)McGrath, Ghogomu, Tsona, Siepmann, Chen,
  Napari, and Vehkamaki]{McGrath2010}
McGrath,~M.~J.; Ghogomu,~J.~N.; Tsona,~N.~T.; Siepmann,~J.~I.; Chen,~B.;
  Napari,~I.; Vehkamaki,~H. \emph{J. Chem. Phys.} \textbf{2010},
  \emph{133}\relax
\mciteBstWouldAddEndPuncttrue
\mciteSetBstMidEndSepPunct{\mcitedefaultmidpunct}
{\mcitedefaultendpunct}{\mcitedefaultseppunct}\relax
\EndOfBibitem
\bibitem[Parker and Sherrill(2015)Parker, and Sherrill]{Parker2015}
Parker,~T.~M.; Sherrill,~C.~D. \emph{J. Chem. Theory Comput.} \textbf{2015},
  \emph{11}, 4197--4204\relax
\mciteBstWouldAddEndPuncttrue
\mciteSetBstMidEndSepPunct{\mcitedefaultmidpunct}
{\mcitedefaultendpunct}{\mcitedefaultseppunct}\relax
\EndOfBibitem
\bibitem[Sherrill \latin{et~al.}(2009)Sherrill, Sumpter, Sinnokrot, Marshall,
  Hohenstein, Walker, and Gould]{Sherrill2009}
Sherrill,~C.~D.; Sumpter,~B.~G.; Sinnokrot,~M.~O.; Marshall,~M.~S.;
  Hohenstein,~E.~G.; Walker,~R.~C.; Gould,~I.~R. \emph{J. Comput. Chem.}
  \textbf{2009}, \emph{30}, 2187--2193\relax
\mciteBstWouldAddEndPuncttrue
\mciteSetBstMidEndSepPunct{\mcitedefaultmidpunct}
{\mcitedefaultendpunct}{\mcitedefaultseppunct}\relax
\EndOfBibitem
\bibitem[Zgarbov{\'{a}} \latin{et~al.}(2010)Zgarbov{\'{a}}, Otyepka, Sponer,
  Hobza, and Jurecka]{Zgarbova2010}
Zgarbov{\'{a}},~M.; Otyepka,~M.; Sponer,~J.; Hobza,~P.; Jurecka,~P. \emph{Phys.
  Chem. Chem. Phys.} \textbf{2010}, \emph{12}, 10476--10493\relax
\mciteBstWouldAddEndPuncttrue
\mciteSetBstMidEndSepPunct{\mcitedefaultmidpunct}
{\mcitedefaultendpunct}{\mcitedefaultseppunct}\relax
\EndOfBibitem
\bibitem[Bastea(2003)]{Bastea2003}
Bastea,~S. \emph{Phys. Rev. E. Stat. Nonlin. Soft Matter Phys.} \textbf{2003},
  \emph{68}, 031204\relax
\mciteBstWouldAddEndPuncttrue
\mciteSetBstMidEndSepPunct{\mcitedefaultmidpunct}
{\mcitedefaultendpunct}{\mcitedefaultseppunct}\relax
\EndOfBibitem
\bibitem[Errington and Panagiotopoulos(1999)Errington, and
  Panagiotopoulos]{Errington1999}
Errington,~J.~R.; Panagiotopoulos,~A.~Z. \emph{J. Phys. Chem. B} \textbf{1999},
  \emph{103}, 6314--6322\relax
\mciteBstWouldAddEndPuncttrue
\mciteSetBstMidEndSepPunct{\mcitedefaultmidpunct}
{\mcitedefaultendpunct}{\mcitedefaultseppunct}\relax
\EndOfBibitem
\bibitem[Ross and Ree(1980)Ross, and Ree]{Ross1980}
Ross,~M.; Ree,~F.~H. \emph{J. Chem. Phys.} \textbf{1980}, \emph{73}, 6146\relax
\mciteBstWouldAddEndPuncttrue
\mciteSetBstMidEndSepPunct{\mcitedefaultmidpunct}
{\mcitedefaultendpunct}{\mcitedefaultseppunct}\relax
\EndOfBibitem
\bibitem[Schmidt \latin{et~al.}(2015)Schmidt, Yu, and McDaniel]{Schmidt2015}
Schmidt,~J.~R.; Yu,~K.; McDaniel,~J.~G. \emph{Acc. Chem. Res.} \textbf{2015},
  \emph{48}, 548--556\relax
\mciteBstWouldAddEndPuncttrue
\mciteSetBstMidEndSepPunct{\mcitedefaultmidpunct}
{\mcitedefaultendpunct}{\mcitedefaultseppunct}\relax
\EndOfBibitem
\bibitem[Abrahamson(1963)]{Abrahamson1963}
Abrahamson,~A.~A. \emph{Phys. Rev.} \textbf{1963}, \emph{130}, 693--707\relax
\mciteBstWouldAddEndPuncttrue
\mciteSetBstMidEndSepPunct{\mcitedefaultmidpunct}
{\mcitedefaultendpunct}{\mcitedefaultseppunct}\relax
\EndOfBibitem
\bibitem[Mackerell(2004)]{Mackerell2004}
Mackerell,~A.~D. \emph{J. Comput. Chem.} \textbf{2004}, \emph{25},
  1584--1604\relax
\mciteBstWouldAddEndPuncttrue
\mciteSetBstMidEndSepPunct{\mcitedefaultmidpunct}
{\mcitedefaultendpunct}{\mcitedefaultseppunct}\relax
\EndOfBibitem
\bibitem[Halgren(1992)]{Halgren1992}
Halgren,~T.~A. \emph{J. Am. Chem. Soc.} \textbf{1992}, \emph{114},
  7827--7843\relax
\mciteBstWouldAddEndPuncttrue
\mciteSetBstMidEndSepPunct{\mcitedefaultmidpunct}
{\mcitedefaultendpunct}{\mcitedefaultseppunct}\relax
\EndOfBibitem
\bibitem[Kim \latin{et~al.}(1981)Kim, Kim, and Lee]{Kim1981}
Kim,~Y.~S.; Kim,~S.~K.; Lee,~W.~D. \emph{Chem. Phys. Lett.} \textbf{1981},
  \emph{80}, 574--575\relax
\mciteBstWouldAddEndPuncttrue
\mciteSetBstMidEndSepPunct{\mcitedefaultmidpunct}
{\mcitedefaultendpunct}{\mcitedefaultseppunct}\relax
\EndOfBibitem
\bibitem[Misquitta and Stone(2015)Misquitta, and Stone]{Misquitta2015a}
Misquitta,~A.~J.; Stone,~A.~J. {Ab initio atom-atom potentials using CamCASP:
  Theory}. 2015; \url{https://arxiv.org/abs/1512.06150v2}\relax
\mciteBstWouldAddEndPuncttrue
\mciteSetBstMidEndSepPunct{\mcitedefaultmidpunct}
{\mcitedefaultendpunct}{\mcitedefaultseppunct}\relax
\EndOfBibitem
\bibitem[Nyeland and Toennies(1986)Nyeland, and Toennies]{Nyeland1986}
Nyeland,~C.; Toennies,~J.~P. \emph{Chem. Phys. Lett.} \textbf{1986},
  \emph{127}, 3--8\relax
\mciteBstWouldAddEndPuncttrue
\mciteSetBstMidEndSepPunct{\mcitedefaultmidpunct}
{\mcitedefaultendpunct}{\mcitedefaultseppunct}\relax
\EndOfBibitem
\bibitem[Ihm \latin{et~al.}(1990)Ihm, Cole, Toigo, and Klein]{Ihm1990}
Ihm,~G.; Cole,~M.~W.; Toigo,~F.; Klein,~J.~R. \emph{Phys. Rev. A}
  \textbf{1990}, \emph{42}, 5244--5252\relax
\mciteBstWouldAddEndPuncttrue
\mciteSetBstMidEndSepPunct{\mcitedefaultmidpunct}
{\mcitedefaultendpunct}{\mcitedefaultseppunct}\relax
\EndOfBibitem
\bibitem[Duke \latin{et~al.}(2014)Duke, Starovoytov, Piquemal, and
  Cisneros]{Duke2014}
Duke,~R.~E.; Starovoytov,~O.~N.; Piquemal,~J.-P.; Cisneros,~G.~A. \emph{J.
  Chem. Theory Comput.} \textbf{2014}, \emph{10}, 1361--1365\relax
\mciteBstWouldAddEndPuncttrue
\mciteSetBstMidEndSepPunct{\mcitedefaultmidpunct}
{\mcitedefaultendpunct}{\mcitedefaultseppunct}\relax
\EndOfBibitem
\bibitem[Elking \latin{et~al.}(2010)Elking, Cisneros, Piquemal, Darden, and
  Pedersen]{Elking2010}
Elking,~D.~M.; Cisneros,~G.~A.; Piquemal,~J.~P.; Darden,~T.~A.; Pedersen,~L.~G.
  \emph{J. Chem. Theory Comput.} \textbf{2010}, \emph{6}, 190--202\relax
\mciteBstWouldAddEndPuncttrue
\mciteSetBstMidEndSepPunct{\mcitedefaultmidpunct}
{\mcitedefaultendpunct}{\mcitedefaultseppunct}\relax
\EndOfBibitem
\bibitem[Cisneros \latin{et~al.}(2006)Cisneros, Piquemal, and
  Darden]{Cisneros2006a}
Cisneros,~G.~A.; Piquemal,~J.~P.; Darden,~T.~A. \emph{J. Chem. Phys.}
  \textbf{2006}, \emph{125}\relax
\mciteBstWouldAddEndPuncttrue
\mciteSetBstMidEndSepPunct{\mcitedefaultmidpunct}
{\mcitedefaultendpunct}{\mcitedefaultseppunct}\relax
\EndOfBibitem
\bibitem[Chaudret \latin{et~al.}(2014)Chaudret, Gresh, Narth, Lagardere,
  Darden, Cisneros, and Piquemal]{Chaudret2014}
Chaudret,~R.; Gresh,~N.; Narth,~C.; Lagardere,~L.; Darden,~T.~A.;
  Cisneros,~G.~A.; Piquemal,~J.~P. \emph{J. Phys. Chem. A} \textbf{2014},
  \emph{118}, 7598--7612\relax
\mciteBstWouldAddEndPuncttrue
\mciteSetBstMidEndSepPunct{\mcitedefaultmidpunct}
{\mcitedefaultendpunct}{\mcitedefaultseppunct}\relax
\EndOfBibitem
\bibitem[Chaudret \latin{et~al.}(2013)Chaudret, Gresh, Cisneros, Scemama, and
  Piquemal]{Chaudret2013}
Chaudret,~R.; Gresh,~N.; Cisneros,~G.~A.; Scemama,~A.; Piquemal,~J.-p.
  \emph{Can. J. Chem.} \textbf{2013}, \emph{91}, 804--810\relax
\mciteBstWouldAddEndPuncttrue
\mciteSetBstMidEndSepPunct{\mcitedefaultmidpunct}
{\mcitedefaultendpunct}{\mcitedefaultseppunct}\relax
\EndOfBibitem
\bibitem[{\"{O}}hrn \latin{et~al.}(2016){\"{O}}hrn, Hermida-Ramon, and
  Karlstr{\"{o}}m]{Ohrn2016}
{\"{O}}hrn,~A.; Hermida-Ramon,~J.~M.; Karlstr{\"{o}}m,~G. \emph{J. Chem. Theory
  Comput.} \textbf{2016}, \emph{12}, 2298--2311\relax
\mciteBstWouldAddEndPuncttrue
\mciteSetBstMidEndSepPunct{\mcitedefaultmidpunct}
{\mcitedefaultendpunct}{\mcitedefaultseppunct}\relax
\EndOfBibitem
\bibitem[Gresh \latin{et~al.}(2007)Gresh, Cisneros, Darden, and
  Piquemal]{Gresh2007}
Gresh,~N.; Cisneros,~G.~A.; Darden,~T.~A.; Piquemal,~J.-P. \emph{J. Chem.
  Theory Comput.} \textbf{2007}, \emph{3}, 1960--1986\relax
\mciteBstWouldAddEndPuncttrue
\mciteSetBstMidEndSepPunct{\mcitedefaultmidpunct}
{\mcitedefaultendpunct}{\mcitedefaultseppunct}\relax
\EndOfBibitem
\bibitem[Gordon \latin{et~al.}(2001)Gordon, Freitag, Bandyopadhyay, Jensen,
  Kairys, and Stevens]{Gordon2001}
Gordon,~M.~S.; Freitag,~M.~A.; Bandyopadhyay,~P.; Jensen,~J.~H.; Kairys,~V.;
  Stevens,~W.~J. \emph{J. Phys. Chem. A} \textbf{2001}, \emph{105},
  293--307\relax
\mciteBstWouldAddEndPuncttrue
\mciteSetBstMidEndSepPunct{\mcitedefaultmidpunct}
{\mcitedefaultendpunct}{\mcitedefaultseppunct}\relax
\EndOfBibitem
\bibitem[Xie and Gao(2007)Xie, and Gao]{Xie2007}
Xie,~W.; Gao,~J. \emph{J. Chem. Theory Comput.} \textbf{2007}, \emph{3},
  1890--1900\relax
\mciteBstWouldAddEndPuncttrue
\mciteSetBstMidEndSepPunct{\mcitedefaultmidpunct}
{\mcitedefaultendpunct}{\mcitedefaultseppunct}\relax
\EndOfBibitem
\bibitem[Xie \latin{et~al.}(2009)Xie, Orozco, Truhlar, and Gao]{Xie2009}
Xie,~W.; Orozco,~M.; Truhlar,~D.~G.; Gao,~J. \emph{J. Chem. Theory Comput.}
  \textbf{2009}, \emph{5}, 459--467\relax
\mciteBstWouldAddEndPuncttrue
\mciteSetBstMidEndSepPunct{\mcitedefaultmidpunct}
{\mcitedefaultendpunct}{\mcitedefaultseppunct}\relax
\EndOfBibitem
\bibitem[Patil and Tang(2000)Patil, and Tang]{PatilT-AsymptoticMethods}
Patil,~S.~H.; Tang,~K.~T. In \emph{{Asymptotic methods in quantum mechanics}};
  Sc{\"{a}}fer,~F.~P., Toennies,~J.~P., Zinth,~W., Eds.; Springer, 2000\relax
\mciteBstWouldAddEndPuncttrue
\mciteSetBstMidEndSepPunct{\mcitedefaultmidpunct}
{\mcitedefaultendpunct}{\mcitedefaultseppunct}\relax
\EndOfBibitem
\bibitem[Hoffmann-Ostenhof and Hoffmann-Ostenhof(1977)Hoffmann-Ostenhof, and
  Hoffmann-Ostenhof]{Hoffmann-Ostenhof1977}
Hoffmann-Ostenhof,~M.; Hoffmann-Ostenhof,~T. \emph{Phys. Rev. A} \textbf{1977},
  \emph{16}, 1782--1785\relax
\mciteBstWouldAddEndPuncttrue
\mciteSetBstMidEndSepPunct{\mcitedefaultmidpunct}
{\mcitedefaultendpunct}{\mcitedefaultseppunct}\relax
\EndOfBibitem
\bibitem[Amovilli and March(2006)Amovilli, and March]{Amovilli2006}
Amovilli,~C.; March,~N.~H. \emph{J. Phys. A. Math. Gen.} \textbf{2006},
  \emph{39}, 7349--7357\relax
\mciteBstWouldAddEndPuncttrue
\mciteSetBstMidEndSepPunct{\mcitedefaultmidpunct}
{\mcitedefaultendpunct}{\mcitedefaultseppunct}\relax
\EndOfBibitem
\bibitem[Bunge and Esquivel(1986)Bunge, and Esquivel]{Bunge1986}
Bunge,~A.~V.; Esquivel,~R.~O. \emph{Phys. Rev. A} \textbf{1986}, \emph{34},
  853\relax
\mciteBstWouldAddEndPuncttrue
\mciteSetBstMidEndSepPunct{\mcitedefaultmidpunct}
{\mcitedefaultendpunct}{\mcitedefaultseppunct}\relax
\EndOfBibitem
\bibitem[Tai(1986)]{Tai1986}
Tai,~H. \emph{Phys. Rev. A} \textbf{1986}, \emph{33}, 3657--3666\relax
\mciteBstWouldAddEndPuncttrue
\mciteSetBstMidEndSepPunct{\mcitedefaultmidpunct}
{\mcitedefaultendpunct}{\mcitedefaultseppunct}\relax
\EndOfBibitem
\bibitem[Rosen(1931)]{Rosen1931}
Rosen,~N. \emph{Phys. Rev. Lett.} \textbf{1931}, \emph{38}, 255--276\relax
\mciteBstWouldAddEndPuncttrue
\mciteSetBstMidEndSepPunct{\mcitedefaultmidpunct}
{\mcitedefaultendpunct}{\mcitedefaultseppunct}\relax
\EndOfBibitem
\bibitem[Rappe \latin{et~al.}(1992)Rappe, Casewit, Colwell, Goddard, and
  Skiff]{Rappe1992}
Rappe,~A.~K.; Casewit,~C.; Colwell,~K.; Goddard,~W. A.~I.; Skiff,~W. \emph{J.
  Am. Chem. Soc.} \textbf{1992}, \emph{114}, 10024--10035\relax
\mciteBstWouldAddEndPuncttrue
\mciteSetBstMidEndSepPunct{\mcitedefaultmidpunct}
{\mcitedefaultendpunct}{\mcitedefaultseppunct}\relax
\EndOfBibitem
\bibitem[Waldman and Hagler(1993)Waldman, and Hagler]{Waldman1993}
Waldman,~M.; Hagler,~A.~T. \emph{J. Comput. Chem.} \textbf{1993}, \emph{14},
  1077--1084\relax
\mciteBstWouldAddEndPuncttrue
\mciteSetBstMidEndSepPunct{\mcitedefaultmidpunct}
{\mcitedefaultendpunct}{\mcitedefaultseppunct}\relax
\EndOfBibitem
\bibitem[McDaniel and Schmidt(2012)McDaniel, and Schmidt]{McDaniel2012}
McDaniel,~J.~G.; Schmidt,~J.~R. \emph{J. Phys. Chem. C} \textbf{2012},
  \emph{116}, 14031--14039\relax
\mciteBstWouldAddEndPuncttrue
\mciteSetBstMidEndSepPunct{\mcitedefaultmidpunct}
{\mcitedefaultendpunct}{\mcitedefaultseppunct}\relax
\EndOfBibitem
\bibitem[Podeszwa \latin{et~al.}(2006)Podeszwa, Bukowski, and
  Szalewicz]{Podeszwa2006}
Podeszwa,~R.; Bukowski,~R.; Szalewicz,~K. \emph{J. Phys. Chem. A}
  \textbf{2006}, \emph{110}, 10345--10354\relax
\mciteBstWouldAddEndPuncttrue
\mciteSetBstMidEndSepPunct{\mcitedefaultmidpunct}
{\mcitedefaultendpunct}{\mcitedefaultseppunct}\relax
\EndOfBibitem
\bibitem[Bukowski \latin{et~al.}(2006)Bukowski, Szalewicz, Groenenboom, and
  van~der Avoird]{Bukowski2006}
Bukowski,~R.; Szalewicz,~K.; Groenenboom,~G.; van~der Avoird,~A. \emph{J. Chem.
  Phys.} \textbf{2006}, \emph{125}, 044301\relax
\mciteBstWouldAddEndPuncttrue
\mciteSetBstMidEndSepPunct{\mcitedefaultmidpunct}
{\mcitedefaultendpunct}{\mcitedefaultseppunct}\relax
\EndOfBibitem
\bibitem[Jeziorska \latin{et~al.}(2007)Jeziorska, Cencek, Patkowski, Jeziorski,
  and Szalewicz]{Jeziorska2007}
Jeziorska,~M.; Cencek,~W.; Patkowski,~K.; Jeziorski,~B.; Szalewicz,~K. \emph{J.
  Chem. Phys.} \textbf{2007}, \emph{127}, 124303\relax
\mciteBstWouldAddEndPuncttrue
\mciteSetBstMidEndSepPunct{\mcitedefaultmidpunct}
{\mcitedefaultendpunct}{\mcitedefaultseppunct}\relax
\EndOfBibitem
\bibitem[Sum \latin{et~al.}(2002)Sum, Sandler, Bukowski, and
  Szalewicz]{Sum2002}
Sum,~A.~K.; Sandler,~S.~I.; Bukowski,~R.; Szalewicz,~K. \emph{J. Chem. Phys.}
  \textbf{2002}, \emph{116}, 7637\relax
\mciteBstWouldAddEndPuncttrue
\mciteSetBstMidEndSepPunct{\mcitedefaultmidpunct}
{\mcitedefaultendpunct}{\mcitedefaultseppunct}\relax
\EndOfBibitem
\bibitem[Konieczny and Sokalski(2015)Konieczny, and Sokalski]{Konieczny2015}
Konieczny,~J.~K.; Sokalski,~W.~A. \emph{J. Mol. Model.} \textbf{2015},
  \emph{21}, 197\relax
\mciteBstWouldAddEndPuncttrue
\mciteSetBstMidEndSepPunct{\mcitedefaultmidpunct}
{\mcitedefaultendpunct}{\mcitedefaultseppunct}\relax
\EndOfBibitem
\bibitem[Kita \latin{et~al.}(1976)Kita, Noda, and Inouye]{Kita1976a}
Kita,~S.; Noda,~K.; Inouye,~H. \emph{J. Chem. Phys.} \textbf{1976}, \emph{64},
  3446--3449\relax
\mciteBstWouldAddEndPuncttrue
\mciteSetBstMidEndSepPunct{\mcitedefaultmidpunct}
{\mcitedefaultendpunct}{\mcitedefaultseppunct}\relax
\EndOfBibitem
\bibitem[Kuechler \latin{et~al.}(2015)Kuechler, Giese, and York]{Kuechler2015}
Kuechler,~E.~R.; Giese,~T.~J.; York,~D.~M. \emph{J. Chem. Phys.} \textbf{2015},
  \emph{143}\relax
\mciteBstWouldAddEndPuncttrue
\mciteSetBstMidEndSepPunct{\mcitedefaultmidpunct}
{\mcitedefaultendpunct}{\mcitedefaultseppunct}\relax
\EndOfBibitem
\bibitem[Giese and York(2007)Giese, and York]{Giese2007}
Giese,~T.~J.; York,~D.~M. \emph{J. Chem. Phys.} \textbf{2007}, \emph{127}\relax
\mciteBstWouldAddEndPuncttrue
\mciteSetBstMidEndSepPunct{\mcitedefaultmidpunct}
{\mcitedefaultendpunct}{\mcitedefaultseppunct}\relax
\EndOfBibitem
\bibitem[Giese \latin{et~al.}(2013)Giese, Chen, Dissanayake, Giambaşu,
  Heldenbrand, Huang, Kuechler, Lee, Panteva, Radak, and York]{Giese2013}
Giese,~T.~J.; Chen,~H.; Dissanayake,~T.; Giambaşu,~G.~M.; Heldenbrand,~H.;
  Huang,~M.; Kuechler,~E.~R.; Lee,~T.~S.; Panteva,~M.~T.; Radak,~B.~K.;
  York,~D.~M. \emph{J. Chem. Theory Comput.} \textbf{2013}, \emph{9},
  1417--1427\relax
\mciteBstWouldAddEndPuncttrue
\mciteSetBstMidEndSepPunct{\mcitedefaultmidpunct}
{\mcitedefaultendpunct}{\mcitedefaultseppunct}\relax
\EndOfBibitem
\bibitem[Day and Price(2003)Day, and Price]{Day2003}
Day,~G.~M.; Price,~S.~L. \emph{J. Am. Chem. Soc.} \textbf{2003}, \emph{125},
  16434--16443\relax
\mciteBstWouldAddEndPuncttrue
\mciteSetBstMidEndSepPunct{\mcitedefaultmidpunct}
{\mcitedefaultendpunct}{\mcitedefaultseppunct}\relax
\EndOfBibitem
\bibitem[Nobeli \latin{et~al.}(1998)Nobeli, Price, and Wheatley]{Nobeli1998}
Nobeli,~I.; Price,~S.~L.; Wheatley,~R.~J. \emph{Mol. Phys.} \textbf{1998},
  \emph{95}, 525--537\relax
\mciteBstWouldAddEndPuncttrue
\mciteSetBstMidEndSepPunct{\mcitedefaultmidpunct}
{\mcitedefaultendpunct}{\mcitedefaultseppunct}\relax
\EndOfBibitem
\bibitem[McDaniel and Schmidt(2013)McDaniel, and Schmidt]{McDaniel2013}
McDaniel,~J.~G.; Schmidt,~J.~R. \emph{J. Phys. Chem. A} \textbf{2013},
  \emph{117}, 2053--2066\relax
\mciteBstWouldAddEndPuncttrue
\mciteSetBstMidEndSepPunct{\mcitedefaultmidpunct}
{\mcitedefaultendpunct}{\mcitedefaultseppunct}\relax
\EndOfBibitem
\bibitem[Totton \latin{et~al.}(2010)Totton, Misquitta, and Kraft]{Totton2010}
Totton,~T.~S.; Misquitta,~A.~J.; Kraft,~M. \emph{J. Chem. Theory Comput.}
  \textbf{2010}, \emph{6}, 683--695\relax
\mciteBstWouldAddEndPuncttrue
\mciteSetBstMidEndSepPunct{\mcitedefaultmidpunct}
{\mcitedefaultendpunct}{\mcitedefaultseppunct}\relax
\EndOfBibitem
\bibitem[Misquitta(2013)]{Misquitta2013}
Misquitta,~A.~J. \emph{J. Chem. Theory Comput.} \textbf{2013}, \emph{9},
  5313--5326\relax
\mciteBstWouldAddEndPuncttrue
\mciteSetBstMidEndSepPunct{\mcitedefaultmidpunct}
{\mcitedefaultendpunct}{\mcitedefaultseppunct}\relax
\EndOfBibitem
\bibitem[Misquitta and Stone(2015)Misquitta, and Stone]{Misquitta2015b}
Misquitta,~A.~J.; Stone,~A.~J. {Ab initio atom-atom potentials using CamCASP:
  Application to Pyridine}. 2015;
  \url{https://arxiv.org/abs/1512.06155v2}\relax
\mciteBstWouldAddEndPuncttrue
\mciteSetBstMidEndSepPunct{\mcitedefaultmidpunct}
{\mcitedefaultendpunct}{\mcitedefaultseppunct}\relax
\EndOfBibitem
\bibitem[Tang and Toennies(1984)Tang, and Toennies]{Tang1984}
Tang,~K.~T.; Toennies,~J.~P. \emph{J. Chem. Phys.} \textbf{1984}, \emph{80},
  3726--3741\relax
\mciteBstWouldAddEndPuncttrue
\mciteSetBstMidEndSepPunct{\mcitedefaultmidpunct}
{\mcitedefaultendpunct}{\mcitedefaultseppunct}\relax
\EndOfBibitem
\bibitem[Tang and Toennies(1992)Tang, and Toennies]{Tang1992}
Tang,~K.~T.; Toennies,~J.~P. \emph{Surf. Sci.} \textbf{1992}, \emph{279},
  L203--L206\relax
\mciteBstWouldAddEndPuncttrue
\mciteSetBstMidEndSepPunct{\mcitedefaultmidpunct}
{\mcitedefaultendpunct}{\mcitedefaultseppunct}\relax
\EndOfBibitem
\bibitem[Wheatley and Price(1990)Wheatley, and Price]{Wheatley1990}
Wheatley,~R.~J.; Price,~S.~L. \emph{Mol. Phys.} \textbf{1990}, \emph{69},
  507--533\relax
\mciteBstWouldAddEndPuncttrue
\mciteSetBstMidEndSepPunct{\mcitedefaultmidpunct}
{\mcitedefaultendpunct}{\mcitedefaultseppunct}\relax
\EndOfBibitem
\bibitem[Mitchell and Price(2000)Mitchell, and Price]{Mitchell2000}
Mitchell,~J. B.~O.; Price,~S.~L. \emph{J. Phys. Chem. A} \textbf{2000},
  \emph{104}, 10958--10971\relax
\mciteBstWouldAddEndPuncttrue
\mciteSetBstMidEndSepPunct{\mcitedefaultmidpunct}
{\mcitedefaultendpunct}{\mcitedefaultseppunct}\relax
\EndOfBibitem
\bibitem[S{\"{o}}derhjelm \latin{et~al.}(2006)S{\"{o}}derhjelm,
  Karlstr{\"{o}}m, and Ryde]{Soderhjelm2006}
S{\"{o}}derhjelm,~P.; Karlstr{\"{o}}m,~G.; Ryde,~U. \emph{J. Chem. Phys.}
  \textbf{2006}, \emph{124}, 244101\relax
\mciteBstWouldAddEndPuncttrue
\mciteSetBstMidEndSepPunct{\mcitedefaultmidpunct}
{\mcitedefaultendpunct}{\mcitedefaultseppunct}\relax
\EndOfBibitem
\bibitem[Tkatchenko \latin{et~al.}(2012)Tkatchenko, Distasio, Car, and
  Scheffler]{Tkatchenko2012}
Tkatchenko,~A.; Distasio,~R.~A.; Car,~R.; Scheffler,~M. \emph{Phys. Rev. Lett.}
  \textbf{2012}, \emph{108}, 1--5\relax
\mciteBstWouldAddEndPuncttrue
\mciteSetBstMidEndSepPunct{\mcitedefaultmidpunct}
{\mcitedefaultendpunct}{\mcitedefaultseppunct}\relax
\EndOfBibitem
\bibitem[Tkatchenko and Scheffler(2009)Tkatchenko, and
  Scheffler]{Tkatchenko2009}
Tkatchenko,~A.; Scheffler,~M. \emph{Phys. Rev. Lett.} \textbf{2009},
  \emph{102}, 6--9\relax
\mciteBstWouldAddEndPuncttrue
\mciteSetBstMidEndSepPunct{\mcitedefaultmidpunct}
{\mcitedefaultendpunct}{\mcitedefaultseppunct}\relax
\EndOfBibitem
\bibitem[Cole \latin{et~al.}(2016)Cole, Vilseck, Tirado-Rives, Payne, and
  Jorgensen]{Cole2016}
Cole,~D.~J.; Vilseck,~J.~Z.; Tirado-Rives,~J.; Payne,~M.~C.; Jorgensen,~W.~L.
  \emph{J. Chem. Theory Comput.} \textbf{2016}, \emph{12}, 2312--2323\relax
\mciteBstWouldAddEndPuncttrue
\mciteSetBstMidEndSepPunct{\mcitedefaultmidpunct}
{\mcitedefaultendpunct}{\mcitedefaultseppunct}\relax
\EndOfBibitem
\bibitem[Manz and Sholl(2010)Manz, and Sholl]{Manz2010}
Manz,~T.~A.; Sholl,~D.~S. \emph{J. Chem. Theory Comput.} \textbf{2010},
  \emph{6}, 2455--2468\relax
\mciteBstWouldAddEndPuncttrue
\mciteSetBstMidEndSepPunct{\mcitedefaultmidpunct}
{\mcitedefaultendpunct}{\mcitedefaultseppunct}\relax
\EndOfBibitem
\bibitem[Manz and Sholl(2012)Manz, and Sholl]{Manz2012}
Manz,~T.~A.; Sholl,~D.~S. \emph{J. Chem. Theory Comput.} \textbf{2012},
  \emph{8}, 2844--2867\relax
\mciteBstWouldAddEndPuncttrue
\mciteSetBstMidEndSepPunct{\mcitedefaultmidpunct}
{\mcitedefaultendpunct}{\mcitedefaultseppunct}\relax
\EndOfBibitem
\bibitem[Yu \latin{et~al.}(2011)Yu, McDaniel, and Schmidt]{Yu2011}
Yu,~K.; McDaniel,~J.~G.; Schmidt,~J.~R. \emph{J. Phys. Chem. B} \textbf{2011},
  \emph{115}, 10054--10063\relax
\mciteBstWouldAddEndPuncttrue
\mciteSetBstMidEndSepPunct{\mcitedefaultmidpunct}
{\mcitedefaultendpunct}{\mcitedefaultseppunct}\relax
\EndOfBibitem
\bibitem[Levy \latin{et~al.}(1984)Levy, Perdew, and Sahni]{Levy1984}
Levy,~M.; Perdew,~J.~P.; Sahni,~V. \emph{Phys. Rev. A} \textbf{1984},
  \emph{30}, 2745--2748\relax
\mciteBstWouldAddEndPuncttrue
\mciteSetBstMidEndSepPunct{\mcitedefaultmidpunct}
{\mcitedefaultendpunct}{\mcitedefaultseppunct}\relax
\EndOfBibitem
\bibitem[Kitaigorodsky(2012)]{kitaigorodsky2012molecular}
Kitaigorodsky,~A. \emph{{Molecular crystals and Molecules}}; Physical
  Chemistry; Elsevier Science: New York, 2012\relax
\mciteBstWouldAddEndPuncttrue
\mciteSetBstMidEndSepPunct{\mcitedefaultmidpunct}
{\mcitedefaultendpunct}{\mcitedefaultseppunct}\relax
\EndOfBibitem
\bibitem[Lillestolen and Wheatley(2008)Lillestolen, and
  Wheatley]{Lillestolen2008}
Lillestolen,~T.~C.; Wheatley,~R.~J. \emph{Chem. Commun.} \textbf{2008},
  \emph{7345}, 5909--5911\relax
\mciteBstWouldAddEndPuncttrue
\mciteSetBstMidEndSepPunct{\mcitedefaultmidpunct}
{\mcitedefaultendpunct}{\mcitedefaultseppunct}\relax
\EndOfBibitem
\bibitem[Lillestolen and Wheatley(2009)Lillestolen, and
  Wheatley]{Lillestolen2009}
Lillestolen,~T.~C.; Wheatley,~R.~J. \emph{J. Chem. Phys.} \textbf{2009},
  \emph{131}, 144101\relax
\mciteBstWouldAddEndPuncttrue
\mciteSetBstMidEndSepPunct{\mcitedefaultmidpunct}
{\mcitedefaultendpunct}{\mcitedefaultseppunct}\relax
\EndOfBibitem
\bibitem[Hirshfeld(1977)]{Hirshfeld1977}
Hirshfeld,~F.~L. \emph{Theor. Chim. Acta} \textbf{1977}, \emph{44},
  129--138\relax
\mciteBstWouldAddEndPuncttrue
\mciteSetBstMidEndSepPunct{\mcitedefaultmidpunct}
{\mcitedefaultendpunct}{\mcitedefaultseppunct}\relax
\EndOfBibitem
\bibitem[Misquitta and Szalewicz(2002)Misquitta, and Szalewicz]{Misquitta2002}
Misquitta,~A.~J.; Szalewicz,~K. \emph{Chem. Phys. Lett.} \textbf{2002},
  \emph{357}, 301--306\relax
\mciteBstWouldAddEndPuncttrue
\mciteSetBstMidEndSepPunct{\mcitedefaultmidpunct}
{\mcitedefaultendpunct}{\mcitedefaultseppunct}\relax
\EndOfBibitem
\bibitem[Misquitta \latin{et~al.}(2003)Misquitta, Jeziorski, and
  Szalewicz]{Misquitta2003}
Misquitta,~A.~J.; Jeziorski,~B.; Szalewicz,~K. \emph{Phys. Rev. Lett.}
  \textbf{2003}, \emph{91}, 033201\relax
\mciteBstWouldAddEndPuncttrue
\mciteSetBstMidEndSepPunct{\mcitedefaultmidpunct}
{\mcitedefaultendpunct}{\mcitedefaultseppunct}\relax
\EndOfBibitem
\bibitem[Misquitta \latin{et~al.}(2005)Misquitta, Podeszwa, Jeziorski, and
  Szalewicz]{Misquitta2005}
Misquitta,~A.~J.; Podeszwa,~R.; Jeziorski,~B.; Szalewicz,~K. \emph{J. Chem.
  Phys.} \textbf{2005}, \emph{123}\relax
\mciteBstWouldAddEndPuncttrue
\mciteSetBstMidEndSepPunct{\mcitedefaultmidpunct}
{\mcitedefaultendpunct}{\mcitedefaultseppunct}\relax
\EndOfBibitem
\bibitem[He{\ss}elmann \latin{et~al.}(2005)He{\ss}elmann, Jansen, and
  Schütz]{Heßelmann2005a}
He{\ss}elmann,~A.; Jansen,~G.; Schütz,~M. \emph{J. Chem. Phys.}
  \textbf{2005}, \emph{122}, 014103\relax
\mciteBstWouldAddEndPuncttrue
\mciteSetBstMidEndSepPunct{\mcitedefaultmidpunct}
{\mcitedefaultendpunct}{\mcitedefaultseppunct}\relax
\EndOfBibitem
\bibitem[Podeszwa \latin{et~al.}(2006)Podeszwa, Bukowski, and
  Szalewicz]{Podeszwa2006a}
Podeszwa,~R.; Bukowski,~R.; Szalewicz,~K. \emph{J. Chem. Theory Comput.}
  \textbf{2006}, \emph{2}, 400--412\relax
\mciteBstWouldAddEndPuncttrue
\mciteSetBstMidEndSepPunct{\mcitedefaultmidpunct}
{\mcitedefaultendpunct}{\mcitedefaultseppunct}\relax
\EndOfBibitem
\bibitem[He{\ss}elmann and Jansen(2002)He{\ss}elmann, and
  Jansen]{Heßelmann2002}
He{\ss}elmann,~A.; Jansen,~G. \emph{Chem. Phys. Lett.} \textbf{2002},
  \emph{362}, 319--325\relax
\mciteBstWouldAddEndPuncttrue
\mciteSetBstMidEndSepPunct{\mcitedefaultmidpunct}
{\mcitedefaultendpunct}{\mcitedefaultseppunct}\relax
\EndOfBibitem
\bibitem[He{\ss}elmann and Jansen(2003)He{\ss}elmann, and
  Jansen]{Heßelmann2003}
He{\ss}elmann,~A.; Jansen,~G. \emph{Chem. Phys. Lett.} \textbf{2003},
  \emph{367}, 778--784\relax
\mciteBstWouldAddEndPuncttrue
\mciteSetBstMidEndSepPunct{\mcitedefaultmidpunct}
{\mcitedefaultendpunct}{\mcitedefaultseppunct}\relax
\EndOfBibitem
\bibitem[He{\ss}elmann and Jansen(2002)He{\ss}elmann, and
  Jansen]{Heßelmann2002a}
He{\ss}elmann,~A.; Jansen,~G. \emph{Chem. Phys. Lett.} \textbf{2002},
  \emph{357}, 464--470\relax
\mciteBstWouldAddEndPuncttrue
\mciteSetBstMidEndSepPunct{\mcitedefaultmidpunct}
{\mcitedefaultendpunct}{\mcitedefaultseppunct}\relax
\EndOfBibitem
\bibitem[Jansen \latin{et~al.}(2001)Jansen, Hesselmann, Williams, and
  Chabalowski]{Jansen2001}
Jansen,~G.; Hesselmann,~A.; Williams,~H.~L.; Chabalowski,~C.~F. \emph{J. Phys.
  Chem. A} \textbf{2001}, \emph{105}, 11156--11158\relax
\mciteBstWouldAddEndPuncttrue
\mciteSetBstMidEndSepPunct{\mcitedefaultmidpunct}
{\mcitedefaultendpunct}{\mcitedefaultseppunct}\relax
\EndOfBibitem
\bibitem[Podeszwa and Szalewicz(2005)Podeszwa, and Szalewicz]{Podeszwa2005a}
Podeszwa,~R.; Szalewicz,~K. {Accurate interaction energies from perturbation
  theory based on Kohn-Sham model}. 2005;
  \url{http://arxiv.org/abs/physics/0501023}\relax
\mciteBstWouldAddEndPuncttrue
\mciteSetBstMidEndSepPunct{\mcitedefaultmidpunct}
{\mcitedefaultendpunct}{\mcitedefaultseppunct}\relax
\EndOfBibitem
\bibitem[Jeziorska \latin{et~al.}(1987)Jeziorska, Jeziorski, and
  {\v{C}}{\'{i}}{\v{z}}ek]{Jeziorska1987}
Jeziorska,~M.; Jeziorski,~B.; {\v{C}}{\'{i}}{\v{z}}ek,~J. \emph{Int. J. Quantum
  Chem.} \textbf{1987}, \emph{32}, 149--164\relax
\mciteBstWouldAddEndPuncttrue
\mciteSetBstMidEndSepPunct{\mcitedefaultmidpunct}
{\mcitedefaultendpunct}{\mcitedefaultseppunct}\relax
\EndOfBibitem
\bibitem[Drude \latin{et~al.}(1902)Drude, Riborg, and
  Millikan]{drude1902theory}
Drude,~P.; Riborg,~C.; Millikan,~R.~A. \emph{{The Theory of Optics...
  Translated from the German by CR Mann and RA Millikan}}; London; New York
  [printed], 1902\relax
\mciteBstWouldAddEndPuncttrue
\mciteSetBstMidEndSepPunct{\mcitedefaultmidpunct}
{\mcitedefaultendpunct}{\mcitedefaultseppunct}\relax
\EndOfBibitem
\bibitem[Lamoureux and Roux(2003)Lamoureux, and Roux]{Lamoureux2003}
Lamoureux,~G.; Roux,~B. \emph{J. Chem. Phys.} \textbf{2003}, \emph{119},
  3025\relax
\mciteBstWouldAddEndPuncttrue
\mciteSetBstMidEndSepPunct{\mcitedefaultmidpunct}
{\mcitedefaultendpunct}{\mcitedefaultseppunct}\relax
\EndOfBibitem
\bibitem[Joh(2015)]{Johnson2015NIST}
{NIST Computational Chemistry Comparison and Benchmark Database, NIST Standard
  Reference Database Number 101}. 2015; \url{http://cccbdb.nist.gov/}\relax
\mciteBstWouldAddEndPuncttrue
\mciteSetBstMidEndSepPunct{\mcitedefaultmidpunct}
{\mcitedefaultendpunct}{\mcitedefaultseppunct}\relax
\EndOfBibitem
\bibitem[Shoemake(1992)]{Shoemake1992}
Shoemake,~K. \emph{Graph. Gems 3}; 1992; Chapter 6, pp 124--132\relax
\mciteBstWouldAddEndPuncttrue
\mciteSetBstMidEndSepPunct{\mcitedefaultmidpunct}
{\mcitedefaultendpunct}{\mcitedefaultseppunct}\relax
\EndOfBibitem
\bibitem[Werner \latin{et~al.}(2012)Werner, Knowles, Knizia, Manby, and
  Sch{\"{u}}tz]{MOLPRO-WIREs}
Werner,~H.-J.; Knowles,~P.~J.; Knizia,~G.; Manby,~F.~R.; Sch{\"{u}}tz,~M.
  \emph{WIREs Comput Mol Sci} \textbf{2012}, \emph{2}, 242--253\relax
\mciteBstWouldAddEndPuncttrue
\mciteSetBstMidEndSepPunct{\mcitedefaultmidpunct}
{\mcitedefaultendpunct}{\mcitedefaultseppunct}\relax
\EndOfBibitem
\bibitem[Misquitta and Stone(2015)Misquitta, and Stone]{camcasp5.8}
Misquitta,~A.~J.; Stone,~A.~J. {CamCASP: a program for studying intermolecular
  interactions and for the calculation of molecular properties in distributed
  form, version 5.8}. University of Cambridge, 2015\relax
\mciteBstWouldAddEndPuncttrue
\mciteSetBstMidEndSepPunct{\mcitedefaultmidpunct}
{\mcitedefaultendpunct}{\mcitedefaultseppunct}\relax
\EndOfBibitem
\bibitem[Aidas \latin{et~al.}(2014)Aidas, Angeli, Bak, Bakken, Bast, Boman,
  Christiansen, Cimiraglia, Coriani, Dahle, Dalskov, Ekstr{\"{o}}m, Enevoldsen,
  Eriksen, Ettenhuber, Fern{\'{a}}ndez, Ferrighi, Fliegl, Frediani, Hald,
  Halkier, H{\"{a}}ttig, Heiberg, Helgaker, Hennum, Hettema, Hjerten{\ae}s,
  H{\o}st, H{\o}yvik, Iozzi, Jans{\'{i}}k, Jensen, Jonsson, J{\o}rgensen,
  Kauczor, Kirpekar, Kj{\ae}rgaard, Klopper, Knecht, Kobayashi, Koch, Kongsted,
  Krapp, Kristensen, Ligabue, Lutn{\ae}s, Melo, Mikkelsen, Myhre, Neiss,
  Nielsen, Norman, Olsen, Olsen, Osted, Packer, Pawlowski, Pedersen, Provasi,
  Reine, Rinkevicius, Ruden, Ruud, Rybkin, Sa{\l}ek, Samson, de~Mer{\'{a}}s,
  Saue, Sauer, Schimmelpfennig, Sneskov, Steindal, Sylvester-Hvid, Taylor,
  Teale, Tellgren, Tew, Thorvaldsen, Th{\o}gersen, Vahtras, Watson, Wilson,
  Ziolkowski, and {\AA}gren]{WCMS:WCMS1172}
Aidas,~K.; Angeli,~C.; Bak,~K.~L.; Bakken,~V.; Bast,~R.; Boman,~L.;
  Christiansen,~O.; Cimiraglia,~R.; Coriani,~S.; Dahle,~P.; Dalskov,~E.~K.;
  Ekstr{\"{o}}m,~U.; Enevoldsen,~T.; Eriksen,~J.~J.; Ettenhuber,~P.;
  Fern{\'{a}}ndez,~B.; Ferrighi,~L.; Fliegl,~H.; Frediani,~L.; Hald,~K.;
  Halkier,~A.; H{\"{a}}ttig,~C.; Heiberg,~H.; Helgaker,~T.; Hennum,~A.~C.;
  Hettema,~H.; Hjerten{\ae}s,~E.; H{\o}st,~S.; H{\o}yvik,~I.-M.; Iozzi,~M.~F.;
  Jans{\'{i}}k,~B.; Jensen,~H. J.~A.; Jonsson,~D.; J{\o}rgensen,~P.;
  Kauczor,~J.; Kirpekar,~S.; Kj{\ae}rgaard,~T.; Klopper,~W.; Knecht,~S.;
  Kobayashi,~R.; Koch,~H.; Kongsted,~J.; Krapp,~A.; Kristensen,~K.;
  Ligabue,~A.; Lutn{\ae}s,~O.~B.; Melo,~J.~I.; Mikkelsen,~K.~V.; Myhre,~R.~H.;
  Neiss,~C.; Nielsen,~C.~B.; Norman,~P.; Olsen,~J.; Olsen,~J. M.~H.; Osted,~A.;
  Packer,~M.~J.; Pawlowski,~F.; Pedersen,~T.~B.; Provasi,~P.~F.; Reine,~S.;
  Rinkevicius,~Z.; Ruden,~T.~A.; Ruud,~K.; Rybkin,~V.~V.; Sa{\l}ek,~P.;
  Samson,~C. C.~M.; de~Mer{\'{a}}s,~A.~S.; Saue,~T.; Sauer,~S. P.~A.;
  Schimmelpfennig,~B.; Sneskov,~K.; Steindal,~A.~H.; Sylvester-Hvid,~K.~O.;
  Taylor,~P.~R.; Teale,~A.~M.; Tellgren,~E.~I.; Tew,~D.~P.; Thorvaldsen,~A.~J.;
  Th{\o}gersen,~L.; Vahtras,~O.; Watson,~M.~A.; Wilson,~D. J.~D.;
  Ziolkowski,~M.; {\AA}gren,~H. \emph{Wiley Interdiscip. Rev. Comput. Mol.
  Sci.} \textbf{2014}, \emph{4}, 269--284\relax
\mciteBstWouldAddEndPuncttrue
\mciteSetBstMidEndSepPunct{\mcitedefaultmidpunct}
{\mcitedefaultendpunct}{\mcitedefaultseppunct}\relax
\EndOfBibitem
\bibitem[Stone \latin{et~al.}(2015)Stone, Dullweber, Engkvist, Fraschini,
  Hodges, Meredith, Nutt, Popelier, and Wales]{orient4.8}
Stone,~A.~J.; Dullweber,~A.; Engkvist,~O.; Fraschini,~E.; Hodges,~M.~P.;
  Meredith,~A.~W.; Nutt,~D.~R.; Popelier,~P. L.~A.; Wales,~D.~J. {ORIENT: a
  program for studying interactions between molecules, version 4.8}. 2015;
  \url{http://www-stone.ch.cam.ac.uk/programs/orient.html}\relax
\mciteBstWouldAddEndPuncttrue
\mciteSetBstMidEndSepPunct{\mcitedefaultmidpunct}
{\mcitedefaultendpunct}{\mcitedefaultseppunct}\relax
\EndOfBibitem
\bibitem[Ferenczy \latin{et~al.}(1997)Ferenczy, Winn, and
  Reynolds]{Ferenczy1997}
Ferenczy,~G.~G.; Winn,~P.~J.; Reynolds,~C.~a. \emph{J. Phys. Chem. A}
  \textbf{1997}, \emph{101}, 5446--5455\relax
\mciteBstWouldAddEndPuncttrue
\mciteSetBstMidEndSepPunct{\mcitedefaultmidpunct}
{\mcitedefaultendpunct}{\mcitedefaultseppunct}\relax
\EndOfBibitem
\bibitem[Eastman \latin{et~al.}(2013)Eastman, Friedrichs, Chodera, Radmer,
  Bruns, Ku, Beauchamp, Lane, Wang, Shukla, Tye, Houston, Stich, Klein, Shirts,
  and Pande]{Eastman2013}
Eastman,~P.; Friedrichs,~M.~S.; Chodera,~J.~D.; Radmer,~R.~J.; Bruns,~C.~M.;
  Ku,~J.~P.; Beauchamp,~K.~A.; Lane,~T.~J.; Wang,~L.~P.; Shukla,~D.; Tye,~T.;
  Houston,~M.; Stich,~T.; Klein,~C.; Shirts,~M.~R.; Pande,~V.~S. \emph{J. Chem.
  Theory Comput.} \textbf{2013}, \emph{9}, 461--469\relax
\mciteBstWouldAddEndPuncttrue
\mciteSetBstMidEndSepPunct{\mcitedefaultmidpunct}
{\mcitedefaultendpunct}{\mcitedefaultseppunct}\relax
\EndOfBibitem
\bibitem[Jorgensen \latin{et~al.}(1996)Jorgensen, Maxwell, and
  Tirado-Rives]{Jorgensen1996}
Jorgensen,~W.~L.; Maxwell,~D.~S.; Tirado-Rives,~J. \emph{J. Am. Chem. Soc.}
  \textbf{1996}, \emph{118}, 11225--11236\relax
\mciteBstWouldAddEndPuncttrue
\mciteSetBstMidEndSepPunct{\mcitedefaultmidpunct}
{\mcitedefaultendpunct}{\mcitedefaultseppunct}\relax
\EndOfBibitem
\bibitem[Sebetci and Beran(2010)Sebetci, and Beran]{Sebetci2010}
Sebetci,~A.; Beran,~G. J.~O. \emph{J. Chem. Theory Comput.} \textbf{2010},
  \emph{6}, 155--167\relax
\mciteBstWouldAddEndPuncttrue
\mciteSetBstMidEndSepPunct{\mcitedefaultmidpunct}
{\mcitedefaultendpunct}{\mcitedefaultseppunct}\relax
\EndOfBibitem
\bibitem[Misquitta \latin{et~al.}(2008)Misquitta, Welch, Stone, and
  Price]{Misquitta2008a}
Misquitta,~A.; Welch,~G.; Stone,~A.; Price,~S. \textbf{2008}, \emph{456},
  105--109\relax
\mciteBstWouldAddEndPuncttrue
\mciteSetBstMidEndSepPunct{\mcitedefaultmidpunct}
{\mcitedefaultendpunct}{\mcitedefaultseppunct}\relax
\EndOfBibitem
\bibitem[Price \latin{et~al.}(2010)Price, Leslie, Welch, Habgood, Price,
  Karamertzanis, and Day]{Price2010}
Price,~S.~L.; Leslie,~M.; Welch,~G. W.~A.; Habgood,~M.; Price,~L.~S.;
  Karamertzanis,~P.~G.; Day,~G.~M. \emph{Phys. Chem. Chem. Phys.}
  \textbf{2010}, \emph{12}, 8478--8490\relax
\mciteBstWouldAddEndPuncttrue
\mciteSetBstMidEndSepPunct{\mcitedefaultmidpunct}
{\mcitedefaultendpunct}{\mcitedefaultseppunct}\relax
\EndOfBibitem
\bibitem[Totton \latin{et~al.}(2010)Totton, Misquitta, and Kraft]{Totton2010a}
Totton,~T.~S.; Misquitta,~A.~J.; Kraft,~M. \emph{J. Chem. Theory Comput.}
  \textbf{2010}, \emph{6}, 683--695\relax
\mciteBstWouldAddEndPuncttrue
\mciteSetBstMidEndSepPunct{\mcitedefaultmidpunct}
{\mcitedefaultendpunct}{\mcitedefaultseppunct}\relax
\EndOfBibitem
\bibitem[Hermida-Ram{\'{o}}n and R{\'{i}}os(2000)Hermida-Ram{\'{o}}n, and
  R{\'{i}}os]{Hermida-Ramon2000}
Hermida-Ram{\'{o}}n,~J.~M.; R{\'{i}}os,~M.~A. \emph{Chem. Phys.} \textbf{2000},
  \emph{262}, 423--436\relax
\mciteBstWouldAddEndPuncttrue
\mciteSetBstMidEndSepPunct{\mcitedefaultmidpunct}
{\mcitedefaultendpunct}{\mcitedefaultseppunct}\relax
\EndOfBibitem
\bibitem[Nyeland(1990)]{Nyeland1990}
Nyeland,~C. \emph{Chem. Phys.} \textbf{1990}, \emph{147}, 229--240\relax
\mciteBstWouldAddEndPuncttrue
\mciteSetBstMidEndSepPunct{\mcitedefaultmidpunct}
{\mcitedefaultendpunct}{\mcitedefaultseppunct}\relax
\EndOfBibitem
\bibitem[Dymond and Smith(1980)Dymond, and Smith]{Dymond1980}
Dymond,~J.~H.; Smith,~E.~B. \emph{{The Virial Coefficients of Pure Gases and
  Mixtures}}, 2nd ed.; Clarendon Press: Berlin Heidelberg, 1980\relax
\mciteBstWouldAddEndPuncttrue
\mciteSetBstMidEndSepPunct{\mcitedefaultmidpunct}
{\mcitedefaultendpunct}{\mcitedefaultseppunct}\relax
\EndOfBibitem
\bibitem[Vogel \latin{et~al.}(2010)Vogel, J{\"{a}}ger, Hellmann, and
  Bich]{Vogel2010}
Vogel,~E.; J{\"{a}}ger,~B.; Hellmann,~R.; Bich,~E. \emph{Mol. Phys.}
  \textbf{2010}, \emph{108}, 3335--3352\relax
\mciteBstWouldAddEndPuncttrue
\mciteSetBstMidEndSepPunct{\mcitedefaultmidpunct}
{\mcitedefaultendpunct}{\mcitedefaultseppunct}\relax
\EndOfBibitem
\bibitem[McDaniel and Schmidt(2014)McDaniel, and Schmidt]{McDaniel2014}
McDaniel,~J.~G.; Schmidt,~J.~R. \emph{J. Phys. Chem. B} \textbf{2014},
  \emph{118}, 8042--8053\relax
\mciteBstWouldAddEndPuncttrue
\mciteSetBstMidEndSepPunct{\mcitedefaultmidpunct}
{\mcitedefaultendpunct}{\mcitedefaultseppunct}\relax
\EndOfBibitem
\bibitem[McDaniel \latin{et~al.}(2012)McDaniel, Yu, and Schmidt]{McDaniel2012a}
McDaniel,~J.~G.; Yu,~K.; Schmidt,~J.~R. \emph{J. Phys. Chem. C} \textbf{2012},
  \emph{116}, 1892--1903\relax
\mciteBstWouldAddEndPuncttrue
\mciteSetBstMidEndSepPunct{\mcitedefaultmidpunct}
{\mcitedefaultendpunct}{\mcitedefaultseppunct}\relax
\EndOfBibitem
\bibitem[Witt and Kemp(1937)Witt, and Kemp]{Witt1937}
Witt,~R.~K.; Kemp,~J.~D. \emph{J. Am. Chem. Soc.} \textbf{1937}, \emph{59},
  273--276\relax
\mciteBstWouldAddEndPuncttrue
\mciteSetBstMidEndSepPunct{\mcitedefaultmidpunct}
{\mcitedefaultendpunct}{\mcitedefaultseppunct}\relax
\EndOfBibitem
\bibitem[Riddick \latin{et~al.}(1986)Riddick, Bunger, and Sakano]{Riddick1986}
Riddick,~J.~A.; Bunger,~W.~B.; Sakano,~T.~K. \emph{{Techniques of Chemistry,
  Vol. II: Organic Solvents: Physical Properties and Methods of Purification}},
  4th ed.; Wiley-Interscience: New York, 1986\relax
\mciteBstWouldAddEndPuncttrue
\mciteSetBstMidEndSepPunct{\mcitedefaultmidpunct}
{\mcitedefaultendpunct}{\mcitedefaultseppunct}\relax
\EndOfBibitem
\bibitem[Mayo \latin{et~al.}(1990)Mayo, Olafson, and Goddard]{Mayo1990}
Mayo,~S.~L.; Olafson,~B.~D.; Goddard,~W. A.~I. \emph{J. Phys. Chem.}
  \textbf{1990}, \emph{101}, 8897--8909\relax
\mciteBstWouldAddEndPuncttrue
\mciteSetBstMidEndSepPunct{\mcitedefaultmidpunct}
{\mcitedefaultendpunct}{\mcitedefaultseppunct}\relax
\EndOfBibitem
\bibitem[Lim(2009)]{Lim2009}
Lim,~T. \emph{Zeitschrift f{\"{u}}r Naturforschung-A} \textbf{2009}, \emph{64},
  200--204\relax
\mciteBstWouldAddEndPuncttrue
\mciteSetBstMidEndSepPunct{\mcitedefaultmidpunct}
{\mcitedefaultendpunct}{\mcitedefaultseppunct}\relax
\EndOfBibitem
\bibitem[{Van Duin} \latin{et~al.}(2001){Van Duin}, Dasgupta, Lorant, and
  Goddard]{VanDuin2001}
{Van Duin},~a. C.~T.; Dasgupta,~S.; Lorant,~F.; Goddard,~W.~a. \emph{J. Phys.
  Chem. A} \textbf{2001}, \emph{105}, 9396--9409\relax
\mciteBstWouldAddEndPuncttrue
\mciteSetBstMidEndSepPunct{\mcitedefaultmidpunct}
{\mcitedefaultendpunct}{\mcitedefaultseppunct}\relax
\EndOfBibitem
\bibitem[Eramian \latin{et~al.}(2013)Eramian, Tian, Fox, Beneberu, and
  Kertesz]{Eramian2013}
Eramian,~H.; Tian,~Y.-H.; Fox,~Z.; Beneberu,~H.~Z.; Kertesz,~M. \emph{J. Phys.
  Chem. A} \textbf{2013}, \emph{117}, 14184--14190\relax
\mciteBstWouldAddEndPuncttrue
\mciteSetBstMidEndSepPunct{\mcitedefaultmidpunct}
{\mcitedefaultendpunct}{\mcitedefaultseppunct}\relax
\EndOfBibitem
\bibitem[Badenhoop and Weinhold(1997)Badenhoop, and Weinhold]{Badenhoop1997}
Badenhoop,~J.~K.; Weinhold,~F. \emph{J. Chem. Phys.} \textbf{1997}, \emph{107},
  5422\relax
\mciteBstWouldAddEndPuncttrue
\mciteSetBstMidEndSepPunct{\mcitedefaultmidpunct}
{\mcitedefaultendpunct}{\mcitedefaultseppunct}\relax
\EndOfBibitem
\bibitem[Kim \latin{et~al.}(2014)Kim, Doan, Cho, Madhav, and Kim]{Kim2014b}
Kim,~H.; Doan,~V.~D.; Cho,~W.~J.; Madhav,~M.~V.; Kim,~K.~S. \emph{Sci. Rep.}
  \textbf{2014}, \emph{4}, 1--8\relax
\mciteBstWouldAddEndPuncttrue
\mciteSetBstMidEndSepPunct{\mcitedefaultmidpunct}
{\mcitedefaultendpunct}{\mcitedefaultseppunct}\relax
\EndOfBibitem
\bibitem[Wheatley and Gopal(2012)Wheatley, and Gopal]{Wheatley2012}
Wheatley,~R.~J.; Gopal,~A.~A. \emph{Phys. Chem. Chem. Phys.} \textbf{2012},
  \emph{14}, 2087--2091\relax
\mciteBstWouldAddEndPuncttrue
\mciteSetBstMidEndSepPunct{\mcitedefaultmidpunct}
{\mcitedefaultendpunct}{\mcitedefaultseppunct}\relax
\EndOfBibitem
\end{mcitethebibliography}
